\documentclass[12pt]{article}
\usepackage{amsfonts}
\usepackage{amsmath, amssymb}
\usepackage{youngtab}
\usepackage{hyperref}
\usepackage{graphicx, subfigure}
\usepackage{psfrag}
\Yboxdim{5pt}

\newcommand{\ra}{\rightarrow}

\newcommand{\be}{\begin{equation}}
\newcommand{\ee}{\end{equation}}
\newcommand{\ba}{\begin{eqnarray}}
\newcommand{\ea}{\end{eqnarray}}
\newcommand{\bi}{\begin{itemize}}  
\newcommand{\ei}{\end{itemize}}

\newcommand{\Tr}{{\rm Tr}}

\newcommand{\p}{\partial}
\newcommand{\ep}{\epsilon}
\newcommand{\ve}{\varepsilon}

\newcommand{\Dcal}{{\mathcal D}}

\newcommand{\Kahler}{K\"{a}hler }

\newcommand{\mo}{{-1}} 

\newcommand{\aslash}[1]{\,\,{\raise.15ex\hbox{/}\mkern-12mu #1}}
\newcommand{\bslash}[1]{\,\,{\raise.15ex\hbox{/}\mkern-11mu #1}}
\newcommand{\cslash}[1]{\,\,{\raise.15ex\hbox{/}\mkern-10mu #1}}
\newcommand{\dslash}[1]{\,\,{\raise.15ex\hbox{/}\mkern-9mu #1}}

\renewcommand{\bar}{\overline}
\renewcommand{\tilde}{\widetilde}
\renewcommand{\hat}{\widehat}

\newcommand{\sfb}{{\mathsf b}}

\renewcommand{\title}[1]{\vbox{\center\LARGE{#1}}\vspace{5mm}}
\renewcommand{\author}[1]{\vbox{\center\large#1}\vspace{5mm}}
\newcommand{\address}[1]{\vbox{\center\em#1}}

\numberwithin{equation}{section} 

\begin{document}
\bibliographystyle{utphys}

\begin{titlepage}
\begin{center}
\vspace{5mm}
\hfill {\tt UT-Komaba/11-12}\\
\hfill {\tt YITP-11-96}\\
\vspace{16mm}

\title{
Line operators on $S^1\times \mathbb R^3$
and
\\ 
quantization of the Hitchin moduli space
}
\vspace{6mm}

Yuto Ito${}^{a}$\footnote{\href{mailto:ito@hep1.c.u-tokyo.ac.jp}
{\tt ito@hep1.c.u-tokyo.ac.jp}},
Takuya Okuda${}^{a}$\footnote{\href{mailto:takuya@hep1.c.u-tokyo.ac.jp}
{\tt takuya@hep1.c.u-tokyo.ac.jp}},
and
Masato Taki${}^{b}$\footnote{\href{mailto:taki@yukawa.kyoto-u.ac.jp}
{\tt taki@yukawa.kyoto-u.ac.jp}}
\vskip 5mm
\address{
${}^a$Institute of Physics, University of Tokyo,\\
Komaba, Meguro-ku, Tokyo 153-8902, Japan
}
\address{
${}^b$Yukawa Institute for Theoretical Physics, \\
Kyoto University, Sakyo-ku, Kyoto 606-8502, Japan}

\end{center}

\vspace{8mm}
\abstract{
\noindent
We perform an exact localization calculation
for the expectation values of Wilson-'t Hooft line operators
in $\mathcal N=2$ gauge theories on $S^1\times \mathbb R^3$.
The expectation values are naturally expressed
in terms of the complexified Fenchel-Nielsen coordinates,
and form a quantum mechanically deformed
algebra of functions on the associated 
Hitchin moduli space by Moyal multiplication.
We propose that these expectation values
are the Weyl transform of the Verlinde operators, which 
 act on Liouville/Toda conformal blocks as difference operators.
We demonstrate our proposal explicitly in $SU(N)$ examples.

}
\vfill

\end{titlepage}

\tableofcontents

\setcounter{footnote}{0}

\section{Introduction and summary}
\label{sec:Introduction}

Wilson loops and their magnetic cousins, 't Hooft loops,
are universal observables in gauge theories
whose properties characterize the phases of each theory.
They represent heavy probe particles with electric and magnetic charges
moving along a closed trajectory in spacetime.
When acting on the Hilbert space, these operators do not commute
if the two loops are linked within the constant time slice
\cite{'tHooft:1977hy}.
Indeed 't Hooft successfully used their non-trivial commutation relations
to classify the possible phases of non-Abelian gauge theories
\cite{'tHooft:1977hy,'tHooft:1979uj,tHooft:1981ht}.

Noncommutativity is also a hallmark of quantization.
The position and the momentum of a particle do not commute with each other;
they cannot be simultaneously diagonalized or precisely measured.
For physicists quantization is usually the process of obtaining
a Hilbert space and noncommuting operators acting on it from a classical system.
In certain situations (especially for mathematicians), however, one is 
primarily interested in the ``operators'' without a Hilbert space.
In such a scheme, called deformation quantization,
the product of two functions on a phase space (Poisson manifold)
is continuously (in $\hbar$) deformed into a noncommutative associative product
whose order $\mathcal O(\hbar)$ correction is given by the Poisson bracket.
It is a non-trivial result that 
any Poisson structure admits a canonical deformation quantization \cite{Kontsevich:1997vb}.

In this paper we study Wilson-'t Hooft line operators in
$\mathcal N=2$ supersymmetric gauge theories on $S^1\times\mathbb R^3$
in the Coulomb phase.
We consider half-BPS line operators 
\cite{Kapustin:2005py} extended along $S^1$
and perform an exact localization calculation of their expectation values
(vevs) following \cite{Pestun:2007rz}.
The vev of the product of operators turns out to
be given by the Moyal product of the vevs of the individual operators.
For an $\mathcal N=2$ theory characterized by a punctured Riemann surface
\cite{Gaiotto:2009we}, the line operators precisely realize
the deformation quantization of the Hitchin moduli space, with respect to the
Poisson structure specified by the complexified Fenchel-Nielsen coordinates
\cite{Nekrasov:2011bc,Dimofte:2011jd}.

Let us summarize our main results in more detail.
The vev of a line operator is a holomorphic function
of $a$ and $b$, which take values
in the complexified Cartan subalgebra $\mathfrak t_{\mathbb C}$
and its dual $\mathfrak t^*_{\mathbb C}$.
The variable $a$ is a combination
of the electric Wilson line $A_\tau$ and a real vector multiplet scalar,
while $b$ combines the magnetic Wilson line
and the other real scalar in the vector multiplet, all evaluated at 
the infinity of $\mathbb R^3$.
These variables $a$ and $b$ parametrize the Coulomb branch of the gauge theory compactified
on $S^1$.
Since the path integral for the vev defines
a supersymmetric index (\ref{trace-def}), the electric and magnetic Wilson lines
can be regarded as chemical potentials for electric and magnetic charges.
The vev also depends holomorphically on the non-dynamical variables
$m_f$, which are complex combinations of masses 
and chemical potentials for flavor symmetries.
Importantly, a real parameter $\lambda$ also enters in the vev.
It is defined as the chemical potential for the simultaneous 
spatial and R-symmetry rotations.

We find that the vev of the Wilson operator in representation $R$
is simply given by 
\begin{equation}
  \langle W_R\rangle= \Tr_R\, e^{2\pi i a}\,,
\end{equation}
where the trace is taken in $R$.
The localization calculation is only non-trivial for 't Hooft and dyonic
line operators.
In particular, they have a non-trivial one-loop determinant as well
as the non-perturbative contributions from Polyakov-'t Hooft monopoles
screening the charges of the singular monopole \cite{Kapustin:2006pk,Gomis:2011pf}.
The 't Hooft operator specified by a coweight $B$ has a vev of the form
\begin{equation}
  \langle T_B\rangle= 
\sum_v e^{2\pi i v\cdot b}
Z_\text{1-loop}(a,m_f,\lambda;v)
Z_\text{mono}(a,m_f,\lambda;B,v)\,.
\label{eq:intro-thooft}
\end{equation}
For simplicity we often suppress the dependence on some of $a$, $m_f$ and $\lambda$.
The function $Z_\text{1-loop}(v=B)$  given in (\ref{1-loop-vm}-\ref{1-loop-total})
 is the one-loop determinant
around the leading saddle point.
The sum is over the magnetic charges $v$ reduced from $B$ due to monopole screening.
There exists a number of non-perturbative 
saddle points that correspond to coweight $v$,
and $Z_\text{1-loop}(v)Z_\text{mono}(B,v)$
is the sum of the fluctuation determinants around the saddle points
in the sector $v$.
The function $Z_\text{mono}(B,v)$, given in 
(\ref{mono-adj}, \ref{mono-fund}, \ref{mono-total})
for $G=U(N)$ and matter in the adjoint or fundamental representation, 
is a monopole analog of the Nekrasov
instanton partition function \cite{Nekrasov:2002qd}.
For a more general dyonic Wilson-'t Hooft operator,
we insert into (\ref{eq:intro-thooft}) 
a Wilson operator in the subgroup of the gauge group unbroken by $B$.

Let us suppose that the spatial rotation associated with $\lambda$
takes place in the 12-plane.
It is useful to think of the 3-axis as the Euclidean time direction,
and consider line operators $L_i$ at various points on the axis.
By the original argument of 't Hooft, we show that
these operators form a noncommutative algebra, generalizing the standard
't Hooft commutation relation $W\cdot T=e^{2\pi i/N} T\cdot W$
for minimal Wilson ($W$) and 't Hooft ($T$) loops in $SU(N)$ gauge theories.\footnote{Non-commutativity can also be understood by thinking of $S^1$ as time.
The electric and magnetic fields produced by the Wilson and 't Hooft operators
generate a non-zero Poyinting vector carrying angular momentum,
and contribute non-trivially to the supersymmetric index with $\lambda \neq 0$
\cite{Gaiotto:2010be}.
}
Moreover, operator multiplication is implemented by
noncommutative associative products, i.e.,
\begin{equation}
  \langle L_1\cdot L_2\cdot\ldots \cdot L_n\rangle
= \langle L_1\rangle *\langle L_2\rangle *
\ldots *
\langle L_n\rangle\,,
\end{equation}
where
\begin{equation}
(f*g)(a,b)\equiv 
\left.
e^{i\frac{\lambda}{4\pi}
(
\partial_{b}\cdot \partial_{a'}
-
\partial_a\cdot \partial_{b'}
)
}f(a,b) g(a',b')
\right|_{a'=a,b'=b}\,.
\end{equation}
This is the Moyal product associated with the 
Poisson structure determined by the holomorphic symplectic form
\begin{equation}
  da\wedge db
\label{eq:sympl-intro}
\end{equation}
and $\hbar=\frac{\lambda}{2\pi}$,
where $a$ and $b$ are contracted in the canonical way.

It is illuminating to be even more explicit, focusing
on the $SU(2)$ $\mathcal N=2^*$ theory.
In this case the expectation values of the minimal 
Wilson ($W$), 't Hooft ($T$), and dyonic ($D$) operators are given by\footnote{For $SU(2)$ we simplify notation by substituting
$a\ra \text{diag}(a,-a),
b\ra \text{diag}(b,-b)$.
When there is only one mass parameter, $m\equiv m_{f=1}$.
}
\begin{equation}
  \begin{aligned}
&  \langle W\rangle=\, e^{2\pi i a}+e^{-2\pi i a}\,,
\quad\langle T\rangle=
\left( e^{2\pi i b}+e^{-2\pi i b}\right)
\left(
\frac{
\sin\left(2\pi a+\pi m \right)
\sin\left(2\pi a-\pi m\right)
}{
\sin\left(2\pi a+\frac{\pi}2 \lambda\right)
\sin\left(2\pi a-\frac{\pi}2 \lambda\right)
}
\right)^{1/2}\hspace{-3mm}
\,,
\\
&\hspace{20mm}
\langle D\rangle
=
\left( e^{2\pi i (b+a)}+e^{-2\pi i (b+a)}\right)
\left(
\frac{
\sin\left(2\pi a+\pi m \right)
\sin\left(2\pi a-\pi m\right)
}{
\sin\left(2\pi a+\frac{\pi}2 \lambda\right)
\sin\left(2\pi a-\frac{\pi}2 \lambda\right)
}
\right)^{1/2}\,.
  \end{aligned}
\end{equation}
For $\lambda=0$, these expressions precisely appeared in \cite{Nekrasov:2011bc}
as the definition of Darboux coordinates $(a,b)$ on the Hitchin
moduli space for a one-punctured torus.
In \cite{Dimofte:2011jd},
they
were identified
as the complexification of the Fenchel-Nielsen coordinates,
which are Darboux coordinates for the real symplectic structure 
on Teichm\"uller space.
Their findings are consistent with our identification 
(\ref{eq:sympl-intro}) of the symplectic structure.
For $\lambda\neq 0$, our results provide quantum deformations.

Thus $\mathcal N=2$ gauge theories on $S^1\times \mathbb R^3$
produce a noncommutative algebra of operators quantizing the Hitchin moduli space.
Is there a Hilbert space on which the line operators
naturally act?
We claim that the space of conformal blocks in Liouville or Toda conformal
field theories
is such a Hilbert space.
This is demonstrated by showing that the Verlinde operators\footnote{Verlinde operators are the difference operators that act on conformal blocks, and
arise from the monodromy of extended conformal blocks with degenerate field insertions.
Verlinde operators in Liouville theory coincide with the geodesic length operators in quantum Teichm\"uller theory \cite{Teschner:2005bz}.
}
\cite{Drukker:2009id,Alday:2009fs},
labeled by closed curves on the Riemann surface
and corresponding to line operators in gauge theories \cite{Drukker:2009tz}, 
are exactly the Weyl transform (also known as the Weyl ordering) of
the vevs of the line operators on $S^1\times \mathbb R^3$,
where $a$ and $b$ are treated as coordinates and momenta, respectively.
The twist/quantization variable $\lambda$ is related to the variable
${\mathsf b}$ that parametrizes the central charge $c=1+6({\mathsf b}+{\mathsf b}^{-1})^2$
as
$  \lambda={\mathsf b}^2 $.
This result is a concrete realization of the proposal
that the algebra of line operators provide quantization
of the Hitchin moduli space
\cite{Nekrasov:2010ka,Teschner:2010je,Gaiotto:2010be,Dimofte:2011jd}.
The connection to Liouville/Toda theories provides a very strong
check of our localization computations.
Moreover, we conjecture that the connection should hold
even when $\mathcal N=2$ gauge theories have no Lagrangian description.
Thus it is now possible to compute the line operator vevs
on $S^1\times \mathbb R^3$ for such theories as
the inverse Weyl transform of the Verlinde operators.

Then the AGT relation \cite{AGT} between Liouville/Toda theories
and four-dimensional gauge theories
would suggest that our analysis should intimately parallel 
the localization computation of 't Hooft loops on $S^4$
\cite{Gomis:2011pf} corresponding to $\lambda={\mathsf b}^2=1$.
Indeed $Z_\text{1-loop}(\lambda;B)$ and $Z_\text{mono}(\lambda;B,v)$
with $\lambda=1$ appeared in
\cite{Gomis:2011pf} as the contributions from the equator $S^1$ of $S^4$,
where a 't Hooft loop was inserted.

Exactly the same physical system on $S^1\times \mathbb R^3$
 was considered
in  \cite{Gaiotto:2010be},
where supersymmetric line operators were
analyzed from the point of view of wall-crossing in the IR effective theories.
Their twist parameter $y$ is given by $y=-e^{\pi i\lambda}$.
Based on the consistency of the wall-crossing formula 
in $\mathcal N=2$ gauge theories and several other assumptions,
they conjectured expressions for the line operator vevs
in terms of the (commutative and noncommutative) Fock-Goncharov coordinates on
the Hitchin moduli space.
It would be desirable to perform more detailed comparisons.
This should help create a bridge between the AGT correspondence
\cite{AGT} and the study of wall-crossing, perhaps along the line of
\cite{Dimofte:2011ju}.

This paper is organized as follows.
Section \ref{sec:set-up} defines the gauge theory setup
and the quantities we wish to compute.
We begin our localization calculations in Section \ref{sec:localization},
where we analyze the symmetries of the system and lay out our strategy.
We also calculate the classical on-shell action
in the supersymmetric background defining a 't Hooft operator.
Section \ref{sec:one-loop} is devoted to the one-loop analysis.
In Section \ref{sec:mono} we compute the non-perturbative contributions due
to monopole screening.
Putting all together the classical, one-loop, and screening contributions,
Section \ref{sec:gauge-results} summarizes the results of our localization calculations and gives explicit expressions in several examples.
We then turn to the quantization aspects of our results.
In Section \ref{sec:non-com} we study the noncommutative structure
in the algebra formed by line operators and show that it implements
the deformation quantization of the Hitchin moduli space.
Next we discuss the relation to gauge theories on $S^4$ and Liouville/Toda
theories in Section \ref{sec:S4-CFT}.
We conclude the paper in Section \ref{sec:discussion}
with a discussion on related works and future directions.
Appendix \ref{sec:spinor-gamma} explains our convention
for spinors and gamma matrices.
In Appendix \ref{sec:mono-inst} we review Kronheimer's correspondence
between singular monopoles and $U(1)$-invariant instantons on a Taub-NUT space.
This relation is used in Sections \ref{sec:one-loop} 
and \ref{sec:mono}.
Appendix \ref{sec:CFT-details} contains technical computations in Liouville and
Toda theories.
In Appendix \ref{sec:cl-holo}
we compute the classical $SL(2,\mathbb C)$ holonomies
on the four-punctured sphere and compare them with
gauge and Liouville calculations.

\section{$\mathcal N=2$ gauge
theories on  $S^1\times \mathbb R^3$
and line operators}
\label{sec:set-up}

In this paper we study four-dimensional gauge theories with
$\mathcal N=2$ supersymmetry on $S^1\times \mathbb R^3$
in the Coulomb branch.
For notational convenience, we will use the 
notation appropriate for $\mathcal N=2^*$ theory,
which can be thought of as a dimensional reduction of
the ten-dimensional super Yang-Mills,
though we will state
 general results applicable to other field contents 
\cite{Pestun:2007rz,Gomis:2011pf}.
The ten-dimensional gauge field $A_M$ ($M=1,\ldots, 9,0$) gives rise
to the four-dimensional gauge field $A_\mu$ ($\mu=1, \ldots, 4$),
hypermultiplet scalars $A_i\equiv \Phi_i$ ($i=5,\ldots, 8$),
and vector multiplet scalars $A_A\equiv \Phi_A$ ($A=0,9$).
The ten-dimensional chiral spinor $\Psi$ also decomposes into
the gaugino $\psi\equiv \frac{1-\Gamma_{5678}}2 \Psi$
and hypermultiplet fermion $\chi\equiv \frac{1+\Gamma_{5678}}2 \Psi$.
Our spinor and gamma matrix conventions are summarized in
Appendix \ref{sec:spinor-gamma}.
Real fields are hermitian matrices,
and the gauge covariant derivative is $D_\mu=\p_\mu+iA_\mu$.
In terms of the coordinates $x^\mu=(x^i,\tau)$ 
($\mu=1,\ldots,4$, $i=1,2,3$),
the metric is simply $ds^2=d\tau^2+d x^i dx^i$.
We denote the radius of the Euclidean time circle by $R$.

The theory is defined by the physical action
\begin{equation}
    S=S_\text{vec}+S_\text{hyp}\,,
\label{total-action}
\end{equation}
where the two terms describing the vector and hypermultiplets are given by
\begin{eqnarray}
  S_\text{vec}
&=&\frac {1}{g^2} \int_{S^1\times \mathbb R^3} 
d^4x\,
\Tr \left  ( 
\frac 1 2
F_{\mu\nu} F^{\mu\nu}
+
D_\mu \Phi_A D^\mu \Phi_A
-[\Phi_0,\Phi_9]^2
-\psi \Gamma^\mu D_\mu\psi
- i \psi \Gamma^A [\Phi_A,  \psi]
\right)
\nonumber\\
&&\quad
+\frac{i\vartheta}{8\pi^2} \int_{S^1\times \mathbb R^3}\Tr \left( F\wedge F\right)
\,,
\label{action-vec}
\end{eqnarray}
and 
\begin{eqnarray}
  S_\text{hyp}
&=&\frac {1}{g^2} \int_{S^1\times \mathbb R^3} 
d^4x\,
\Tr \bigg( 
D_\mu \Phi_i D^\mu \Phi_i
-\frac 1 2[\Phi_i,\Phi_j]^2
-\left(
[\Phi_A, \Phi_i]
-i M_{A\,ij}\Phi_j
\right)^2
-\chi \Gamma^\mu D_\mu\chi
\nonumber\\
&&\quad\quad\quad\quad\quad\quad\quad\quad\quad
- i \chi 
\Gamma^A \left([\Phi_A,  \chi] -\frac{i}{4} M_{A\,ij}\Gamma^{ij} \chi
\right)
- i \chi \Gamma^i [\Phi_i,  \chi]
\bigg)
\,.
\label{action-hyp}
\end{eqnarray}
Here $\Tr$ denotes an invariant metric on the Lie algebra of
the gauge group $G$, $\vartheta$ is the theta angle,
and $i,j=5,6,7,8$ denote the hypermultiplet
scalar directions.
The two real anti-symmetric matrices $M_{ij}\equiv M_{0\,ij}$ and $M_{9\,ij}$ are
proportional to a single pure-imaginary anti-symmetric matrix
$F_{ij}$%
\footnote{%
The flavor symmetry generator $F_{ij}$ ($i,j=5,\ldots,7$)
should not be confused with the field strength
$F_{MN}=-i [D_M,D_N]$ ($M,N=1,\ldots,9,0$).
}, which is normalized
as $F_{ij}F_{ji}=4$ and is taken to be anti-self-dual in the $5678$ directions
so that only the hypermultiplet fermions get massive.
The flavor generator $F$ is represented as $F_{ij}$
on the scalars and as $\frac 1 4 F_{ij}\Gamma^{ij}$ on spinors.
The real mass parameters $M\equiv M_0$ and $M_9$ are defined by
$M_{A\,ij}=i M_A F_{ij}$ ($A=0,9$).
The massless limit is $\mathcal N=4$ super Yang-Mills.

Our aim is to compute the expectation value
of half-BPS line operators along $S^1$,
placed at a point on the 3-axis of $\mathbb R^3$.
The most basic line operator
is the Wilson operator defined as
\begin{eqnarray}
 W_R=\Tr_R P \exp\oint_{S^1}\left(-iA+\Phi_0\right) d\tau\,.
\label{wilson-def}
\end{eqnarray}
This is labeled by the representation $R$ of the gauge group, or equivalently
its highest weight.
The supersymmetric 't Hooft operator with charge $B$ is defined
by integrating over the fluctuations of the fields
around the configuration \cite{Kapustin:2005py}
\begin{equation}
  \begin{aligned}
  A&\equiv A_\mu dx^\mu
=\left(ig^2\vartheta\frac{B}{16\pi^2}\frac{1}{r} +{A_{\tau}^{(\infty)}} \right)d\tau 
+\frac B 2  \cos\theta d\varphi  \\
\Phi_0&=
- g^2\vartheta \frac{B}{16\pi^2}\frac{1}{r}+\Phi_0^{(\infty)}\,,
\quad\quad
\Phi_9= \frac B{2r}+\Phi_9^{(\infty)}
  \end{aligned}
\quad\quad\text{ in the background.}
\label{thooft-background}
\end{equation}
We recall that $\tau\equiv x^4$ and that $\vartheta$ is the gauge theory theta angle.  We have also introduced polar coordinates
$(r\equiv |\vec x|,\theta, \varphi)$ for $\mathbb R^3$.
Our choice of scalars in (\ref{wilson-def}) and (\ref{thooft-background})
ensures that the Wilson and 't Hooft operators
preserve the same sets of supercharges.
The action of the $U(1)$ R-symmetry 
rotates $\Phi_0+i\Phi_9$ and changes
the set of preserved supercharges.
Note that we define the electric Wilson line $A_\tau^{(\infty)}$
in the local trivialization such that the $d\varphi$ term is
given by $(B/2)\cos\theta d\varphi$ rather than the more familiar
$-(B/2)(\pm 1-\cos\theta)d\varphi$.
Our choice guarantees that when $\lambda\neq 0$,
the holonomy at the spatial infinity with $\theta=\pi/2$
is $\exp(-2\pi i R A_\tau^{(\infty)})$.
This will play a role in Section \ref{sec:non-com}.

More general line operators are dyonic and carry both electric and magnetic
charges.  
Such operators are defined by a path integral for a 't Hooft operator
with charge $B$, with the insertion of a Wilson operator
for the stabilizer of $B$ in $G$.
The dyonic charges are elements of the sum of coweight and weight lattices of $G$
\begin{equation}
  \Lambda_{cw}\oplus \Lambda_w\,,
\end{equation}
and the charges related by a simultaneous action of the Weyl group 
the two lattices are equivalent
\cite{Kapustin:2005py}.
Due to Dirac quantization, the magnetic charge must be a coweight 
which has integer inner products with all the 
weights in the matter representation.\footnote{In the theories whose gauge group is a product of $SU(2)$'s, 
the electric and magnetic charges with these constraints and equivalence
relations  match the homotopy classes of non-self-intersecting
curves on the corresponding Riemann surface \cite{Drukker:2009tz}.
}

Having defined the line operators whose vevs we wish to compute,
let us explain the parameters of the theory those vevs will depend on.
We are studying the theory in the Coulomb branch, so
the real scalars in the vector multiplet have the expectation values
\begin{equation}
  \langle \Phi_A\rangle \equiv \Phi_A^{(\infty)}\in \mathfrak t
\quad\quad
 A=0\,,9\,,
\end{equation}
which are the asymptotic values at $|\vec{x}|=\infty$.
Since we compactify the theory on $S^1$,
we also have the electric and magnetic Wilson lines.
The electric Wilson line is the asymptotic value of the $\tau$-component of the 
gauge field
\begin{equation}
 A_\tau^{(\infty)}\in \mathfrak t\,.
\end{equation}
Due to potential terms in the action (\ref{action-vec}),
$\Phi_A^{(\infty)}$ and $A_\tau^{(\infty)}$ can be 
simultaneously diagonalized, i.e., we can let them take values in the Cartan subalgebra $\mathfrak t$.

We also need to consider the magnetic Wilson line.
In the IR theory this is the vev of the scalar dual to the gauge field
in three dimensions.
In the UV theory we define it as follows.
At a generic point of the Coulomb branch, the scalar vevs $\Phi_A^{(\infty)}$ classically break the gauge group $G$ to the maximal torus $T$.  
The path integral includes infinitely many sectors
classified by the magnetic charges at infinity.
The general boundary condition is such that
asymptotically as $|\vec x|\rightarrow \infty$,  
we allow $\Phi_A(\vec x)$ to take
any values that are gauge equivalent to $\Phi_A^{(\infty)}$, i.e.,
there is a map $g: S^2\rightarrow G$ such that
\begin{eqnarray}
  \Phi_A(\vec x)\rightarrow g(\vec n)\cdot \Phi_A^{(\infty)}\cdot
g^\mo(\vec n) \quad \text{ as }\quad |\vec x| \rightarrow \infty
\end{eqnarray}
with $\vec n\equiv \vec x /|\vec x|\in S^2$.
Then the scalars $\Phi_A(\vec x)|_{|\vec x|=\infty}$ themselves
define a map from $S^2$ to the orbit 
$\{g
(\Phi_0^{(\infty)}, \Phi_9^{(\infty)})
g^{-1}|
g\in G\}$,
which is diffeomorphic to $G/T$ 
because the stabilizer of a generic element of $\mathfrak t\times \mathfrak t$
is $T$.
We can demand that $g=1$ at the north pole of $S^2$, 
so that $\Phi_A$ at $|\vec x|=\infty$ define
a homotopy class in $\pi_2(G/T)$ with a base point at the north pole.
If $G$ is simply connected,
the maximal torus can be identified
with the quotient of the Cartan subalgebra 
by the coroot lattice\footnote{See \cite{Gukov:2006jk} for a review of lattices in the Cartan subalgebra 
$\mathfrak t$ and
its dual $\mathfrak t^*$.
} 
$T\simeq \mathfrak{t}/\Lambda_{cr}$,
so $\pi_2(G/T)\simeq \pi_1(T)  =\Lambda_{cr}$.
In fact $G/T$ depends only on the Lie algebra of $G$,
so $\pi_2(G/T)=  \Lambda_{cr}$ for any $G$.
The infinitely many topological sectors are therefore classified by
$ \Lambda_{cr}$.
Physically this makes sense because $ \Lambda_{cr}$ is the lattice of magnetic charges carried by Polyakov-'t Hooft monopoles.
This lattice is more coarse than the coweight lattice
 $\Lambda_{cw}$ in which the magnetic charge $B$ of the 't Hooft operator
takes values,
$\Lambda_{cr}\subset \Lambda_{cw}$.
With generic matter representations, the lattice of 
't Hooft charges $B$ allowed by Dirac quantization
would be smaller than $\Lambda_{cw}$.

Let us now insert a 't Hooft operator with magnetic charge $B\in \Lambda_{cw}$ 
at the origin. The insertion of the 't Hooft operator 
changes the topology of the vector bundles in which the
fields take values, and in particular 
the structure of the boundary conditions at spatial infinity.
One can classify the allowed configurations by the asymptotic
magnetic charges taking values in the shifted lattice
$\Lambda_{cr}+B\subset \Lambda_{cw}$.
We define the magnetic Wilson line $\Theta\in \mathfrak t^*$ as
the chemical potential for the magnetic charges.
The expectation value of the 't Hooft operator
is given by the sum
\begin{eqnarray}
\langle T_B\rangle=  \sum_{v\in \Lambda_{cr}+B}
 e^{i v\cdot \Theta}
\int_v \mathcal D A
\mathcal D\Psi
\, e^{-S}\,,
\label{vev-sum-v}
\end{eqnarray}
where the path integral in each summand is performed with the boundary condition specified by $v$.
In the three-dimensional low-energy Abelian gauge theory that arises
via dimensional reduction, $\Theta$
is identified with 
the expectation values of scalars dual to the photons
\cite{Affleck:1982as},
and the UV and IR definitions of $\Theta$ are consistent.

Along the circle $S^1$ we can impose various
twisted boundary conditions on the fields.
It is convenient to exhibit them by representing the line operator vev as a supersymmetric index, taking $S^1$ as a time direction.
The line operator $L$
modifies the Hilbert space of the theory, rather than
acts on the original Hilbert space as a linear transformation.
We define our observable,
the expectation value of the line operator $L$,
to be a trace in the modified Hilbert space $\mathcal H_L$
\begin{eqnarray}
\langle L\rangle= \Tr_{\mathcal H_L} 
(-1)^{F}e^{-2\pi R H}e^{2\pi i\lambda(J_3+I_3)}
e^{2\pi i \mu_f F_f}\,,
\label{trace-def}
\end{eqnarray}
where $J_3$ and $I_3$ are the generators of the Lorentz $SU(2)$
and the R-symmetry $SU(2)$.
Here $J_3$ generates a rotation along the 3-axis:
$i J_3= 
x^1 \partial_2-x^2 \partial_1$
when acting on a scalar.
As we will see below, the combination $J_3+I_3$ commutes with the supercharge we use for localization.
We have also included the twist by the flavor symmetries
with generators $F_f$
and chemical potentials $\mu_f$, $f=1,\ldots, N_\text{F}$.
The definition (\ref{trace-def}) of the line operator vev
coincides with the one used in \cite{Gaiotto:2010be}.
The system may be realized in terms of a path integral
over the fields with appropriate twisted boundary conditions along $S^1$.
In this paper we adopt the equivalent formulation
where everywhere in the action (\ref{total-action}) on $\mathbb R^4$ the time derivative
is shifted as
\begin{equation}
  \partial_\tau \ra
\partial_\tau 
- \frac{i}{R} \lambda(J_3+I_3)
- \frac{i}{R}\sum_{f=1}^{N_\text{F}} \mu_f F_f
\label{der-shift}
\end{equation}
and the fields are periodic in $\tau$.
The electric and magnetic Wilson lines can also be regarded as the chemical potentials for the corresponding charges.

As we will see all the parameters except $\lambda$ will enter the line operator vevs in specific complex combinations.
These are the moduli
\begin{equation}
  a\equiv R\,  (A_\tau^{(\infty)}+i\Phi_0^{(\infty)})
\in \mathfrak t_{\mathbb C}\,,
\quad
 b\equiv
\frac{\Theta}{2\pi }-  \frac{4\pi i R}{g^2}\Phi_9^{(\infty)}
+ \frac{i \vartheta}{2\pi} R\Phi_0^{(\infty)}
\in \mathfrak t^*_{\mathbb C}
\,.\label{def-ab}
\end{equation}
and the complexified mass parameters
\begin{equation}
 m_f\equiv -\mu_f+iR M_f\in \mathbb C
\quad\quad
f=1,\ldots, N_\text{F}\,.
\end{equation}
We use the Lie algebra metric $\Tr$
in the action to regard $\Phi_9^{(\infty)}$
and $a$
as elements of $\mathfrak t^*_{\mathbb C}$.

General $\mathcal N=2$ theories
have several mass parameters $M_{Af}$ with
$A=0,9$ and $f=1,\ldots, N_\text{F}$.
These can be thought of as the vevs of the scalars
in the vector multiplets
that weakly gauge the flavor symmetries.
Only $M_f\equiv M_{A=0,f}$, which are the analog of $\Phi_0$,
will enter the line operator vevs.

\section{Localization for gauge theories on $S^1\times \mathbb R^3$}
\label{sec:localization}

We apply the localization technique
introduced for calculations in gauge theory on $S^4$ \cite{Pestun:2007rz}.
In this formalism, one adds a new term $t Q\cdot V$ to the action,
so that the path integral takes the form
\begin{eqnarray}
  \int \Dcal A\Dcal\Psi e^{-S-t Q\cdot V}\,.
\end{eqnarray}
Here $A$ and $\Psi$ include all the bosons and fermions, respectively.
We will also need to add ghost fields after gauge-fixing.
For observables that are invariant under the supercharge $Q$ of choice,
the path integral is independent of the parameter $t$.
The localization action is chosen to be
$V=(\Psi,\overline{Q\cdot \Psi})=(\psi,\overline{Q\cdot\psi})
+(\chi,\overline{Q\cdot\chi})$, where $\psi$ and $\chi$ denote
the fermions in the vector multiplet and the hypermultiplet.
Since the bosonic part of $Q\cdot V$ is a positive definite term
$||Q\cdot \Psi||^2$, the path integral is dominated by the solutions
of $Q\cdot \Psi=0$ in the limit $t\ra +\infty$ and can be calculated exactly by summing the fluctuation determinants at all the saddle points.

\subsection{Symmetries}

For localization we need to close off-shell the relevant subalgebra
of the whole superalgebra.
For this we introduce seven auxiliary fields $K_j$ as  in \cite{Pestun:2007rz}.
The supersymmetry transformations in $\mathcal N=2^*$ theory are given by
\begin{eqnarray}
Q\cdot A_M&=& \ep \Gamma_M \Psi\,,
\\
  Q\cdot\Psi&=&\frac 1 2 F_{MN} \Gamma^{MN} \ep+i K^i\nu_i\,,
\\
Q\cdot K_j &=& i\nu_j \Gamma^M D_M \Psi\,.
\end{eqnarray}
The gamma matrices and the constant spinors $\nu_i$ ($i=1,\ldots,7$) are defined in Appendix \ref{sec:spinor-gamma}.
The gauge fields in $F_{MN}$ and $D_M$ include mass matrices 
$M_{Aij}=i M_A F_{ij}$ through the Scherk-Schwarz mechanism \cite{Pestun:2007rz}.
The spinor $\ep$ must be chosen so that the line operators are invariant under the supersymmetry transformation $Q$.  We will use the same spinor as used in
\cite{Gomis:2011pf}
\begin{equation}
\ep
 =  \frac{1}{\sqrt 2}(1,0^7,1,0^7) \,,
\label{spinor-chosen}
\end{equation}
where the power indicates the number of repeated entries.
It satisfies\footnote{\label{foot:Nek}%
The third condition implies that $Q$ corresponds to the fermionic symmetry for the Donaldson-Witten twist \cite{Witten:1988ze} in the 1239-directions.  Thus $\langle L\rangle$ is a limit of the five-dimensional Nekrasov partition function \cite{Nekrasov:2002qd} for a theory on $S^1\times \mathbb R^4$ with a line operator insertion, where one of the equivariant parameter for the rotation in the 39 plane
is set to zero and a direction in $\mathbb R^4$ is compactified on an
infinitely small circle.
}
\begin{equation}
\Gamma_{5678}\ep=-\ep\,,
\quad
  \Gamma_{04}\ep=-i\ep\,,
\quad
\Gamma_{1239}\ep=\ep\,,
\quad
(2\Gamma_{12}+\Gamma_{56}+\Gamma_{78})\ep=0\,.
\end{equation}
The last condition implies that the supercharge commutes with the combination $J_3+I_3$ of spatial and R-symmetry rotations.
This explains why this particular combination entered the definition (\ref{trace-def}) of the vev.

We will need later the square of the supersymmetry transformation
given by
the spinor $\ep$ in (\ref{spinor-chosen}),
Using the vector
\begin{eqnarray}
  v^M\equiv \ep \Gamma^M \ep=(i,0^3,1,0^5)\quad\quad
M=0,1,\ldots,9\,,
\end{eqnarray}
we find that $Q^2$ generates time translation,
minus the complexified gauge transformation $G_\Lambda$ 
with gauge parameter $\Lambda=A_\tau+i\Phi_0$,
and the flavor symmetry transformation $iMF$:
\begin{eqnarray} 
  Q^2 \cdot A_M&=&-  F_{\tau M}-[i \Phi_0, D_M] -i \delta_M^i M_{ij}\Phi_{j}
\,,\nonumber\\
Q^2\cdot \Psi&=&- \partial_\tau \Psi-i [ A_\tau+i\Phi_0,\Psi]
-\frac i 4 M_{ij} \Gamma^{ij}\Psi
\,,\label{Q2APsiK}\\
Q^2\cdot K_i&=&- \partial_\tau K^i-i[A_\tau+i\Phi_0,K_i]
\,.\nonumber
\end{eqnarray}
See Appendix C  and (2.27) of  \cite{Pestun:2007rz}.

\subsection{Localization equations}
\label{sec:loc-eq}

Let us study the localization equations $Q\cdot \Psi=0$,
whose solutions the path integral localizes to.
We decompose $\Psi$ as
\begin{eqnarray}
  \Psi=\sum_{M=1}^9\Psi_M\tilde \Gamma^M\bar \ep
+i\sum_{j=1}^7  \Upsilon_j \nu^j\,.
\end{eqnarray}
Noting that
\begin{eqnarray}
  \Psi_M=\ep\Gamma_M \Psi\,,~~~~~
i \Upsilon_j=\bar \nu_j \Psi\,.
\end{eqnarray}
we obtain
\begin{eqnarray}
 0&=& Q\cdot\Psi_M= \frac 1 2 F_{PQ}\,\ep \Gamma_M \Gamma^{PQ} \ep
\quad\quad
M=1,\ldots,9\,,
\label{QPsiM}
\\
 0&=& i Q\cdot \Upsilon_j= \frac 12 F_{MN}\, \bar\nu_j \Gamma^{MN}\ep
+i K_j
\quad\quad\quad
j=1,\ldots,7\,.
\label{QUpsilon}
\end{eqnarray}
The equations (\ref{QPsiM}) reduce to\footnote{To show this we used the identities  $\Gamma_{M} \tilde \Gamma_{[P} \Gamma_{Q]}=\Gamma_{[M} \tilde \Gamma_P \Gamma_{Q]} +2 \delta_{M[P} \Gamma_{Q]}$ and $\ep \Gamma_{[M} \tilde \Gamma_P \Gamma_{Q]}\ep=0$.
} 
\begin{eqnarray}
0=  Q\cdot \Psi_M= -v^N F_{NM}\,.
\end{eqnarray}
According to (\ref{Q2APsiK}), these are equivalent to $Q^2$-invariance,
i.e., invariance under a combination of $\tau$-translation,
gauge transformations, and flavor transformations.
Due to the replacement of the $\tau$-derivative in (\ref{der-shift}),
for generic $\lambda$ the bosonic fields must also be invariant under the combination $J_3+I_3$ of spatial and R-symmetry rotations.
Among the various components of (\ref{QUpsilon}),
the most important equations are\footnote{We used the following facts:
$\bar \nu_j \Gamma^{kl}\ep=-\epsilon_{jkl}$ for $j,k,l\in\{1,2,3\}$,
$\bar\nu_j \Gamma^{kl}\ep=0$ for $j,k\in\{1,2,3\}$ and $l\in\{5,6,7,8\}$,
$\bar\nu_j \Gamma^{k9}\ep=\delta_{jk}$ for $j,k\in\{1,2,3\}$, and
$\bar\nu_j \Gamma^{9l}\ep=0$ for $j\in\{1,2,3\}$ and $l\in\{5,6,7,8\}$.
We also went ahead and set the hypermultiplets to zero.
This is justified below by $Q^2$-invariance.
}
\begin{equation}
0=i  Q\cdot \Upsilon_j= D_j \Phi_9- \frac{1}{2}\sum_{k,l=1}^3 \ep_{jkl}F_{kl}
+i K_j
\quad\quad
j,k,l=1,2,3\,.
\end{equation}
The imaginary part sets $K_j$ to zero.
The real part is precisely the Bogomolny equations
\begin{equation}
  *_3 F=D \Phi_9
\label{eq:Bogo}
\end{equation}
 that describe monopoles on $\mathbb R^3$!  Thus we conclude that the path integral localizes to the fixed points on the monopole moduli space with respect to spatial rotations and gauge transformations.

Four other components of  (\ref{QUpsilon}) read\begin{equation}
  \begin{aligned}
0=
i Q\cdot \Upsilon_j= \sum_{k=1}^3 \sum_{l=5}^8(\bar\nu_j \Gamma^{kl} \ep) D_k \Phi_l
+ \sum_{l=5}^8 (\bar\nu_j \Gamma^{9l} \ep)i[\Phi_9, \Phi_{l}]
+i K_j
\quad\quad
j=4,5,6,7\,.
  \end{aligned}
\label{QPsij4-7}
\end{equation}
Again the imaginary part requires $K_j$ to vanish.
The real part of (\ref{QPsij4-7}) is in fact the ``realification''
of the Dirac-Higgs equation
\begin{equation}
\sum_{i=1}^3
\sigma^i D_{i} q +
 [\Phi_9, q]=0\,,
\label{DH-eq}
\end{equation}
where the two-component ``spinor'' $q$ is a linear combination of $\Phi_i$
with $i=5,6,7,8$.
See Appendix \ref{sec:diff-op} for a related discussion.
As in topological twist, the hypermultiplet scalars behave as a spinor under the combination $J_3+I_3$.
Though generically (\ref{DH-eq}) itself admits non-zero solutions,
the $Q^2$-invariance, in particular the invariance under flavor transformations, requires $q$ to vanish.

Thus localization on $S^1\times \mathbb R^3$ leaves no bosonic zero-mode to be integrated over, and the final answer for the vev will
be expressed as a finite sum.  This is in contrast with the results for $S^4$ \cite{Pestun:2007rz,Gomis:2011pf} where the path integral reduced to a finite dimensional matrix integral.

\subsection{On-shell action}
\label{sec:classical}
Let us work out the classical contribution $e^{-S_\text{cl}}$,
given by the on-shell action evaluated in the background
(\ref{thooft-background}).
The on-shell action for the hypermultiplet simply vanishes,
therefore we focus on the action (\ref{action-vec}) for the vector multiplet.
For the background (\ref{thooft-background}), we also have
\begin{eqnarray}
  F&=&ig^2\vartheta \frac{B}{16\pi^2 } \frac{d\tau\wedge dr}{r^2}-\frac B 2
 \sin\theta d\theta\wedge d\varphi
\,,\\
  *F&=&-\frac B {2r^2} d\tau\wedge dr +ig^2\vartheta \frac{B}{16\pi^2 }
 \sin\theta d\theta\wedge d\varphi
\,.
\end{eqnarray}
Our orientation is such that the volume form is 
$d\tau\wedge dx^1\wedge dx^2\wedge dx^3$.
The action (\ref{action-vec}) is divergent in the presence of 
such a singular dyonic background.
We can render the action finite
by cutting off the spacetime at $\Sigma_3\equiv\{r=\delta\}$ and
by adding the boundary term \cite{Giombi:2009ek,Gomis:2011pf}
\begin{eqnarray}
S_\text{bdry}
&=&
\frac 2{g^2}\int_{\Sigma_3} \Tr \left( \Phi_9 F -i \Phi_0 * F\right)
\wedge d\tau
\,.
\label{boundary-action}
\end{eqnarray}
We find that
\begin{equation}
  \begin{aligned}
S_\text{vec}=& 
\frac{1}{g^2 \delta}\left(4\pi^2 R+\frac{g^2\vartheta^2 R}{16\pi^2}\right) \Tr B^2  \,,\\
S_\text{bdry}=&    
-\frac{1}{g^2 \delta}\left(4\pi^2 R+\frac{g^2\vartheta^2 R}{16\pi^2}\right)
\Tr B^2
-\frac{8\pi^2 R}{g^2}\Tr\left(\Phi_9^{(\infty)} B\right)
+\vartheta R \Tr\left( \Phi_0^{(\infty)} B\right)\,.
  \end{aligned}
\end{equation}
Thus the classical on-shell action is given by
\begin{eqnarray}
 S_\text{cl}(B)\equiv S_\text{vec}+S_\text{bdry}
&=&
-\frac{8\pi^2 R }{g^2}
\Tr
\left[
\Phi_9^{(\infty)}
B
\right]
+\vartheta R\ \Tr \left[  \Phi_0^{(\infty)} B \right]
\,.
\label{on-shell-action-result}
\end{eqnarray}
The on-shell action nicely combines with the weight $e^{i B\cdot \Theta}$
for the magnetic charge in (\ref{vev-sum-v}) so that
\begin{equation}
  \langle T_B\rangle\sim e^{i B\cdot\Theta} e^{-S_\text{cl}(B)}
=e^{2\pi i B\cdot b}\,,
\end{equation}
where $b$ was defined in (\ref{def-ab}).%
\footnote{%
In terms of $\Theta'=\Theta - \vartheta R A_\tau^{(\infty)}$, we can also write $b = \frac{\Theta'}{2\pi }-  \frac{4\pi i R}{g^2}\Phi_9^{(\infty)}
+ \frac{ \vartheta}{2\pi} a$.
By carefully performing dimensional reduction and 3d abelian duality, one sees that $2\pi R A_\tau^{(\infty)}$ and $\Theta'$ are the vevs of linear combinations of scalars that diagonalize the kinetic terms.
}
This is the leading classical approximation to the 't Hooft operator vev.
We will compute one-loop and non-perturbative corrections in the following sections.

\section{One-loop determinants}
\label{sec:one-loop}

Having computed the classical contribution to the 't Hooft operator vev, in this section we will compute the one-loop correction 
following \cite{Pestun:2007rz} and in parallel with \cite{Gomis:2011pf}.  As we saw in the previous section, the path integral reduces to a sum over saddle points.  For each saddle point we need to compute the fluctuation determinants.  The methods here will also be used in Section \ref{sec:mono} for the computation of such non-perturbative corrections.

\subsection{Gauge fixing}
The  gauge fixing action in the $R_\xi$-gauge is 
\begin{eqnarray}
  S_\text{gf}=\int d^4x\,
\Tr
\left (
-i\,
\tilde c\, 
\sum_{M=1,2,3,9}D_{(0)}^M D_M c
+ \tilde b\left(i
\sum_{M=1,2,3,9}D_{(0)}^M 
\tilde A_M+\frac \xi 2 \tilde b\right)
\right)\,.
\end{eqnarray}
We have defined $\tilde A_M\equiv A_M-A_{(0)M}$ where
$A_{(0)M}$ is the background configuration given in (\ref{thooft-background}).
The ghost fields $c, \tilde c$ are fermionic, and $\tilde b$ is bosonic.
By defining the BRST transformations\footnote{To compare with Pestun's formalism in \cite{Pestun:2007rz},
set $\tilde a_0, b_0,c_0, \tilde c_0$ to zero.
Then separate his BRST transformation $\delta$ into our $Q_\text{B}$
and the part $\delta_0$ proportional to $a_0$: $\delta= Q_\text{B}+\delta_0$.
Then our $Q$ can be written as $s+\delta_0$ with $a_0=-\Phi_{(0)0}$,
where $s$ denotes the supersymmetry transformation in \cite{Pestun:2007rz}.
}
\begin{equation}
  \begin{aligned}
&  Q_\text{B}\cdot A_M=- [c, D_M]\,,\quad
  Q_\text{B}\cdot \Psi=-i [c, \Psi]\,,\quad
  Q_\text{B}\cdot K_i=- i[c, K_i]\,,\quad
\\
&
Q_\text{B}\cdot c=-\frac i 2 [c,c]\,,\quad
Q_\text{B} \cdot\tilde c=\tilde b\,,\quad
Q_\text{B}\cdot \tilde b=0\,,
  \end{aligned}\end{equation}
we can write
\begin{eqnarray}
  S_\text{gf}=
Q_\text{B}\cdot V_\text{gh}\,,
\quad
V_\text{gh}\equiv
\int d^4x 
 \Tr
\left(
\tilde c\left(i\, 
\sum_{M=1,2,3,9}D_{(0)}^M \tilde A_M
+\frac \xi 2 \tilde b\right)
\right)\,.
\end{eqnarray}
The BRST transformation squares to zero, $\{Q_\text{B},Q_\text{B}\}=0$.
Unlike the case of $S^4$ \cite{Pestun:2007rz}
where the spacetime is compact,
we do not need to introduce ghosts-for-ghosts to deal with
constant gauge transformations.

We define the action of the supercharge $Q$ on the ghosts by

\begin{equation}
  \begin{aligned}
&  Q\cdot c=- v^M \tilde A_M\equiv -\tilde \Phi=-i \tilde \Phi_0- \tilde A_\tau\,,\quad
Q\cdot \tilde c=0\,,
\\
&
Q\cdot \tilde b=- v^M D_M\tilde c=-\partial_\tau \tilde c
-i [A_\tau+i\Phi_0,\tilde c]\,.
  \end{aligned}
\end{equation}
In the background $Q$ annihilates all the fermions, therefore the background
is supersymmetric.
We have $\{Q,Q\}(\text{ghost})=0$.

\subsection{One-loop determinants and the index theorem}

After gauge fixing, the total fermionic symmetry we use for localization is
\begin{equation}
  \hat Q\equiv Q+Q_\text{B}\,.
\end{equation}
While $Q^2$ in (\ref{Q2APsiK}) involves a gauge transformation $G_\Lambda$
with a dynamical gauge parameter $\Lambda=A_\tau+i\Phi_0$,
the gauge transformation that appears in $\hat Q^2=Q^2+\{Q,Q_\text{B}\}$ 
turns out to have a fixed parameter  
$\Lambda=A_{(0)\tau}+i\Phi_{(0)0}=A^{(\infty)}_\tau+i\Phi^{(\infty)}_0$:\footnote{For the gauge field
$\hat Q^2\cdot A_M=
-\p_\tau \tilde A_M -i[A^{(\infty)}_\tau+i\Phi^{(\infty)}_0, \tilde A_M]$.
}
\begin{eqnarray}
  \hat Q^2=-
\partial_\tau
-i (A_\tau^{(\infty)}+i\Phi_0^{(\infty)})
+M F\,.
\label{Qhat-squared}
\end{eqnarray}
Saddle points of the path integral remain the same
after we replace $Q\cdot V$ by $\hat Q\cdot \hat V$.
Recall that $M\equiv M_0$ is one of the mass parameters
defined below (\ref{action-hyp})
and that $F$ is the flavor symmetry generator.
The path integral to consider is
\begin{eqnarray}
  \int \mathcal D A \mathcal D\Psi \mathcal DK
\mathcal D \tilde b
\mathcal D c
\mathcal D \tilde c\,
 e^{-S-t \hat Q\cdot \hat V}\,,
\end{eqnarray}
where
\begin{equation}
  \hat  V 
=
\left\langle \Psi\,,\,
\overline{\hat Q\cdot \Psi}
\right\rangle
+V_\text{gh}\,.
\end{equation}
In order to evaluate the path integral in the limit $t\ra \infty$, we need to compute the superdeterminant of the kinetic operator in $\hat Q_{(0)}\cdot \hat V^{(2)}$, where $\hat Q_{(0)}$ is the linearization of $\hat Q$, and $ \hat V^{(2)}$ is the quadratic part of $\hat V$.
Following \cite{Pestun:2007rz} let us define
\begin{eqnarray}
  X_0=(\tilde A_M)_{M=1}^9\,,\quad
X_1=(\Upsilon_i, c, \tilde c)
\label{X-def}
\end{eqnarray}
and their partners
\begin{equation}
\begin{aligned}
  X_0'\equiv& \hat Q_{(0)}\cdot X_0=(\Psi_M-[c, D_{(0)M}])_{M=1}^9\,,
\\
X_1'\equiv& \hat Q_{(0)}\cdot X_1=
\left(
 K_i-i(\bar\nu_i\Gamma^{MN}\ep)
 D_{(0)M} \tilde A_N
\,
, -\tilde\Phi,b\right)\,.
\end{aligned}
\label{Xprime-def}
\end{equation}
Now $\hat V^{(2)}$ takes the form
\begin{eqnarray}
V^{(2)}
=
\left\langle
\left(
  \begin{array}{cc}
    X'_0& X_1
  \end{array}
\right)
\,,
\left(
  \begin{array}{cc}
   D_{00}&D_{01}\\
   D_{10}&D_{11}
  \end{array}
\right)
\left(
  \begin{array}{c}
X_0\\
X'_1
  \end{array}
\right)
\right\rangle
\,,
\label{eq:V2}
\end{eqnarray}
 where $D_{00}$ and others are certain differential operators.
Then ${\hat Q}_{(0)}\cdot V^{(2)}$ is given by
\begin{equation}
  \begin{aligned}
  \hat{Q}_{(0)}\cdot V^{(2)}=&
\left\langle
  \begin{pmatrix}
    X_0,X_1'
  \end{pmatrix}
\begin{pmatrix}
-\mathcal R_{00}&\\
&1
\end{pmatrix}
\,,
\begin{pmatrix}
D_{00}&D_{01}\\  
D_{10}&D_{11}
\end{pmatrix}
\begin{pmatrix}
X_0\\
X_1'  
\end{pmatrix}
\right\rangle
\\
&
+
\left\langle
  \begin{pmatrix}
    X_0',X_1
  \end{pmatrix}
\,,
\begin{pmatrix}
D_{00}&D_{01}\\  
D_{10}&D_{11}
\end{pmatrix}
\begin{pmatrix}
-1&\\
&-\mathcal R_{11}
\end{pmatrix}
\begin{pmatrix}
X_0'\\
X_1  
\end{pmatrix}
\right\rangle
\,,
  \end{aligned}
\end{equation}
where $\hat Q_{(0)}^2\cdot X_0=\mathcal R_{00}\cdot X_0$
and
$\hat Q_{(0)}^2\cdot X_1=\mathcal R_{11}\cdot X_1$.
Thus the one-loop determinant is given by
\begin{eqnarray}
  Z_\text{1-loop}&=&
\frac{\det^{1/2}\left[
\begin{pmatrix}
D_{00}&D_{01}\\  
D_{10}&D_{11}
\end{pmatrix}
      \begin{pmatrix}
       - 1&\\
&-\mathcal R_{11}
      \end{pmatrix}
\right]}{\det^{1/2}
\left[
\begin{pmatrix}
-\mathcal R_{00}&\\
&1
\end{pmatrix}
\begin{pmatrix}
D_{00}&D_{01}\\  
D_{10}&D_{11}
\end{pmatrix}
\right]
}
=
\frac{\det^{1/2} \mathcal R_{11}}
{\det^{1/2}\mathcal R_{00}}
\nonumber\\
&=&
\frac{\det^{1/2}_{\text{Coker} D_{10}} \mathcal R}
{\det^{1/2}_{\text{Ker} D_{10}}\mathcal R}
\,.
\label{Z-detR}
\end{eqnarray}
In the final line we have introduced notation $\mathcal R=\hat Q^2_{(0)}$ and used the fact that $\mathcal R$ commutes with $D_{10}$ as guaranteed by $\mathcal R$-invariance of $\hat V$.
Thus we only need the differential operator $D_{10}$, which can be obtained by explicitly computing $\hat V^{(2)}$.
It is easy to see what to expect from the results in Section \ref{sec:loc-eq}.
There we saw that the localization equations are given by the Bogomolny and Dirac-Higgs equations.  In Appendix \ref{sec:diff-op}, we will show that $D_{10}$ involves the linearization of these equations as well as the dual of the gauge transformation.

The symmetry generator $\mathcal R=\hat Q^2_{(0)}$ is given in 
(\ref{Qhat-squared}).
In a general $\mathcal N=2$ theory, we replace the last term $M F$ by $\sum_f M_f F_f$, where $F_f$ are the flavor symmetry generators in (\ref{trace-def}).
We also perform the shift (\ref{der-shift}) of the $\tau$ derivative.
It is also useful to rescale $\mathcal R$ as $\mathcal R\ra -R\, \mathcal R$.
This does not affect the value of the one-loop determinant
(\ref{Z-detR}) due to cancellations between the numerator and the denominator.
Then $\mathcal R$ takes a simple expression
\begin{equation}
  \mathcal R
= \ve R\p_\tau  -i \lambda(J_3+I_3) + i a  +i \sum_{f=1}^{N_\text{F}} m_f F_f\,.
\end{equation}
We have introduced a formal parameter $\ve$ 
that should be set to one at the end of calculation.
A Fourier mode $e^{in\tau/R}$ along $S^1$ contributes $i n\ve$
to $\mathcal R$.

The form (\ref{Z-detR}) of the one-loop determinant implies that
it can be obtained from the equivariant index of the operator $D_{10}$
\begin{equation}
  \text{ind}\, D_{10}\equiv
\Tr_{\text{Ker} D_{10}} e^{2\pi \mathcal R}
-
\Tr_{\text{Coker} D_{10}} e^{2\pi \mathcal R}\,.
\end{equation}
Indeed if it is given in terms of weights $w_j$ and multiplicities
$c_j$ as
\begin{equation}
  \text{ind}\, D_{10}=\sum c_j e^{w_j}\,,
\end{equation}
the one-loop determinant is given by
$
  Z_\text{1-loop}=\left(\prod_j w_j^{c_j}\right)^{-1/2}.
$
In the following we will separately define the indices for differential operators acting on vector and hypermultiplets.  We will also adopt a normalization 
for $\text{ind}$ 
that corresponds to $ \text{ind}( D_{10})\rightarrow - \frac 1 2 \text{ind}( D_{10})$,
so that the translation from the index to the one-loop determinant 
is simply given by the rule $\sum_j c_j e^{w_j}
\ra \prod_j w_j^{c_j}$.
Then
\begin{equation}
    Z_\text{1-loop}=\prod_j w_j^{c_j}\,.
\end{equation}

Thus we need to compute the weights under
 the gauge transformation with parameter 
$a\equiv R( A_\tau^{(\infty)}+i\Phi_0^{(\infty)})$,
a time translation by $\ve$, and
a spatial rotation along the 3-axis with angle $2\pi\lambda$,
and flavor transformations with parameters $m_f$.

\subsection{Calculation of the equivariant index}
\label{sec:calc-index}

Before we delve into the details of the calculations, let us
summarize our methodology
that extends the techniques developed in \cite{Gomis:2011pf}, 
listing at the same time
the relevant complexes and their interrelations.
We showed above that the vector multiplet contribution to the one-loop determinant
can be computed from the index of the
complex that linearizes the Bogomolny equations in $\mathbb R^3$
\begin{equation}
D_\text{Bogo}:  0 \ra \Omega^0(\text{ad}\,E)
\stackrel{
\text{\raisebox{-1mm}[0mm][0mm]{$(D\,, [i\Phi_9,\,\bullet\,])$}}
}{
\xrightarrow{\hspace*{1.6cm}}} 
\Omega^1(\text{ad}\,E)\oplus \Omega^0(\text{ad}\,E)
\ra\Omega^1(\text{ad}\,E)\ra 0\,,
\label{complex-Bogo}
\end{equation}
where $\text{ad}\,E$ is the adjoint gauge bundle.
The second arrow is the gauge transformation whose conjugate\footnote{The equivariant index remains the same when we ``fold''
(\ref{complex-Bogo}) into
$  0\ra \Omega^0 \oplus \Omega^1 \ra
\Omega^1\oplus \Omega^0\ra 0$,
where twisting by $\text{ad}\,E$ is implicit, and
the second arrow is the linearized Bogomolny equations
plus the dual of a gauge transformation (\ref{dual-Bogo}).
The same remark applies to the self-dual complex (\ref{complex-SD}).
It is the folded form of the complexes that naturally arises from gauge-fixing.
}
appear in (\ref{dual-Bogo}),
and the third is the map 
 $(\delta A,\delta\Phi_9)
\mapsto *D\delta A-D\delta\Phi_9+i[\Phi_9, \delta A]$
in (\ref{DBogo}).
As reviewed in Appendix \ref{sec:mono-inst},
the Bogomolny equations with a single
singularity  on $\mathbb R^3$ are equivalent to the
anti-self-duality equations on the (single-centered) Taub-NUT space
with invariance under the action of
the group that we call $U(1)_K$.
Linearizing the correspondence, 
we will obtain the index of the Bogomolny complex\footnote{We will refer to
(\ref{complex-Bogo}) and
(\ref{complex-DH})
as the Bogomolny and Dirac-Higgs (DH) complexes.
} (\ref{complex-Bogo})
from the index of the self-dual complex
\begin{equation}
D_\text{SD}: 
0\ra 
\Omega^0
(\text{ad}\,E)
\stackrel{D}{\ra}
\Omega^1(\text{ad}\,E)
\stackrel{
\text{\raisebox{-1mm}[0mm][0mm]{$(1+*) D$}}
}{
\xrightarrow{\hspace*{12mm}}} 
\Omega^{2+}(\text{ad}\,E)
\ra
0
  \label{complex-SD}
\end{equation}
on the four-dimensional space
by taking an invariant part under the
 $U(1)_K$ action \cite{MR1624279,Gomis:2011pf}.
Similarly the hypermultiplet contribution
will be derived from the index of the complex
\begin{equation}
D_{\text{DH},R}:  0\ra
 \Gamma(S\otimes R(E))
\stackrel{\text{\raisebox{-1mm}[0mm][0mm]{$\sigma^jD_j+\Phi_9$}}}{
 \xrightarrow{\hspace*{13mm}} 
} 
 \Gamma(S\otimes R(E))
 \ra 0\,,
\label{complex-DH}
\end{equation}
where $S$ is the spinor bundle over $\mathbb R^3$,
and $\Phi_9$ acts on $q\in \Gamma(S\otimes R(E))$ in the matter representation $R$.
Its index will be obtained from the $U(1)_K$ invariant
part of the index of the twisted Dirac complex
\cite{Gomis:2011pf}
\begin{equation}
D_{\text{Dirac},R}:  0\ra \Gamma(S^+\otimes R(E)) \stackrel{
\text{\raisebox{-1mm}[0mm][0mm]{$ \bar \sigma^\mu D_\mu$}}
 }{
 \xrightarrow{\hspace*{9mm}} 
} \Gamma(S^- \otimes R(E)) \ra 0
\label{complex-Dirac}
\end{equation}
in four dimensions.

Both the self-dual and Dirac complexes are related to the Dolbeault complex
\begin{equation}
  \bar D_R: 0\ra \Omega^{0,0}(R(E))\ra \Omega^{0,1}(R(E))
\ra \Omega^{0,2}(R(E))
\ra 0\,.
\label{complex-Dolb}
\end{equation}
To see this note that upon complexification we have
$\Omega^0_\mathbb{C}=\Omega^{0,0}$,
$\Omega^1_\mathbb{C}=\Omega^{1,0}\oplus \Omega^{0,1}$
and
$\Omega^{2+}_{ \mathbb C}=\Omega^{2,0}\oplus \Omega^{0,0}\omega 
\oplus \Omega^{0,2}$, where $\omega$ is the \Kahler form.  
See, e.g., \cite{MR1079726}.
Since by Hodge duality 
$\Omega^{2,2}=\Omega^{0,0}$ and $\Omega^{2,1}=\Omega^{1,0}$,
the complexification of the self-dual complex (\ref{complex-SD}) is isomorphic
to the Dolbeault complex (\ref{complex-Dolb}) with $R=\text{ad}$
twisted by $\Omega^{0,0}\oplus \Omega^{2,0}$.
For spinors recall that $\Omega^{p,q}=\Gamma(\Lambda^{p,q})$
and that $K=\Lambda^{2,0}$ is the canonical line bundle.
We have
\begin{equation}
  S^+=K^{1/2}\otimes( \Lambda^{0,0}\oplus \Lambda^{0,2})\,,
\quad
S^-=K^{1/2}\otimes \Lambda^{0,1}\,.
\end{equation}
Thus the Dirac complex (\ref{complex-Dirac}) 
is isomorphic to the Dolbeault complex (\ref{complex-Dolb})
twisted by $(\Omega^{2,0})^{1/2}$.

Let us now review the index of the Dolbeault complex.
We will compute the index of the Dolbeault complex 
on Taub-NUT space by applying the Atiyah-Bott fixed point formula.
Taub-NUT space is holomorphically isomorphic to flat $\mathbb C^2$
 with local coordinates $(z_1,z_2)$,
for which the $U(1)\times U(1)$-equivariant index of the (untwisted)
Dolbeault complex
 is given by
\begin{eqnarray}
\text{ind}(\bar\partial)
=
\frac{t_1 t_2}{(1-t_1)(1-t_2)}\,.
\label{index-Dolbeault}
\end{eqnarray}
Let us denote by $ U(1)_{J+R}$ the group generated
by $J_3+I_3$, the simultaneous spatial and R-symmetry rotations.
The action of $(t_1, t_2)$ on $\mathbb C^2$
is standard, $(z_1,z_2)\mapsto (t_1 z_1, t_2 z_2)$,
and is related to $U(1)_K\times U(1)_{J+R}$ as
\begin{eqnarray}
  t_1=e^{-2\pi i\nu+\pi i\lambda }\,,~~~~~t_2=e^{2\pi i\nu+\pi i\lambda}\,,
\label{tnuep}
\end{eqnarray}
as can be seen from (\ref{z1z2}).
Here $e^{2\pi i\nu}$ parametrizes $U(1)_K$,
while $2\pi \lambda$ is
the angle of rotation
 along the 3-axis of $\mathbb R^3$,
which is the base of the circle fibration in Taub-NUT space
(\ref{TN-metric}).
The $SU(2)$ R-symmetry action on the fields is also parametrized by $\lambda$.

For our purposes
the best way to understand the formula
(\ref{index-Dolbeault})
is to consider the group action
on the basis of sections.  
For example an element of $\Omega^{0,0}$
can be expanded as
\begin{eqnarray}
\sum_{
k,l,m,n
} 
c_{klmn}
z_1^{k}    
\bar z_1^{l}
z_2^{m}   \bar z_2^{n} \,,
\end{eqnarray}
where $k,l,m,n\in \mathbb Z_\geq 0$ and the coefficients transform as 
$c_{klmn}\mapsto  t_1^{-k+l} t_2^{-m+n} c_{klmn}$.
Elements of $\Omega^{0,1}$ and 
$\Omega^{0,2}$ 
admit similar expansions.
Summing up the weights with appropriate signs determined by the degrees
in the complex, we obtain
\begin{eqnarray}
\text{ind}_\delta(\bar\partial)
&=&\sum_{k,l,m,n\geq 0}
(1-t_1-t_2+t_1 t_2) 
 t_1^{-k+l} t_2^{-m+n}
\nonumber\\
&=&
\frac{(1-t_1)(1-t_2)}
{
(1- e^{-\delta} t_1^{-1})(1- e^{-\delta} t_1)
(1-e^{-\delta} t_2^{-1})(1-e^{-\delta} t_2)
}\,.
\label{index-Dolbeault-reg}
\end{eqnarray}
Factors $e^{-\delta}$ with small $\delta>0$ 
are inserted to keep track of how we expand the numerator.
We obtain (\ref{index-Dolbeault})
from the regularized index (\ref{index-Dolbeault-reg})
by taking the limit $\delta \ra 0$.
Including the gauge group action, we obtain
the index for the Dolbeault operator twisted by $R(E)$
\begin{equation}
  \text{ind}_\delta(\bar D_R)
=  
\frac{(1-t_1)(1-t_2)}
{
(1- e^{-\delta} t_1^{-1})(1- e^{-\delta} t_1)
(1-e^{-\delta} t_2^{-1})(1-e^{-\delta} t_2)
}
\sum_{w\in R} e^{2\pi i w\cdot a}
\,.
\end{equation}

The relationships of the self-dual and Dirac complexes
to the Dolbeault complex described above imply
that
\begin{eqnarray}
\text{ind}_\delta(D_{\text{SD},\mathbb C})  
&=&(1+t_1^{-1}t_2^{-1})
\text{ind}_\delta(\bar D_\text{adj})\,,
\\    
\text{ind}_\delta(D_{\text{Dirac},R})  
&=&t_1^{-1/2}t_2^{-1/2}
\text{ind}_\delta(\bar D_R)\,.
\end{eqnarray}
Furthermore, the indices
of the Bogomolny and Dirac-Higgs complexes are obtained
by taking the $U(1)_K$-invariant parts.
This can be implemented by substituting
(\ref{tnuep}) and $a\ra a+B\nu$ and
then integrating over $\nu$:
 \begin{eqnarray}
 \text{ind}(D_{\text{Bogo},\mathbb C})  
&=&
\lim_{\delta \ra 0}
\int_0^{1}d\nu
\left.
\text{ind}_\delta(D_{\text{SD},\mathbb C})\right|_{a\ra a+B\nu}
\,,
\\    
 \text{ind}(D_{\text{DH},R})
&=&
\lim_{\delta \ra 0}
\int_0^{1}d\nu
\left.
\text{ind}_\delta(D_{\text{Dirac},R})\right|_{a\ra a+B\nu}
\,.
 \end{eqnarray}
The factors $e^{-\delta}$ in the integrands
specify which poles to pick in the contour integrals.
We also need to take into account  the Fourier modes on 
$S^1$ that give rise to an infinite sum $\sum_n e^{in\ve}$.
The formal parameter $\ve$ for time translation 
should be set to one at the end of the calculation.

Finally, the one-loop determinant $Z^\text{vm}_\text{1-loop}$
for the vector multiplet
is obtained by the rule $\sum_j c_j e^{w_j}
\ra \prod_j w_j^{c_j}$
from
\begin{equation}
  \text{ind}(D^\text{vm})
=\frac 1 2
\sum_{n\in \mathbb Z}e^{2\pi i n \ve}
\text{ind}(D_{\text{Bogo},\mathbb C})  \,.
\label{ind-Dvm}
\end{equation}
The factor of $1/2$ in (\ref{ind-Dvm}) accounts for
the complexification of the Bogomolny complex.

For the hypermultiplet, 
the one-loop determinant $Z_\text{1-loop}^\text{hm}$
arises if the same rule is applied to \cite{Gomis:2011pf}
\begin{equation}
  \text{ind}(D^\text{hm}_R)
=
-\frac{1}{2}
\sum_{n\in \mathbb Z}e^{2\pi i n \ve}
\sum_{f=1}^{N_\text{F}}
\Big(
e^{-2\pi im_f}
\text{ind}(D_{\text{DH},R})
+
e^{2\pi im_f}
\text{ind}(D_{\text{DH},R})|_{a\ra-a}
\Big)
\,.
\label{ind-Dhm}
\end{equation}
Let us explain the meaning of this expression (\ref{ind-Dhm}).
The precise flavor symmetry of a massless theory is best described
in terms of half-hypermultiplets.
If an irreducible representation $R$ is real,
half-hypermultiplets can only appear in an even number $2N_\text{F}$,
and the flavor symmetry $G_\text{F}$ is $Sp(2N_\text{F})$.
The symplectic group $Sp(2N_\text{F})$ has rank $N_\text{F}$ in our convention.
For a complex irreducible representation $R$,
half-hypermultiplets always appear in conjugate pairs $R\oplus \bar R$.
With $N_\text{F}$ such pairs, the flavor symmetry is $U(N_\text{F})$.
When an irreducible representation $R$ is pseudo-real,
the theory is anomalous unless an even number $2N_\text{F}$ of half-hypermultiplets
are present \cite{Witten:1982fp}.
The flavor symmetry group in this case is $SO(2N_\text{F})$.
Parameters $m_f$ in (\ref{ind-Dhm}) are the equivariant parameters
for the flavor group $G_\text{F}$ of the massless theory,
and are related to the physical masses $M_f$ 
and the flavor chemical potentials $\mu_f$ as
\begin{equation}
  m_f=-\mu_f+iR M_f\,.
\end{equation}
The particular combination of terms in (\ref{ind-Dhm}) was derived
in \cite{Gomis:2011pf} based on Higgsing which produces various
types of matter representations.

The indices $\text{ind}(D_{\text{Bogo},\mathbb C})$
and $\text{ind}(D_{\text{DH},R})$ were computed in \cite{Gomis:2011pf}:
\begin{eqnarray}
  \text{ind}(D_{\text{Bogo},\mathbb C})
\hspace{-2mm}
&=&
\hspace{-2mm}
-
  \frac{e^{\pi i\lambda}+e^{-\pi i\lambda }}2
\sum_{\alpha}
e^{2\pi i\alpha\cdot  a}
\left(
e^{(|\alpha\cdot B|-1) \pi i\lambda}
+
e^{(|\alpha\cdot B|-3)\pi i\lambda}
+\ldots
+
e^{-(|\alpha\cdot B|-1)\pi i\lambda}
\right)\,,
\nonumber
\\
\text{ind}(D_{\text{DH},R})
&=&
- \frac 1 2
 \sum_{w\in R}
 e^{2\pi i w\cdot  a}
\left(
e^{(|w\cdot B|-1)\pi i \lambda}
+
e^{(|w\cdot B|-3)\pi i \lambda}
+
\ldots
+
e^{-(|w\cdot B|-1)\pi i \lambda}
\right)
\,.
\end{eqnarray}
By applying the rule to (\ref{ind-Dvm}) and (\ref{ind-Dhm}),
we find the one-loop determinant
\begin{eqnarray}
&&\prod_{n\in \mathbb Z}\prod_{\alpha} \prod_{k=0}^{|\alpha \cdot B|-1}
\left[
n \ve+\frac 1 2 \lambda
+\alpha \cdot a
+\left(\frac{|\alpha\cdot B|-1}2-k\right)\lambda
\right]^{-1/2}
\nonumber\\
&\sim&
\prod_{\alpha>0}
\prod_{k=0}^{|\alpha\cdot B|-1}
\prod_\pm
\sin^{-1/2}\left[\pi\left( 
\alpha\cdot a
\pm\left(\frac{|\alpha\cdot B|}2-k\right)\lambda
\right)\right]
\nonumber\\
&=:&  Z_\text{1-loop}^\text{vm}(a,\lambda;B)
\,,
\label{1-loop-vm}
\end{eqnarray}
for the vector multiplet and
\begin{eqnarray}
&&
\prod_{n\in \mathbb Z}\prod_{f=1}^{N_\text{F}}\prod_{w\in R}
\prod_{k=0}^{|w\cdot B|-1}
\left[n \ve+ w\cdot a-m_f 
+\left(\frac{|w\cdot B|-1}2-k\right)\lambda
\right]^{1/2}
\nonumber\\
&\sim &
\prod_{f=1}^{N_\text{F}}\prod_{w\in R}
\prod_{k=0}^{|w\cdot B|-1}
\sin^{1/2}
\left[
\pi\left(
w\cdot a-m_f +\left(\frac{|w\cdot B|-1}2 -k\right)\lambda
\right)
\right]
\nonumber\\
&=:&  Z^\text{hm}_\text{1-loop}(a,m_f,\lambda;B)
\label{1-loop-hm}
\end{eqnarray}
for the hypermultiplet.
In the final expressions we set $\ve$ to one.
When there is more than one matter irreducible representation
we need to take a product over them.
Combining the vector multiplet and hypermultiplet contributions,
the one-loop factor is given by
\begin{equation}
  Z_\text{1-loop}(a,m_f,\lambda;B)
:=
  Z_\text{1-loop}^\text{vm}(a,\lambda;B)
Z_\text{1-loop}^\text{hm}(a,m_f,\lambda;B)\,.
\label{1-loop-total}
\end{equation}

\section{Contributions from monopole screening}
\label{sec:mono}

In this section we calculate the contributions from non-perturbative
saddle points of the localization action $Q\cdot V$.
Since the bosonic part of $Q\cdot V$ is given by $||Q\cdot \Psi||^2$,
these saddle points are the solutions of the equation $Q\cdot \Psi=0$.
As we saw in Section \ref{sec:loc-eq}, the solutions of $Q\cdot \Psi=0$ are 
the fixed points of the Bogomolny equations with a prescribed singularity.

\subsection{Definition of $Z_\text{mono}$}

The moduli space of the solutions of the Bogomolny equations
with a singularity prescribed by $B$ has infinitely many components.
For example, even for $B=0$ there exist the components 
whose elements are smooth monopoles with charges labeled by all
$v\in\Lambda_{cr}$.
In our localization calculation only the components that contain
fixed points of the $U(1)_{J+R}\times T$-action are relevant,
where $T$ is the maximal torus of the gauge group.
Invariance under $U(1)_{J+R}\times T$-action is a strong constraint,
because the $T$-invariance for generic $a\in \mathfrak t$
requires the adjoint fields to be Abelian, {\it i.e.}, that they belong
to $\mathfrak t$.
The only Abelian solutions to the Bogomolny equations
are the singular Dirac monopole solutions,
and the singularity must be located at the point where
the 't Hooft operator is inserted.
This argument almost shows that the background configuration 
(\ref{thooft-background}) is the only saddle point of the path
integral.
Abelian solutions of the Dirac form (\ref{thooft-background}),
where $B$ is replaced by some other coefficient $v\in \Lambda_{cr}+B$,
can however arise as a limit in the family of solutions
whose singularity has coefficient $B$ \cite{Kapustin:2006pk}.
Such solutions represent smooth monopoles that approach 
the singular monopole and screen its charge.
See \cite{Cherkis:2007jm} for an explicit example.
For our calculation we only need to consider the components
of the moduli space that contain such solutions.
Under Kronheimer's correspondence, mentioned in Section \ref{sec:calc-index}
and reviewed in Appendix \ref{sec:mono-inst},
the Abelian solution specified by $v$ uplift to a
small instanton located at the point on Taub-NUT space where
the $S^1$ fiber degenerates.
Since our calculation needs only the local behavior of the fields
near this point, we can replace Taub-NUT space by $\mathbb C^2$.
A more satisfying justification for this replacement is
the fact that 
such a small instanton solution belongs to 
a component of the instanton moduli space 
that is isomorphic as a complex variety
to a component of the instanton moduli space for $\mathbb C^2$ 
\cite{Cherkis:2008ip}.
See also \cite{Witten:2009xu}.
We denote by $\mathcal M(B,v)$ the moduli space for the Bogomolny equations
that descend from the component of the instanton moduli space.
A generic point of
$\mathcal M(B,v)$ is a solution that approaches
the background (\ref{thooft-background}) near the origin,
and the same expression with $B$ replaced by $v$ asymptotically at 
infinity.
It can be shown that we need $||v||\leq ||B||$
for  $\mathcal M(B,v)$ to be non-empty  \cite{Gomis:2011pf}.

Since 
all the fixed points in $\mathcal (B,v)$ take the form of the 't Hooft background (\ref{thooft-background})
except that $B$ is replaced by $v$,
each contributes a factor $e^{-S_\text{cl}(v)}$ 
computed in Section \ref{sec:classical}.
This classical contribution depends only on $v$
and is universal among the fixed points in $\mathcal M(B,v)$.
We also need to include the fluctuation determinant
$ \prod_j w_j^{c_j}$ from each fixed point, which
can be computed from the indices
of the Bogomolny and Dirac-Higgs complexes via the rule
$\sum_j c_j e^{w_j}\rightarrow \prod_j w_j^{c_j}$,
 as in the one-loop analysis in Section \ref{sec:one-loop}.
By factoring out $Z_\text{1-loop}(v)$ that was computed in 
Section \ref{sec:one-loop},
we denote the sum of such determinants by
\begin{equation}
Z_\text{1-loop}(v)  Z_\text{mono}(B,v)
\equiv 
\mathop{\sum_\text{fixed points}}_{\text{in }\mathcal M(B,v)} \prod_j w_j^{c_j}\,.
\end{equation}
This equation defines $Z_\text{mono}(B,v)$ as a function of
$B, v, a, b, m_f$, and $\lambda$.
As mentioned in footnote \ref{foot:Nek},
$\langle L\rangle$ may be thought of as a dimensional reduction
of the five-dimensional instanton partition function
with an operator insertion.
Thus $Z_\text{mono}(B,v)$ can be interpreted in terms of appropriate
characteristic classes on $\mathcal M(B,v)$.

\subsection{Monopole moduli space for $G=U(N)$}

In order to compute $Z_\text{mono}(B,v)$ explicitly,
we need a method to describe the component $\mathcal M(B,v)$
of the monopole moduli space and their fixed points.
Let us now review the ADHM construction of $\mathcal M(B,v)$ in
 the case $G=U(N)$ \cite{Kapustin:2006pk}.

We consider the flat space $\mathbb C^2$
parametrized by coordinates $z=(z_1,z_2)$.
Let us set $W:=\mathbb C^N$ and $V:=\mathbb C^k$.
The instanton bundle over $\mathbb C^2$ with instanton number $k$
is described by a family of complexes
\begin{equation}
  V \stackrel{\alpha(z)}\longrightarrow \mathbb C^2\otimes V\oplus W
\stackrel{\beta(z)}\longrightarrow V\,,
\end{equation}
where the maps depend on $z$ as
\begin{equation}
  \alpha(z)=
  \begin{pmatrix}
z_2-B_2\\
-z_1+B_1\\
-J
  \end{pmatrix}
\,,
\quad
\beta(z)=
\begin{pmatrix}
 z_1-B_1&z_2-B_2&-I
\end{pmatrix}
\,.
\end{equation}
When the complex ADHM equation
\begin{equation}
  [B_1,B_2]+IJ=0
\end{equation}
which is equivalent to $\beta(z)\alpha(z)=0$
is satisfied,
the cohomology groups
\begin{equation}
  H^0_z=\text{Ker} [\alpha(z)]\,,\quad
  H^1_z=\text{Ker}[\beta(z)]/\text{Im}[\alpha(z)]\,,\quad
  H^2_z=V/\text{Im} [\beta(z)]
\end{equation}
can be defined.
If $H^0_z=H^2_z=0$, $E_z=H^1_z$ describes the fiber of a smooth irreducible
instanton bundle over $\mathbb C^2$.
We are also interested in singular configurations that arise as a limit
of smooth ones, therefore we set 
$E_z=H^1_z-H^0_z-H^2_z$ in general.
The Euler characteristic $\dim H^0_z-\dim H^1_z+\dim H^2_z=-\dim E_z=-N$
is independent of $z$.

A monopole solution in $\mathcal M(B,v)$ descends from
a $U(1)_K$-invariant instanton.
The group acts geometrically on $(z_1,z_2)$ as 
$(z_1,z_2)\mapsto (e^{-2\pi i\nu}z_1, e^{2\pi i \nu}z_2)$ as
in
(\ref{tnuep}).
Since $(B_1,B_2)$ represent the positions of the instantons,
they transform as
$(B_1,B_2)\mapsto (e^{-2\pi i\nu}B_1, e^{2\pi i \nu}B_2)$.
The group $U(1)_K$ also acts on the gauge bundle.
The fiber $E_0$ at $z=0$ is mapped to itself,
and its character for $U(1)_K$ is given by $e^{2\pi i B\nu}$
where $e^{2\pi \nu}\in U(1)_K$ and the charge $B$
of the 't Hooft operator is regarded
as a $N\times N$ diagonal matrix.
The group $U(1)_K$ also acts on $W$ and $V$.
Since $W$ represents the fiber $E_\infty$ at $z=\infty$,
its character is $\Tr e^{2\pi i v \nu}$.
The character of $V$ can be written as $e^{2\pi i K\nu}$
with a $k\times k$ diagonal matrix $K$.
The identification of $E_{z}$
with $H^1_z-H^0_z-H^2_z$ implies that $K$ is determined by\footnote{A warning on notation.  The ``$K$'' in $U(1)_K$ stands for Kronheimer.
The matrix $K$ is the weight of $U(1)_K$ acting on the $k$-dimensional
vector space on which $B_1$ and $B_2$ act as endomorphisms. 
}
\begin{equation}
  \Tr e^{2\pi i B\nu}
=\Tr e^{2\pi i v\nu}+(e^{2\pi i\nu}+e^{-2\pi i\nu}-2) \Tr e^{2\pi i K\nu}
\quad
\quad e^{2\pi i\nu} \in U(1)_K
\label{B-v-K-condition-2}
\end{equation}
up to conjugation.

To describe $\mathcal M(B,v)$, we impose $U(1)_K$ invariance on the ADHM data.
Namely the ADHM data must satisfy the conditions
\begin{equation}
 - B_1+[K,B_1]=0\,,\quad
B_2+[K,B_2]=0\,,\quad
KI-Iv=0\,,\quad
v J-J K=0\,.
\label{U(1)K-inv-eq}
\end{equation}
For the instanton moduli space, one would take a quotient by $GL(k,\mathbb C)$.
The matrix $K$ breaks the $GL(k,\mathbb C)$ into its commutant
$\prod_rGL(k_r,\mathbb C)$.
Two combinations of such data are considered equivalent if they are related by an action of $\prod_r GL(k_r,\mathbb C)$:
\begin{equation}
(B_1, B_2, I,J)
\sim
(g B_1 g^{-1}, g B_2 g^{-1}, g I,J g^{-1})
\quad \quad
g\in \prod_r GL(k_r,\mathbb C)\,.
\end{equation}
Thus the complex variety $\mathcal M(B,v)$ is given by
the holomorphic quotient
 \begin{eqnarray}
\mathcal M(B, v)=
\left   \{
 (B_1,B_2,I,J)
 \left|
\begin{array}{ccc}
 -  B_1+[K,B_1]&=&0\\
   B_2+[K,B_2]&=&0\\
 K I - I M&=&0\\
 M J - J K&=&0\\
 \end{array}
\right.
\right\}
{\Big /}
\prod_r GL(k_r,\mathbb C)\,.
 \end{eqnarray}

The notion of fixed points requires a regularization
of singularities in $\mathcal M(B,v)$.
In this paper we do not attempt to describe the regularization
in detail though we believe that this is important for
the precise definition of the 't Hooft loop with a given magnetic charge
$B$.
See Section \ref{sec:discussion} for a further discussion
on this point.
We will use a partial regularization that descends from the moduli space of non-commutative instantons that smooth the small instanton singularities.
This led to a prescription,  based on contour integrals, 
for how to take into account the fixed point contributions in \cite{Gomis:2011pf}.
Here we give an alternative prescription
for the calculation
of the fixed points and their contributions.

\subsection{Fixed points and their contributions}

Next we turn to the description of fixed points.
We need to know which fixed point $\vec Y$ on the instanton
moduli space descends to
the specific component $\mathcal M(B,v)$ of the monopole moduli space.
The fixed points are given by the ADHM data $(B_1,B_2,I,J)$
that satisfy
\begin{equation}
  \begin{aligned}
    \ve_1 B_1+[\phi, B_1]=0\,,
\quad
    \ve_2 B_2+[\phi, B_2]=0\,,
\\
\phi I-I a=0\,,
\quad
(\ve_1+\ve_2) J+ a J- J\phi=0
  \end{aligned}
\label{inst-fixed-point-eq}
\end{equation}
for any $(\ve_1,\ve_2,a)\in \text{Lie}\left[U(1)\times U(1)\times T\right]$
for some $\phi=\text{diag}(\phi_1,\ldots,\phi_k)$
parametrizing the Cartan subalgebra
of $\prod_r U(k_r)\subset U(k)$.
Solutions to these equations are known
\cite{Nekrasov:2002qd, Nakajima:2003pg}
and are expressed in terms of Young diagrams $\vec Y$.
See \cite{Bruzzo:2002xf} for explicit expressions for $(B_1, B_2, I,J)$
at the fixed point $\vec Y$.
Here we only need the expressions\footnote{In this subsection we use Greek alphabets $\alpha,\beta,\ldots$
 to denote the $U(N)$ indices,
and use $(i,j)$ to denote the location of a box
in a Young diagram.
} 
for $\phi_s$ \cite{Nekrasov:2002qd}
\begin{equation}
\phi_s= (i_s-1)\ve_1+(j_s-1)\ve_2+a_{\alpha(s)}
\text{ where $\alpha(s)\in \{1,\ldots, N\}$ is such that } s\in Y_{\alpha(s)}\,.
\label{phi-fixed-point}
\end{equation}
Since the fixed point $\vec Y$ in the instanton moduli space
satisfies the general $U(1)^2\times T$-invariance condition
(\ref{inst-fixed-point-eq}) together with (\ref{phi-fixed-point}),
it also satisfies the $U(1)_K$ invariance condition
(\ref{U(1)K-inv-eq}) if $U(1)_K$ is embedded in
$U(1)^2\times U(N)$ in such a way that their actions are compatible.
Since the embedding is given by the substitution
\begin{equation}
  \ve_1\ra -\nu,,
\quad
  \ve_2\ra \nu\,,
\quad
a \ra a+v\nu\,,
\quad
\end{equation}
the $U(1)_K$-invariant fixed points correspond to
\begin{equation}
\vec Y \quad \text{ such that} \quad
K_s=  v_{\alpha(s)}+j_{\alpha(s)}-i_{\alpha(s)}
\label{vec-Y-cond}
\end{equation}
up to a permutation of $s\in \{1,\ldots, k\}$.

To obtain the weights $w_j$ each fixed point
contributes, we can combine the method in 
Section \ref{sec:one-loop} with the known result
for the Dolbeault index at the fixed point.
We recall from that section that
the Dolbeault index on $\mathbb C^2$,
defined by a formal application of the Atiyah-Bott formula,
is given by
\begin{equation}
\text{ind}(\bar D_\text{adj})
=
\sum_{\alpha,\beta=1}^N e_\alpha e_\beta^{-1} \frac{1}{(1-t_1^{-1})
(1-t_2^{-1})}  \,,
\end{equation}
where $e_\alpha = e^{2\pi i a_\alpha}$. 
Let us define
\begin{equation}
  \chi(Y) = \sum_{(i,j) \in Y} t_1^{i-1} t_2 ^{j-1}
\end{equation}
and the conjugate
\begin{equation}
  \chi(Y)^* = \sum_{(i,j) \in Y} t_1^{1 - i} t_2 ^{1 - j}\,.
\end{equation}
Then the local index for the Dolbeault operator at the fixed point $\vec Y$ is
given by
\begin{equation}
\begin{aligned}
\text{ind}(\bar D_\text{adj})_{\vec Y}
 =& \sum_{\alpha,\beta = 1}^{N} e_\alpha e_\beta^{-1}
\left ( \frac{ 1}{(1 - t_1)(1- t_2)} - \chi(Y_{\alpha}) \right) 
\\
&
\quad\quad
\times
\left ( \frac{ 1}{(1 - t_1^{-1})(1- t_2^{-1})} - \chi(Y_\beta)^*
\right)(1-t_1)(1-t_2)\,.
\end{aligned}
\label{eq:character}
\end{equation}
As shown in Section \ref{sec:calc-index},
the one-loop determinant is obtained from 
the non-polynomial part $\text{ind}(\bar D_\text{adj})^\text{1-loop}\equiv
\sum_{\alpha,\beta} e_\alpha e_\beta^{-1} (1-t_1^{-1})^{-1}(1-t_2^{-1})^{-1}
$
of (\ref{eq:character}).
The rest of (\ref{eq:character}) is a Laurent polynomial, which we denote by
$\text{ind}(\bar D_\text{adj})_{\vec Y}^\text{inst}$.
It is nothing but (minus) the character of the tangent space to
the moduli space, and can be rewritten as \cite{Nakajima:2003pg} 
\begin{equation}
\text{ind}(\bar D_\text{adj})_{\vec Y}^\text{inst}
=-\sum_{\alpha,\beta = 1}^{N} e_\alpha e_\beta^{-1}
\left(
\sum_{s\in Y_\alpha} t_1^\text{$-L_{Y_\beta}(s)$} t_2^{A_{Y_\alpha}(s)+1}
+\sum_{t\in Y_\beta} 
t_1^{L_{Y_\alpha}(t)+1} t_2^\text{\raisebox{.5mm}[0mm][0mm]{$-A_{Y_\beta}(t)$}}
\right)\,.
\end{equation}
We have introduced the arm- and leg-lengths
\begin{equation}
  A_Y(s)=\lambda_i-j\,,\quad L_Y(s)=\lambda^T_j-i\,,
\end{equation}
where $\lambda_i$ and $\lambda^T_i$ are the numbers of boxes in the $i$-th row
and column of $Y$, respectively.

Let us denote 
the $U(1)_K$-invariant part  of
$\sum_{n\in\mathbb Z} e^{2\pi in\ve}
((1+t_1^{-1}t_2^{-1})/2) \text{ind}(\bar D_\text{adj})^{\text{1-loop}}$
by $\text{ind}(D^\text{vm})^\text{1-loop}$,
where $U(1)_K$ acts on the gauge bundle with a generator $v$.
It gives rise to
$Z_\text{1-loop}^\text{vm}(v)$
via the rule $\sum_j c_j e^{w_j}\rightarrow
\prod_j w_j^{c_j}$ as in Section \ref{sec:one-loop}.
Similarly we define  
\begin{equation}
\text{ind}(D^\text{vm})^\text{mono}_{\vec Y}
\equiv \text{ $U(1)_K$-invariant part of }
\sum_{n\in\mathbb Z} e^{2\pi in\ve}
\frac{1+t_1^{-1}t_2^{-1}}{2} \text{ind}(\bar D_\text{adj})^\text{inst}_{\vec Y}\,.
\end{equation}
The same rule applied to this gives a contribution to
$Z_\text{mono}^\text{vm}(B,v)$.

The $U(1)_K$-invariant terms arise from the triples
\begin{equation}
(\alpha, \beta, s\in Y_\alpha)
\quad \text{ such that }
\quad
  v_\alpha-v_\beta+L_{Y_\beta}(s)+A_{Y_\alpha}(s)+1=0
\label{alpha-beta-s-cond}
\end{equation}
which contribute
\begin{equation}
-\frac 1{2}
\sum_{n\in \mathbb Z} e^{2\pi in\ve}
(1+e^{-2\pi i\lambda})
  e^{2\pi i(a_\alpha-a_\beta)} e^{\pi i(A_{Y_\alpha}(s)-L_{Y_\beta}(s)+1) \lambda}\,,
\end{equation}
and also from the triples 
\begin{equation}
(\alpha, \beta, t\in Y_\beta) \quad \text{ such that }
\quad
  v_\alpha-v_\beta-L_{Y_\alpha}(t)-A_{Y_\beta}(t)-1=0
\label{alpha-beta-t-cond}
\end{equation}
which contribute
\begin{equation}
-\frac 1{2}
\sum_{n\in \mathbb Z} e^{2\pi in\ve}
(1+e^{-2\pi i\lambda})
  e^{2\pi i(a_\alpha-a_\beta)} e^{\pi i(L_{Y_\alpha}(s)-A_{Y_\beta}(s)+1) \lambda}
\end{equation}
to $\text{ind}(D^\text{vm})_{\vec Y}^\text{mono}$.
By applying the rule $\sum_j c_j e^{w_j}\rightarrow
\prod_j w_j^{c_j}$,
we find the vector multiplet contribution to $Z_\text{mono}^\text{vm}$
\begin{equation}
  \begin{aligned}
z^\text{vec}_{\vec Y}=
&
\prod_{(\alpha, \beta, s)}
\prod_\pm
\left(\sin\left[
\pi 
\left(
a_\alpha-a_\beta+\frac{1}{2}(A_{Y_\alpha}(s)-L_{Y_\beta}(s)\pm 1
)\lambda \right)
\right]\right)^{-1}
\,.
\label{mono-vec}
  \end{aligned}
\end{equation}
We emphasize that
the products are over the triples $(\alpha, \beta, s)$
 satisfying
(\ref{alpha-beta-s-cond}).
The contributions from $(\alpha,\beta,t)$ in
(\ref{alpha-beta-t-cond}) are identical to those from $(\alpha,\beta,s)$
in (\ref{alpha-beta-s-cond}).
Thus the power in (\ref{mono-vec}) is $-1$, not $-1/2$.
The same remark applies to (\ref{mono-adj}) below.

For a single hypermultiplet in the adjoint representation,
we need to consider 
\begin{equation}
  \begin{aligned}
\text{ind}(D^\text{hm}_\text{adj})_{\vec Y}^\text{mono}
&\equiv \text{$U(1)_K$-invariant part of }
 \\
&\quad\quad
-\frac{e^{2\pi im}+ e^{-2\pi im}}{2}
\sum_{n\in\mathbb Z} e^{2\pi in\ve}\,
t_1^{-1/2}t_2^{-1/2} \text{ind}(\bar D_\text{adj})_{\text{inst}}\,.
  \end{aligned}
\end{equation}
From this we get the contribution of an adjoint hypermultiplet
\begin{equation}
  \begin{aligned}
z^\text{adj}_{\vec Y}=
&
\prod_{(\alpha, \beta, s)}
\prod_\pm
\sin\left[
\pi 
\left(
a_\alpha-a_\beta+\frac{1}{2}(A_{Y_\alpha}(s)-L_{Y_\beta}(s)
)\lambda
\pm m \right)
\right]
\,.
  \end{aligned}
\label{mono-adj}
\end{equation}
The product is over the same triples as above.

For a hypermultiplet in the fundamental representation,
we need the Dolbeault index for the corresponding bundle\footnote{When used in instanton counting, this
leads to the contribution of a fundamental hypermultiplet.
}
\begin{equation}
\begin{aligned}
\text{ind}(\bar D_\text{fund})_{\vec Y}
 =& \sum_{\alpha = 1}^{N}  e_\alpha t_1 t_2
\left ( \frac{ 1}{(1 - t_1)(1- t_2)} - \chi(Y_{\alpha}) \right) \,.
\end{aligned}
\label{eq:character-fund}
\end{equation}
Thus the Dirac index is
\begin{equation}
\begin{aligned}
\text{ind}(\bar D_\text{Dirac,fund})_{\vec Y}
 =& \sum_{\alpha = 1}^{N}  e_\alpha t_1^{1/2} t_2^{1/2}
\left ( \frac{ 1}{(1 - t_1)(1- t_2)} - \chi(Y_{\alpha}) \right) \,.
\end{aligned}
\label{eq:character-fund2}
\end{equation}
For a pair
\begin{equation}
(\alpha, s\in Y_\alpha)
\quad \text{ such that} 
\quad
 v_\alpha-i_s+j_s=0\,,
\label{alpha-s-cond}
\end{equation}
it contributes
\begin{equation}
-  e^{2\pi i a_\alpha} e^{\pi i(i_s+j_s-1)\lambda}
\end{equation}
to $\text{ind}(D_\text{DH,fund})_{\vec Y}^\text{mono}$ and thus
\begin{equation}
  \frac 1 2 \sum_{n\in \mathbb Z} e^{2\pi in \ve}
(
e^{2\pi i a_\alpha} e^{\pi i(i_s+j_s-1)\lambda}e^{-2\pi i m}
+
e^{-2\pi i a_\alpha} e^{-\pi i(i_s+j_s-1)\lambda}e^{2\pi i m}
)
\end{equation}
to $\text{ind}(D^\text{hm}_\text{fund})_{\vec Y}^\text{mono}$.
Then
\begin{equation}
  z_{\vec Y}^\text{fund}
(a,m,\lambda;B,v)
=
\prod_{(\alpha,s)}
 \sin\left[\pi\left(
a_\alpha -m
+ \frac{1}{2}\left( i_s+j_s-1
\right)\lambda
\right)\right]\,.
\label{mono-fund}
\end{equation}
Again we stress that the product is over the pairs
$(\alpha,s)$ satisfying
(\ref{alpha-s-cond}).

The total monopole screening contribution is then given by
\begin{equation}
  Z_\text{mono}(a,m_f,\lambda;B,v)
=\sum_{\vec Y} z_{\vec Y}^\text{vec}(a,\lambda;B,v) 
\prod_R\prod_f z_{\vec Y}^R(a,m_f;B,v)\,,
\label{mono-total}
\end{equation}
where the sum is over $N$-tuples of Young diagrams $\vec Y$
satisfying (\ref{vec-Y-cond}), and the product is over the
matter representations $R$.
Explicit expressions for $Z_\text{mono}(B,v)$ will appear
as part of the operator vevs
in Section \ref{sec:gauge-results}.

\section{Gauge theory results}
\label{sec:gauge-results}

For a Wilson operator in an arbitrary representation $R$,
the on-shell action vanishes.
The only saddle point in the path integral is the trivial one,
and the one-loop determinant is $1$ due to Bose-Fermi cancellations.
Thus the expectation value is given by evaluating the holonomy
(\ref{wilson-def}) in the background:
\begin{equation}
  \langle W_R\rangle=\Tr_R \exp\left[
 2\pi i R \left(A_\tau^{(\infty)}+i\Phi_0^{(\infty)}\right)
\right]
=\Tr_R e^{2\pi i a}\,,
\end{equation}
where $a$ was defined in (\ref{def-ab}).

For the 't Hooft operator, we combine the classical, one-loop,
and monopole screening contributions from the previous sections:
\begin{equation}
  \langle T_B\rangle= \sum_{v} e^{2\pi i v\cdot b}
Z_\text{1-loop}(v)
Z_\text{mono}(B,v)\,.
\end{equation}

\subsection{$SU(2)$ $\mathcal N=2^*$}

For $SU(2)$, it is convenient to substitute
\begin{equation}
  a\ra\left(
    \begin{array}{cc}
      a&\\
&-a
    \end{array}
\right)
\,,
\quad\quad
  b\ra\left(
    \begin{array}{cc}
      b&\\
&-b
    \end{array}
\right)
\end{equation}
with the understanding that in the following the symbols $a$ and $b$
are complex numbers rather than matrices.
For this gauge group we can label the line operators
by a pair of integers $(p,q)$, where $p$ and $q$ are
magnetic and electric charges respectively 
\cite{Kapustin:2006hi,Kapustin:2007wm,Drukker:2009tz},
and they are related to the coweight and 
the highest weight of the representation as
\begin{equation}
  \begin{aligned}
 & B=(p/2,-p/2)\equiv \text{diag}(p/2,-p/2)\in \Lambda_{cw}\,,
\\
&(q/2,-q/2)\equiv \text{diag}(q/2,-q/2)\in \Lambda_w\quad
\leftrightarrow \quad\text{spin $q/2$ representation}\,.
  \end{aligned}
\label{charge-dict}
\end{equation}
The most basic Wilson operator $W_{1/2}=L_{0,1}$ 
corresponding to spin $1/2$ has an expectation value
\begin{equation}
  \langle W_{1/2}\rangle=
  \langle L_{0,1}\rangle=
 e^{2\pi i a}+e^{-2\pi i a}\,.
\label{SU22starL01}
\end{equation}
For the minimal 't Hooft operator $T_{1/2}=L_{1,0}$ that is S-dual to $W_{1/2}$,
we find
\begin{equation}
  \langle T_{1/2}\rangle=
  \langle L_{1,0}\rangle=
( e^{2\pi i b}+e^{-2\pi i b})
\left(
\frac{
\sin\left(2\pi a+\pi m\right)
\sin\left(2\pi a-\pi m\right)
}{
\sin\left(2\pi a+\frac{\pi}2 \lambda\right)
\sin\left(2\pi a-\frac{\pi}2 \lambda\right)
}
\right)^{1/2}\,.
\label{SU22starL10}
\end{equation}
For the minimal dyonic loops $L_{1,\pm 1}$,
\begin{equation}
  \langle L_{1,\pm 1}\rangle=
( e^{2\pi i (b\pm a)}+e^{-2\pi i (b\pm a)})
\left(
\frac{
\sin\left(2\pi a+\pi m\right)
\sin\left(2\pi a-\pi m\right)
}{
\sin\left(2\pi a+\frac{\pi}2 \lambda\right)
\sin\left(2\pi a-\frac{\pi}2 \lambda\right)
}
\right)^{1/2}\,.
\label{SU22starL1pm1}
\end{equation}
The simplest example with monopole screening contribution is given by
\begin{equation}
  \begin{aligned}
\langle  L_{2,0}\rangle
=&
(e^{4\pi i b}
+
e^{-4\pi i b})
\frac{
\prod_{s_1,s_2=\pm 1}\sin^{1/2}\left(2\pi a+s_1 \pi m
+s_2 \frac{\pi}{2}\lambda\right)
}{
\sin^{1/2}\left(
2\pi a+\pi\lambda
\right)
\sin^{1/2}\left(
2\pi a-\pi\lambda
\right)
\sin\left(
2\pi a
\right)
}
 \\
&
\quad
+
\sum_{s=\pm}
\frac{\prod_\pm\sin \pi (2a\pm m+ s\lambda/2) }
{\sin( 2\pi a )\sin \pi(2a+s \lambda)}
\,,
  \end{aligned}
\label{SU22starL20}
\end{equation}
where we used (\ref{1-loop-total}) and (\ref{mono-total}).
We observe that this is the Moyal product of the minimal
't Hooft operator vev with itself,
\begin{equation}
  \langle L_{2,0}\rangle
=
\langle L_{1,0}\rangle 
*
\langle L_{1,0}\rangle \,.
\end{equation}
In the $SU(2)$ case $*$ is defined by
\begin{equation}
(f*g)(a,b)\equiv 
e^{i\frac{\lambda}{8\pi}
(
\partial_{b} \partial_{a'}
-
\partial_a \partial_{b'}
)
}f(a,b) g(a',b')
|_{a'=a,b'=b}
\end{equation}
 with a different coefficient due to the factor of 2
in the inner product 
\begin{equation}
a\cdot b\ra \Tr[\text{diag}(a,-a)\cdot \text{diag}(b,-b) ]=2ab\,.
\label{ab-prod}
\end{equation}
In Section \ref{sec:non-com}, we will explain how the Moyal product appears
from the structure of the path integral.

The precise choice of signs and  relative numerical normalizations among terms
is difficult to fix purely in gauge theory without additional assumptions.
In the examples considered in this paper we choose to be pragmatic and 
make the choice by assuming physically reasonable structures such as
Moyal multiplication, correspondence with the Verlinde operators,
as well as  agreement with classical $SL(2,\mathbb C)$ holonomies in the 
$\lambda \rightarrow 0$ limit.

\subsection{$U(N)$  $\mathcal N=2^*$}
For the gauge group $U(N)$, the minimal 't Hooft operators,
with charges\footnote{The Cartan subalgebra of $U(N)$ is spanned by
real diagonal matrices.
For $SU(N)$ they must be traceless.
We often drop ``diag''
in $a=\text{diag}(a_1,\ldots,a_N)$ to simplify notation.
The inner product is defined by the trace $a \cdot a'=\Tr\, a a'$,
and this is used to identify the Cartan algebra with its dual.
}
 $B=(\pm 1, 0^{N-1})$ (the power indicates the number of repeated entries) corresponding to the
fundamental and anti-fundamental representations of the Langlands dual group,
have the expectation values
\begin{equation}
  \langle T_{B=(\pm 1, 0^{N-1})}\rangle
=
\sum_{l=1}^N 
e^{\pm 2\pi i   b_l}
\left(
\prod_\pm \prod_{j\neq l} 
\frac{
\sin\pi(a_l-a_j\pm m)
}{
\sin\pi(a_l-a_j\pm \lambda/2)
}
\right)^{1/2}\,.
\label{UN2starBpm10N-1}
\end{equation}
For the magnetic charge $B=(1,-1,0^{N-2})$, corresponding
to the adjoint representation,\begin{equation}
\begin{aligned}
&  \langle T_{B=(1,-1,0^{N-2})}\rangle
\\
=&
\sum_{k\neq l}
e^{2\pi i (b_k-b_l)}
\left[
\frac{
\left[
{\displaystyle \prod_{\pm,\pm}} \sin\pi\left( a_{kl}\pm m\pm  \lambda/ 2\right)
\right]
\left[
{\displaystyle
\prod_{\pm}
\prod_{j\neq k,l}
}
\sin\pi\left( a_{kj}\pm m\right)
\sin\pi\left( a_{lj}\pm m\right)
\right]
}{
\sin^2\pi a_{kl}
{\displaystyle \prod_{\pm}}
\sin\pi\left(a_{kl}\pm \lambda\right)
\left[
{\displaystyle
\prod_{\pm}
\prod_{j\neq k,l}
}
\sin\pi\left( a_{kj}\pm \lambda/  2\right)
\sin\pi\left( a_{lj}\pm\lambda/ 2\right)
\right]
}
\right]^{1/2}
\\
&
\quad\quad
+
\sum_{l=1}^N
\prod_{j\neq l}
\frac{
\prod_\pm
\sin\pi( a_{lj}\pm m+\lambda/2)
}{
\sin \pi a_{lj}\sin\pi (a_{lj}+\lambda)
}
\,.
\end{aligned}
\label{UN2starB1-10N-2}
\end{equation}
From (\ref{UN2starBpm10N-1}) and (\ref{UN2starB1-10N-2}) we find that
  \begin{equation}
  \langle T_{B=(1,-1,0^{N-2})}\rangle=
  \langle T_{B=(-1,0^{N-1})}\rangle*
  \langle T_{B=(1,0^{N-1})}\rangle\,.
  \end{equation}
For $B=(2,0^{N-1})$,
\begin{equation}
\begin{aligned}
&\qquad  \langle T_{B=(2,0^{N-1})}\rangle
\\
&=
\sum_{k=1}^N e^{4\pi i b_k}
\left(
\prod_{j\neq k}
\frac{
\prod_{\pm,\pm}
\sin\pi (a_{kj}\pm m \pm \lambda/2)
}
{
\left[
\sin^2\pi a_{kj}
\prod_\pm\sin\pi (a_{kj}\pm \lambda)
\right]
}
\right)^{1/2}
\\
&
\quad +
 \sum_{k\neq l} e^{2\pi i (b_k+b_l)}
  \left(
\frac{
\prod_{j\neq k,l}
\prod_\pm 
\sin\pi( a_{kj}\pm m)
\sin\pi( a_{lj}\pm m)
}{
\prod_{j\neq k,l}
\prod_\pm 
\sin\pi( a_{kj}\pm \lambda/2)
\sin\pi( a_{lj}\pm \lambda/2)
}
\right)^{1/2}
\\
&
\qquad \qquad\times
\frac{
\prod_\pm \sin\pi (a_{kl}\pm m+\lambda/2)
}
{
\sin\pi (a_{kl}+\lambda)
\sin\pi a_{kl}
}
\,.
\end{aligned}
\end{equation}
For this we find
\begin{equation}
  \langle T_{B=(2,0^{N-1})}\rangle
=
  \langle T_{B=(1,0^{N-1})}\rangle
*
  \langle T_{B=(1,0^{N-1})}\rangle\,.
\end{equation}
Results for the gauge group $SU(N)$ can be obtained by taking
$a$ and $b$  traceless.

\subsection{$U(2)$  $N_\text{F}=4$ }

For the minimal 't Hooft operator in this theory, we have
\begin{equation}
  \begin{aligned}
  \langle T\rangle 
=&
e^{\pi i b_{12}}
\left(
\frac{
\prod_{f=1}^4 \sin\pi(a_1-m_f)\sin\pi(a_2-m_f)}
{
\sin^2\pi a_{12} \prod_\pm \sin\pi(a_{12}\pm\lambda)
}
\right)^{1/2}
\\
&+
e^{-\pi i b_{12}}
\left(
\frac{
\prod_{f=1}^4 \sin\pi(a_1+m_f)\sin\pi(a_2+m_f)}
{
\sin^2\pi a_{12} \prod_\pm \sin\pi(a_{12}\pm\lambda)
}
\right)^{1/2}
\\
&+
  \frac{
\prod_{f=1}^{4}
\sin \pi
\left(
a_1-m_f +\frac{\lambda}{2}
\right)
}
{
\sin \pi a_{12}
\sin \pi \left(-a_{12}-\lambda\right)
}
+
  \frac{
\prod_{f=1}^{4}
\sin \pi
\left(
a_2-m_f +\frac{\lambda}{2}
\right)
}
{
\sin \pi a_{21}
\sin \pi \left(-a_{21}-\lambda\right)
}\,.
  \end{aligned}
\label{U2NF4}
\end{equation}
We have defined $a_{jk}=a_j-a_k$.

\subsection{$U(N)$  $N_\text{F}=2N$ }
\label{sec:UNNF=2N}

For the minimal 't Hooft operator given by
the magnetic charge $B=\text{diag}(1,-1,0^{N-2})$ corresponding to
the adjoint representation,
 we obtain
\begin{equation}
  \begin{aligned}
&  \langle T_B\rangle
\\
=&
\mathop{\sum_{1\leq k,l\leq N}}_{k\neq l}
 e^{\pi i(b_k-b_l)}
\frac{
\left[
\prod_{f=1}^N 
\sin\pi (a_k-m_f)
\sin\pi (a_l-m_f)
\right]^{1/2}
}{
\sin\pi a_{kl}
{\displaystyle \prod_\pm }
\left[
\sin \pi (a_{kl}\pm\lambda)
{\displaystyle
\prod_{j\neq k,l}}
\sin\pi(a_{kj}\pm \lambda/2)
\sin\pi(a_{jl}\pm \lambda/2)
\right]
^{1/2}
}
\\
&\quad\quad
+
\sum_{l=1}^N
  \frac{
\prod_{f=1}^{2N}
\sin \pi
\left(
a_l-m_f +\frac{\lambda}{2}
\right)
}
{
\prod_{j\neq l} 
\sin \pi a_{lj}
\sin \pi\left(- a_{lj}-\lambda\right)
}\,.
  \end{aligned}
\label{UNSQCD}
\end{equation}
We have introduced the notation $a_{jk}\equiv a_j-a_k$.

We emphasize that  (\ref{UNSQCD}) and (\ref{U2NF4}) are the vev of the 't Hooft operator
in the $U(N)$ and $U(2)$ theories, not in the $SU(N)$ and $SU(2)$ theories.
We will compare (\ref{UNSQCD}) and (\ref{U2NF4})
with the Verlinde operators in Toda and Liouville theories
in Section \ref{sec:S4-CFT} that we will propose to
be related to the line operators in the $SU(N)$ and $SU(2)$ theories.
While we do not have a computational method intrinsic to $SU(N)$,
we will see that (\ref{UNSQCD}) and (\ref{U2NF4}), when $a$ is restricted
to be traceless, do reproduce $a$-dependent
terms in the CFT results.

\section{Noncommutative algebra and quantization}
\label{sec:non-com}

By using the structure of the path integral  we have
found, in this section we show that the vevs of the line operators
on $S^1\times \mathbb R^3$, inserted on the 3-axis ($x^1=x^2=0$),
 form a non-commutative algebra,
when the axis is considered as time
and the operators are time-ordered.
We will begin with the $U(1)$ case and then discuss the general
gauge group.
\subsection{Maxwell theory}
\label{sec:non-com-U1}
Let us explain how non-commutativity arises
in the algebra of Wilson-'t Hooft
operators in Maxwell theory on $S^1\times \mathbb R^3$
upon twisting by a spatial rotation along the $S^1$.

We begin with an intuitive explanation based on classical fields
\cite{Gaiotto:2010be}.
By taking $S^1$ as time,
the expectation value of the product of Wilson ($W$) 
and 't Hooft ($T$) operators
can be thought of as the trace
\begin{equation}
     \langle W\cdot T\rangle= \Tr_{\mathcal H(W\cdot T)} (-1)^F e^{-2\pi R H}
e^{2\pi i\lambda J_3}
\end{equation}
taken in the Hilbert space $\mathcal H(W\cdot T)$
defined by the line operators.
The space $\mathcal H(W\cdot T)$ differs from
the simple product $\mathcal H(W)\otimes \mathcal H(T)$
because when both $W$ and $T$ are present,
their electric and magnetic fields
produce the Poynting vector $\vec E\times \vec B$ that carries
a non-zero angular momentum.
The orientation of the Poyinting vector,
and therefore the phase  $e^{2\pi i\lambda J_3}$,
 depends on the relative
 positions of the operators on the 3-axis.

Next we present an approach
suitable for localization.
For simplicity let us turn off the theta angle.
The line operator $L_{p,q}$ with magnetic and electric charges $(p,q)$ 
at the origin $\vec x=0$
is
defined by the path integral over the fluctuations around the 
singular background
\begin{equation}
  A
=
A^{(\infty)}_\tau d\tau
+p \frac{\cos\theta}2
d\varphi
\label{U1-background}
\end{equation}
with the insertion of the holonomy
\begin{equation}
  e^{-i q\oint_{S^1}A}\,.
\label{U1-Wilson}
\end{equation}
We note here that the expression for the monopole field 
in (\ref{U1-background})
has Dirac strings in two directions ($\theta=0,\pi$).
The expectation value $\langle L_{p,q}\rangle$
is a function of $(a,b)$, which are normalized
 electric and magnetic background Wilson lines
\begin{equation}
  a\equiv R A_\tau^{(\infty)}\,,
\quad\quad
b\equiv\frac{\Theta}{2\pi}\,.
\end{equation}
We claim that the path integral yields the expectation value
\begin{equation}
  \langle L_{p,q}\rangle=e^{-2\pi i (q a+ p b)}\,.
\end{equation}
The magnetic part is essentially the definition of the magnetic
Wilson line $\Theta$, which is defined as the chemical potential for the
magnetic charge at infinity.
The electric part arises because the holonomy
(\ref{U1-Wilson}) is evaluated against the background Wilson line.

Let us introduce a twist along the $S^1$.  
If we think of the circle as the time direction,
we can write
\begin{equation}
   \langle L_{p,q}\rangle= \Tr_{\mathcal H(L_{p,q})} (-1)^F e^{-2\pi R H}
e^{2\pi i\lambda J_3}\,,
\end{equation}
where $J_3$ is the Cartan
generator of the spatial rotation group $SU(2)$.
The twist by $J_3$ means that we rotate the system by angle
$2\pi \lambda$ as we go along $S^1$,
i.e., we introduce the identification
\begin{equation}
  (\tau+2\pi R, \varphi)\sim (\tau, \varphi+2\pi\lambda)\,.
\end{equation}
In terms of the new coordinates 
$(\tau',\varphi')=(\tau,\varphi+
\frac{\lambda}{R}\tau
)$, the identification is simply
\begin{equation}
  (\tau'+2\pi R,\varphi')\sim (\tau', \varphi'
)\,.
\end{equation}
The components of the gauge field are related as
\begin{equation}
  A_{\tau'}=A_{\tau}-\frac{\lambda}{R} A_{\varphi}\,,
\quad
A_{\varphi'}=A_\varphi\,.
\end{equation}
Note that $A_\varphi$ represents a holonomy around the Dirac strings.
In our choice of local trivialization $A_\varphi(\theta=\pi/2)=0$,
so we have a simple relation
\begin{equation}
 \oint_{S^1} A=  2\pi a\quad\text{ at $\quad\theta=\pi/2$.}
\end{equation}
Thus the monopole field does not contribute to the holonomy as claimed above,
in fact even after twisting.
The holonomies at $\theta\neq \pi/2$ are, however, shifted from $a$.
Indeed we find
\begin{equation}
 \oint_{S^1} A=  \int_0^{2\pi R} d\tau' A_{\tau'}=2\pi
\left( a \mp  \frac{p}{2}\lambda\right)
\quad
\text{ at }
\quad
\theta=
\left\{
  \begin{array}{lll}
0\,,\\
\pi\,.
  \end{array}
\right.
\end{equation}
One can picture the shift as arising from the holonomy winding around
the Dirac strings.
Then for the product of Wilson and 't Hooft
operators $W\equiv L_{0,1}, T\equiv L_{1,0}$
\begin{equation}
 W(\vec x=(0,0, z))
\cdot
T(\vec x=0)\,,
\end{equation}
its expectation value is given by
\begin{equation}
\langle W(\vec x=(0,0, z))
\cdot
T(\vec x=0)\rangle
=
e^{-2\pi i (a\mp \frac 1 2 \lambda)} e^{-2\pi i b}
\text{ for }
\left\{
  \begin{array}{lll}
z>0\,,\\
z<0\,.
  \end{array}
\right.
\end{equation}
The Wilson line operator (\ref{U1-Wilson}) for $z>0$
is evaluated at $\theta=0$,
and for $z<0$ at $\theta=\pi$.
The difference $\lambda$ between the shifts in $a$ at $z>0$ and $z<0$
 is independent of the choice of local trivialization.
We can also see that the expectation value 
of the product of operators is given by the Moyal product
of the expectation values:
\begin{equation}
\langle W(z)
\cdot
T(0)\rangle
=
\left\{
  \begin{array}{lll}
\langle W\rangle *
\langle T\rangle
\quad
\text{ for $z>0$}\,,
 & 
    \\
\langle T\rangle
*
\langle W\rangle
\quad
\text{ for $z<0$}\,,
  \end{array}
\right.
\end{equation}
where the Moyal product $*$ is defined by
\begin{equation}
(f*g)(a,b)\equiv  \lim_{a'\ra a,\,b'\ra b} e^{i\frac{\lambda}{4\pi}
(
\partial_{b} \partial_{a'}
-
\partial_a \partial_{b'}
)
}f(a,b) g(a',b')\,.
\end{equation}
This is the special case of the more general result
for an arbitrary gauge group that we now turn to.

\subsection{Non-Abelian gauge theories}

Here we consider a general $\mathcal N=2$ gauge theory with
arbitrary matter content.
Let us suppose that we have multiple line operators
$L_i\equiv L_{B_i,R_i}(\vec x=(0,0,z_i))$ located at various points
$\vec x=(0,0,z_i)$ on the 3-axis, ordered so that
\begin{equation}
  z_1>z_2>\ldots > z_n\,.
\end{equation}
In the localization calculation, it suffices to consider
the Abelian configurations with magnetic charges $v_i$
associated with $B_i$ as only these
contribute to the path integral.
As is clear from the Maxwell case, 
the holonomy at $z_i$ around $S^1$ 
is shifted by the magnetic fields $\propto v_j$ created by
$L_j$ for $j\neq i$:
\begin{equation}
  a \ra 
 a +\frac \lambda 2 \left(
\sum_{j<i} v_j-\sum_{j>i} v_j
\right)\,.
\end{equation}
Let us assume that the individual operator vevs are given by
\begin{equation}
\langle L\rangle
=
\sum_{v,w}
  Z_{L,\text{total}}(a,b;v,w)
\equiv
\sum_{v,w}
e^{2\pi i(w\cdot a+v \cdot b)}
Z_L(a,m_f,\lambda;v,w)
\end{equation}
for some functions $Z_L(a,m_f,\lambda;v,w)$.
Then localization calculation yields
\begin{equation}
  \langle L_1\cdot L_2\cdot \ldots \cdot L_n\rangle
=
\prod_{i=1}^n
\sum_{w_i}
\sum_{v_i}
Z_{L_i,\text{total}}
\left(a +\frac \lambda 2 \left(
\sum_{j<i} v_j-\sum_{j>i} v_j\right),
b\,
;v_i,w_i\right)\,,
\label{product-vev}
\end{equation}
One can easily see that (\ref{product-vev}) is the Moyal product
of the expectation values of individual operators
\begin{equation}
    \langle L_1\cdot L_2\cdot \ldots \cdot L_n\rangle
=   \langle L_1\rangle *\langle L_2\rangle * \ldots *
\langle L_n\rangle\,,
\end{equation}
where $*$ is defined by 
\begin{equation}
(f*g)(a,b)\equiv 
\left.
e^{i\frac{\lambda}{4\pi}
(
\partial_{b}\cdot \partial_{a'}
-
\partial_a\cdot \partial_{b'}
)
}f(a,b) g(a',b')
\right|_{a'=a,b'=b}
\end{equation}
with the natural product $\cdot$ between the derivatives inside the exponential.

As a concrete example, let us consider $SU(2)$ $\mathcal N=2^*$ theory.
We computed the vev of the charge-two 't Hooft operator in
(\ref{SU22starL20}).
As explained in \cite{Gomis:2011pf},
this operator corresponds to the product
of two minimal 't Hooft operators.
This is because the resolution of the singular
moduli space corresponds to separating the charge-two 't Hooft
operator into two minimal ones \cite{Kapustin:2006pk}.
Indeed one can check that the expression
(\ref{SU22starL20}) is precisely the Moyal product of
(\ref{SU22starL10}) with itself.

\subsection{Deformation quantization of the Hitchin moduli space}
\label{sec:Hitchin}

We are now going to explain that the noncommutative algebra structure
given by the Moyal multiplication above realizes a deformation quantization
of the Hitchin moduli space associated with the gauge theory.

In \cite{Gaiotto:2009we,Gaiotto:2009hg}, a correspondence between
certain $\mathcal N=2$ gauge theories and
punctured Riemann surfaces $C$ was discovered.
The correspondence is a main ingredient of the relation
\cite{AGT} between gauge theories and two-dimensional conformal field theories.
The correspondence is also manifested in the relation between
the gauge theories and the Hitchin systems on the Riemann surfaces.
This made it possible to study the integrable structure
\cite{Nekrasov:2009uh,2009AIPC.1134..154N,Nekrasov:2009ui}
as well as the low-energy dynamics
of these theories using the Hitchin system
on the Riemann surfaces \cite{Gaiotto:2009hg},
generalizing \cite{Donagi:1995cf}.

Let $A
=A_z dz+A_{\bar z} d\bar z
$ be a connection of a $G$-bundle over $C$,
and 
$\varphi=\varphi_z dz+\bar\varphi_{\bar z} d\bar z$
 an adjoint-valued 1-form.
They are assumed to possess prescribed singularities at the punctures.
The Hitchin moduli space is the space of solutions to
\begin{equation}
  \begin{aligned}
&
\hspace{6mm}
F_{z\bar z}=[\varphi_z,\bar\varphi_{\bar z}]\,,    \\
& D_{\bar z}\varphi_z=0\,,\quad
 D_z \bar \varphi_{\bar z}=0\,,
  \end{aligned}
\label{Hitchin-eq}
\end{equation}
up to $G$-gauge transformations.
The Hitchin moduli space is hyper\Kahler, 
and therefore has a $\mathbb{CP}^1$ of complex structures $\mathcal J$,
each being a linear combination of three complex structures 
$\mathcal J=I, J$, and $K$.
Each complex structure $\mathcal J$ is associated with a real symplectic form
$\omega_\mathcal J:=g \mathcal J$, as well as a holomorphic 
symplectic form $\Omega_{\mathcal J}$.
For $\mathcal J=I, J, K$, these are given by
$\Omega_I=\omega_J+i\omega_K,
\Omega_J=\omega_K+i\omega_I,
\Omega_K=\omega_I+i\omega_J$.

In the original assignment of $I,J,K$ by Hitchin \cite{MR887284},
we are particularly interested in the complex structure $J$.
The combination $\mathcal A\equiv A+i\varphi$ is then holomorphic,
and  (\ref{Hitchin-eq}) implies that\footnote{More precisely, the first of (\ref{Hitchin-eq}) combined with
the difference of the second and the third
is equivalent to the flatness of $\mathcal A$.
The $J$-holomorphic structure of the Hitchin moduli space
can be described by dropping the sum
of the second and the third equations,
and by taking the quotient with respect to $G_{\mathbb C}$
gauge transformations.
}
$\mathcal A$ is a flat $G_{\mathbb C}$ connection.
In terms of $\mathcal A$, $\Omega_J$ is given by
\begin{equation}
  \Omega_J\propto\int_C \Tr\, \delta \mathcal A\wedge\delta \mathcal A\,.
\end{equation}
The $U(1)$ R-symmetry rotates the phases of $\varphi_z,
\bar\varphi_{\bar z}$, and $\Phi_0+i\Phi_9$,
and $\Omega_J$ transforms accordingly \cite{Gaiotto:2009hg}.

We focus on the one-punctured torus, which corresponds to
$SU(2)$ $\mathcal N=2^*$ theory.
Let us define generators of the first homology 
 so that the holonomy matrices $(A, B,M)$
along them satisfy the relation
\begin{equation}
  AB=MBA\,.
\end{equation}
Here $M$ is the holonomy around a small circle surrounding the puncture,
and $A$ and $B$ are the holonomy matrices for the 
usual A- and B-cycles.
Dehn's theorem \cite{dehn-breslau,MR956596}
allows us to label the non-self-intersecting closed curves
 by two integers $(p,q)$ with equivalence
$(p,q)\sim (-p,-q)$.
They can be naturally identified with
the charges of
line operators in (\ref{charge-dict}) \cite{Drukker:2009tz}.
In particular, we have the correspondence
\begin{eqnarray}
  \langle L_{0,1}\rangle &\leftrightarrow& \Tr A\,,\\
  \langle L_{1,0}\rangle &\leftrightarrow& \Tr B\,,\\
  \langle L_{1,\pm 1}\rangle &\leftrightarrow& \Tr A^{\pm 1}B\,.
\label{SU22star-corresp}
\end{eqnarray}

Let us consider the case $\lambda=0$.
From (\ref{SU22starL01}-\ref{SU22starL1pm1}) we find that
\begin{eqnarray}
  \langle L_{0,1}\rangle_{\lambda=0}&=&
 e^{2\pi i a}+e^{-2\pi i a}\,,
\label{2star01}
\\
  \langle L_{1,0}\rangle_{\lambda=0}&=&
( e^{2\pi i b}+e^{-2\pi i b})
\left(
\frac{
\sin\left(2\pi a+\pi m\right)
\sin\left(2\pi a-\pi m\right)
}{
\sin^2\left(2\pi a\right)
}
\right)^{1/2}\,,
\label{2star10}
\\
  \langle L_{1,\pm 1}\rangle_{\lambda=0}
&=&
( e^{2\pi i (b\pm a)}+e^{-2\pi i (b\pm a)})
\left(
\frac{
\sin\left(2\pi a+\pi m\right)
\sin\left(2\pi a-\pi m\right)
}{
\sin^2\left(2\pi a\right)
}
\right)^{1/2}\,.
\label{2star1pm1}
\end{eqnarray}
Replacing the arrows in (\ref{SU22star-corresp})
by equalities,
these expressions were exactly given as the definition of
the Darboux coordinates\footnote{In \cite{Nekrasov:2011bc} the Darboux coordinates were denoted by
 $(\alpha, \beta)$, and are related to our $(a,b)$ 
 by a trivial rescaling.
We also have $\Tr M=2\cos \pi m$.
} $(a,b)$ on the Hitchin moduli space with respect to
the symplectic structure $\Omega_J$!
Later in \cite{Dimofte:2011jd}, $(a,b)$
were identified with the complexification of the Fenchel-Nielsen
coordinates of Teichm\"uller space.
Here we see that both the coordinates $(a,b)$ and the symplectic 
structure $\Omega_J$ arise naturally from the gauge theory on $S^1\times 
\mathbb R^3$.

For $SU(2)$ $N_\text{F}=4$ theory, our gauge theory calculation
of the 't Hooft and dyonic operator vevs is not complete
due to the difficulty with monopole screening contributions.
The relation with Liouville theory and the formula
(\ref{Zmono-SU2NF4}) below suggests, however, that
 $(a,b)$ are the complexified Fenchel-Nielsen coordinates
on the Hitchin moduli space
associated with the four-punctured sphere \cite{Cherkis:2000ft}.

\section{Gauge theory on  $S^4$
and  Liouville/Toda theories}
\label{sec:S4-CFT}

In this section we propose a precise relation between
the line operator vevs on $S^1\times \mathbb R^3$
and the corresponding difference operators
that act on the conformal blocks of Liouville and Toda
field theories.
We first motivate the correspondence by gauge theory considerations.
Then we will give an algorithm for computing 
the line operator vevs on $S^1\times \mathbb R^3$
using two-dimensional CFT.

Let us consider the Liouville theory on 
a genus $g$ Riemann surface with $n$ punctures
$C_{g,n}$.
The correlation function of primary fields $V_{\alpha_e}$
($e=1,\ldots, n$)
with momenta $\alpha_e$,
inserted at the punctures,
 takes the form
\begin{equation}
\left \langle \prod V_{\alpha_e}
\right\rangle_{C_{g,n}}=
\int \left[\prod d\alpha_i \right]
\mathcal C(\alpha_i\,;\alpha_e)
|\mathcal F(\alpha_i;\alpha_e)|^2\,,
\label{CFT-corr}
\end{equation}
where the integral is over internal momenta $\alpha_i$
($i=1,\ldots, 3g-3+n$) and the function
$\mathcal C(\alpha_i\,;\alpha_e)$ is
a product of DOZZ three-point functions \cite{Dorn:1994xn,Zamolodchikov:1995aa}.
The conformal block $\mathcal F (\alpha_i;\alpha_e)$
depends on $\alpha_i, \alpha_e$, and the gluing parameters
$q_i$ holomorphically.
The central charge $c$ of Liouville theory is parametrized
as
\begin{equation}
  c=1+6Q^2\,,
\quad\quad
Q={\mathsf b}+ {\mathsf b}^{-1}\,.
\end{equation}

The AGT correspondence \cite{AGT} states that
for ${\mathsf b}=1$ the correlation function (\ref{CFT-corr}) 
coincides with
the partition function of the corresponding $\mathcal N=2$
gauge theory on the round sphere $S^4$ as defined by Pestun in \cite{Pestun:2007rz}.
The gluing parameters $q_i$ are related
to the complexified couplings 
$\tau_i=\frac{\theta_i}{2\pi}+ \frac{4\pi i}{g_i^2}$
as $q_i=e^{2\pi i\tau_i}$.
Pestun's partition function contains as the north and south pole contributions
the Nekrasov instanton partition functions
defined in the Omega background \cite{Nekrasov:2002qd} .
The parameter ${\mathsf b}$ is related to
the equivariant parameters $\ve_1, \ve_2$
of the Omega background 
as ${\mathsf b}^2= \ve_1/\ve_2$.
The path integral formulation
of the deformation to ${\mathsf b}\neq 1$ is unknown at the time of writing.
Even for ${\mathsf b}\neq 1$,
it is expected that
(\ref{CFT-corr}) will be reproduced by
the partition function of the $\mathcal N=2$
gauge theory on a deformed sphere $S^4_{\mathsf b}$, deformed in
a certain way by a parameter $\mathsf b$.

For an $\mathcal N=2$ gauge theory with $SU(2)$ gauge groups
associated with a punctured Riemann surface $C_{g,n}$,
there is a correspondence between the charges $(B,R)$
of Wilson-'t Hooft operators and 
a collection $\gamma$ of non-self-intersecting 
closed curves on $C_{g,n}$ \cite{Drukker:2009tz}.
In \cite{Alday:2009fs,Drukker:2009id}
it was shown that there exists a difference operator $\Lambda_\gamma$,
the Verlinde operator, whose action
on $\mathcal F(\alpha_i;\alpha_e)$ we denote by\footnote{For the Verlinde operator in Liouville theory,
corresponding to a connected closed curve on the Riemann surface,
our normalization of the operator
agrees with \cite{Alday:2009fs}.
Our operator is $2\cos \sfb Q$ times those in \cite{Drukker:2009id,Gomis:2010kv}.
}
\begin{equation}
  \mathcal F(\alpha_i;\alpha_e)
\ra
[\Lambda_\gamma\cdot  \mathcal F]
(\alpha_i;\alpha_e)
\,.
\end{equation}
The same papers demonstrated that, for $\mathsf b=1$,  the expectation value of the
Verlinde operator defined as
\begin{equation}
\int \left[\prod d\alpha_i \right]
\mathcal C(\alpha_i\,;\alpha_e)
\overline{\mathcal F(\alpha_i;\alpha_e)}
[\Lambda_\gamma\cdot  \mathcal F]
(\alpha_i;\alpha_e)
\label{Verlinde-vev-F}
\end{equation}
reproduces the expectation value of the Wilson loop on $S^4$
computed by Pestun \cite{Pestun:2007rz}.

The agreement of (\ref{Verlinde-vev-F}) with the 't Hooft loop
expectation value, again for $\mathsf b=1$,  was
more recently verified in \cite{Gomis:2011pf}.
This was done by performing a localization calculation for 't Hooft loops
placed along a large circle, called the equator, of $S^4$.
The neighborhood of the equator is approximately $S^1\times \mathbb R^3$,
 therefore much of the analysis overlaps the present paper.
Because of the curvature, however, 
in the orthonormal frame of the metric such that
the conformal Killing spinor is periodic,
the hypermultiplet becomes antiperiodic. 
We expect that this property persists for ${\mathsf b}\neq 1$.
The effect of antiperiodicity is to multiply
the right hand side of (\ref{ind-Dhm}) by an extra factor  $e^{ i \ve/2}$.
Thus we conjecture, and the AGT correspondence suggests,
that the 't Hooft loop vev on $S^4_{\mathsf b}$
for general $\mathsf b$ is given by
\begin{equation}
  \begin{aligned}
  \langle T_B\rangle_{S^4_{\mathsf b}}&
= \int_{i \mathfrak t} da
\sum_v
Z_\text{pole}(a+{\mathsf b}^2v/2,\bar q)
Z_\text{equator}(a;B,v)
 Z_\text{north}(a-{\mathsf b}^2v/2,q)
\\
&= \int_{i \mathfrak t} da\,
Z_\text{pole}(a,\bar q)
\sum_v
e^{-(\mathsf b^2/2) v\cdot \partial_a}
Z_\text{equator}(a;B,v)
e^{-(\mathsf b^2/2) v\cdot \partial_a}
 Z_\text{north}(a,q)
\,,
  \end{aligned}
\label{thooft-S4}
\end{equation}
where
\begin{equation}
Z_\text{pole}(a,q)=e^{-S_\text{cl}(a)}
Z_\text{1-loop}(a)
Z_\text{inst}(a,q)
\label{poles}
\end{equation}
is the Nekrasov partition function
including the classical, one-loop, and instanton contributions
on $\mathbb C^2$,\footnote{Since we are primarily interested in 
the equator contributions we suppress the dependence
on
$\ve_1$, $\ve_2$, $\mathsf m_f$, and $\lambda$
in (\ref{poles}).
The mass parameters $\mathsf m_f$
as well as the Coulomb moduli $a$ are pure imaginary.
} 
 and the equator contribution
\begin{equation}
  Z_\text{equator}(a;B,v)
= Z_\text{1-loop}(a,{\mathsf m}_f+1/2,{\mathsf b}^2;v)
 Z_\text{mono}(a,{\mathsf m}_f+1/2,{\mathsf b}^2;B,v)\,.
\label{eq-cont}
\end{equation}
is given in terms of the one-loop determinant in (\ref{1-loop-total}) 
and the monopole screening contribution in (\ref{mono-total}).
The shift in mass is due to the antiperiodicity of hypermultiplets
mentioned above.
For ${\mathsf b}=1$ (\ref{thooft-S4}) was established in \cite{Gomis:2011pf},
where the definitions of $Z_\text{1-loop}$ and $Z_\text{mono}$
were slightly different due to the the shift in the mass.
The second equality in (\ref{thooft-S4}) involves 
a shift of integration contours
and integration by parts.
For some examples in Liouville theory, 
it was checked that the shift of contours
does not encounter poles \cite{Gomis:2011pf}.

Based on the conjectured relation (\ref{thooft-S4}) between the line operator vevs on $S^4_{\mathsf b}$
and $S^1\times \mathbb R^3$
for general $\lambda={\mathsf b}^2$,
we propose that
the vev in the theory on $S^1\times \mathbb R^3$ can be
obtained from the Verlinde operator in Liouville theory
by the following algorithm.
This algorithm was used in \cite{Gomis:2011pf} in the case ${\mathsf b}=1$
to read off $Z_\text{equator}$ from the Verlinde operator.
First we  change the normalization of the conformal block and define\footnote{The Verlinde operator in \cite{Alday:2009fs}
was computed in the standard normalization for conformal blocks
\cite{Belavin:1984vu}.
Yet another normalization introduced in \cite{Ponsot:1999uf} was
used to calculate the Verlinde operators in \cite{Drukker:2009id},
and in this basis the operators are free of square-roots \cite{Dimofte:2011jd}.
}
\begin{equation}
  \mathcal B(\alpha_i;\alpha_e)\equiv \mathcal C(\alpha_i;\alpha_e)^{1/2}
\mathcal F(\alpha_i;\alpha_e)
\label{B-block}
\end{equation}
using the square root of the function $\mathcal C(\alpha_i;\alpha_e)$
that appears in the correlation function (\ref{CFT-corr}).
With the one-loop factor in (\ref{poles}) whose precise definition
was given in  \cite{Gomis:2011pf},
we expect that
$\mathcal B(\alpha)=Z_\text{pole}(a)$
with the identification $\alpha_i=Q/2+a_i/{\mathsf b}$.
In this normalization, the Liouville correlation function is 
simply given by
\begin{equation}
\left \langle \prod V_{\alpha_e}
\right\rangle_{C_{g,n}}=
\int \left[\prod d\alpha_i \right]
\,
|\mathcal B(\alpha_i;\alpha_e)|^2\,,
\label{CFT-corr2}
\end{equation}
where we used the fact that in the physical range of
Liouville momenta, the function $\mathcal C(\alpha_i;\alpha_e)$ is real.
The Verlinde operator acts on $\mathcal B(\alpha)$ as
the difference operator defined by
\begin{equation}
  [\mathcal L_\gamma\cdot  \mathcal B]
(\alpha_i;\alpha_e)
\equiv
\mathcal C(\alpha_i;\alpha_e)^{1/2}
[\Lambda_\gamma\cdot \mathcal F](\alpha_i;\alpha_e)\,.
\label{Ver-on-B}
\end{equation}
Its vev is then given by
\begin{equation}
\int \left[\prod d\alpha_i \right]
\overline{\mathcal B(\alpha_i;\alpha_e)}
[\mathcal L_\gamma\cdot  \mathcal B]
(\alpha_i;\alpha_e)\,.
\label{Verlinde-vev-B}
\end{equation}
The operator algebra of $\mathcal L_\gamma$
is isomorphic to that of $\Lambda_\gamma$.
In the case $\gamma$ is purely magnetic,
we conjecture that $\mathcal L_\gamma$ is related to the 't Hooft loop
$T_B$ above as
\begin{equation}
\mathcal L_\gamma= \sum_v
e^{-(\mathsf b^2/2) v\cdot \partial_a}
Z_\text{equator}(a;B,v)
e^{-(\mathsf b^2/2) v\cdot \partial_a}
\label{ver-general}
\end{equation}
up to an overall constant.
For more general dyonic charges
the Verlinde operator takes the form
\begin{equation}
\mathcal L
=\sum_{v,w}
e^{-(\mathsf b^2/2) v\cdot \partial_a}
 e^{2\pi i w\cdot a}
Z_L\left(a,{\mathsf m}_f+\frac 1 2, \sfb^2;v,w\right)
e^{-(\mathsf b^2/2) v\cdot \partial_a}
\,,
\label{VZZ}
\end{equation}
with some functions $Z_L(a,m_f, \lambda;v,w)$.
For a product of $SU(2)$'s, our Lie algebra convention is such that
$v\cdot \partial_a=\sum_i v_i\frac{\partial}{\partial a_i}$ and
$w\cdot a =\sum_i\Tr\left[
\text{diag}(w_i,-w_i)
\text{diag}(a_i,-a_i)
\right]=2 \sum_i w_i a_i$,
with $v_i$ and $w_i$ being half-integers.
The ``highest'' $v$ (corresponding to the highest weight of the Langlands-dual
representation) and $w$ have $v_i=p_i/2$ and $w_i=q_i/2$,
where $(p_i,q_i)$ are the Dehn-Thurston 
parameters \cite{Drukker:2009tz}.
We conjecture that the line operator vevs are
given in terms of these functions as
\begin{equation}
\hspace{-2mm}
  \langle L\rangle_{S^4_{\mathsf b}}=
\hspace{-1mm}
 \int_{i \mathfrak t} da\,
Z_\text{pole}(a,\bar q)
\sum_{v,w}
e^{-\frac{\mathsf b^2}{2} v\cdot \p_a}
 e^{2\pi i w\cdot a}
Z_L\left(a,{\mathsf m}_f+\frac 1 2, \sfb^2;v,w\right)
e^{-\frac{\mathsf b^2}{2} v\cdot \p_a}
Z_\text{pole}(a,q)
\label{L-S4-conj}
\end{equation}
on $S^4_{\mathsf b}$ and
\begin{equation}
  \langle L\rangle_{S^1\times \mathbb R^3}=
\sum_{v,w}
 e^{2\pi i( w\cdot a+v\cdot b)}
Z_L(a, m_f, \lambda;v,w)
\label{L-S1R3-conj}
\end{equation}
on $S^1\times \mathbb R^3$.

We have focused so far on the correspondence \cite{AGT} between
the gauge theories whose gauge group is a product of $SU(2)$'s
and Liouville theory on the corresponding Riemann surface,
but we also propose that the relation (\ref{thooft-S4})
should hold for more general gauge groups and Toda theories \cite{AGT,Wyllard:2009hg}.
Some examples of Verlinde operators in Toda theories were computed in
\cite{Passerini:2010pr,Gomis:2010kv}.
We conjecture that the Verlinde operators
in Toda theories
are precisely 
related to the line operator vevs on $S^4$ and $S^1\times\mathbb R^3$
via the equations (\ref{VZZ}), (\ref{L-S4-conj}), and
(\ref{L-S1R3-conj}).

We observe that the Verlinde operator (\ref{VZZ})
is related to the vev
 (\ref{L-S1R3-conj})
on $S^1\times \mathbb R^3$
 precisely  by the  Weyl transform (ordering)\footnote{For a 2-dimensional phase space parametrized by $(q,p)$,
the operator $\mathcal O$ and its inverse Weyl transform $f$
 are 
related by
$
f(q,p)
=\int d\sigma e^{-\frac{i}\hbar p \sigma}
\langle q
|
e^{\frac{i} {2\hbar} \sigma \hat p}
\mathcal O(\hat q,\hat p)
e^{\frac{i} {2\hbar} \sigma \hat p}
|
q
\rangle
\,,
\mathcal O
=
\frac{1}{(2\pi)^2 \hbar}
\int d\sigma d\tau dq dp
e^{-i\tau(\hat q-q)-\frac{i}\hbar\sigma(\hat p-p)}
 f(q,p)\,,
$
where $[\hat q,\hat p]=i\hbar$, $\hat q|q\rangle=q|q\rangle$,
$\langle q|q'\rangle=\delta(q-q')$ \cite{Harvey:2001yn}.
}
\begin{equation}
\langle L\rangle_{S^1\times \mathbb R^3}
\quad
\stackrel{\text{Weyl}}{\Longrightarrow}
\quad
\mathcal L\,.
\end{equation}
The parameter $-b$ plays the role of the canonical momentum:
\begin{equation}
  \begin{aligned}
b&\leftrightarrow i\frac{\lambda}{2\pi}\frac{\partial}{\partial a}
\hspace{11mm}
\text{in general,}
\\
b_i&\leftrightarrow i\frac{\lambda}{4\pi}\frac{\partial}{\partial a_i}
\hspace{10mm}
\text{for $SU(2)$ and Liouville. }
  \end{aligned}
\end{equation}
Thus our proposal (\ref{thooft-S4}) implies that the Verlinde operators
are the Weyl transform of the line operator vevs on $S^1\times \mathbb R^3$,
when the gauge theory has a Lagrangian description.
It is very natural to conjecture that this relation
should hold even when the gauge theory does not admit a Lagrangian
description \cite{Argyres:2007cn,Gaiotto:2009we}.

The mass shift in (\ref{eq-cont}) and (\ref{ver-general})
is consistent with a somewhat confusing aspect of the correspondence
\cite{Gaiotto:2009we}
between $\mathcal N=2$ gauge theories and Riemann surfaces.
Namely the massless limit of a gauge theory
corresponds to removing a puncture in the Hitchin system
\cite{Gaiotto:2009hg,Gaiotto:2010be},
while it corresponds to tuning external momenta
to special values in Liouville/Toda theories \cite{AGT,Wyllard:2009hg,Okuda:2010ke} keeping the puncture.
In gauge theories the shift arises due to the difference
in the geometries where the theories live.

Below we demonstrate our proposal with several examples.

\subsection{$SU(2)$ $\mathcal N=2^*$}
This theory corresponds to the Liouville theory on the one-punctured torus
\cite{AGT}.
Let $C(\alpha_1,\alpha_2,\alpha_3)$ be the DOZZ three-point function
of Liouville theory \cite{Dorn:1994xn,Zamolodchikov:1995aa}.
We  denote the internal and external Liouville momenta
 by $\alpha$ and $\alpha_e$ respectively.
The Verlinde loop operator that corresponds to the minimal
't Hooft operator acts on the conformal block 
as
\begin{equation}
[  \mathcal L_{1,0}\cdot \mathcal F](\alpha,\alpha_e)=
\sum_\pm
H_\pm(\alpha)
\mathcal F\left(\alpha\pm {\mathsf b}/{2}\,,\alpha_e\right)\,.
\end{equation}
This implies the following  expression, conjectural for $\mathsf b\neq 1$,
of the minimal 't Hooft operator vev in the $\mathcal{N}=2^*$ theory on $S^4_{\mathsf b}$:
\begin{align}
\langle\,
L_{1,0}\,
\rangle_{S^4_{\mathsf b}}
=
\int_{Q/2+i\mathbb R} d\alpha\,
C(\alpha,\alpha_e,Q-\alpha)
\sum_\pm
\overline{\mathcal F (\alpha,\alpha_e)}
H_\pm(\alpha)
\mathcal F\left(\alpha\pm {\mathsf b}/{2}\,,\alpha_e\right)\,.
\label{tHooftLiou2star}
\end{align}
The map between the Liouville and gauge theory parameters is given by 
\begin{align}
\alpha=\frac{Q}{2}+ \frac{a}{\sfb},\quad \alpha_e=\frac{Q}{2}+ \frac{{\mathsf m}}{\sfb}.
\end{align}
The coefficients $H_\pm$ are known to be \cite{Alday:2009fs}
\begin{align}
H_\pm(\alpha)
=\frac{\Gamma(\pm 2 a)\Gamma(\pm 2 a+\sfb Q)}
{\Gamma(\pm 2 a + {\mathsf m}+\sfb Q/2)\Gamma(\pm 2 a-{\mathsf m}+\sfb Q/2)}.
\end{align}
By performing the manipulations explained above,
in Appendix \ref{app:Liou-2star} we obtain
\begin{align}
\mathcal L_{1,0}
=
\sum_\pm e^{\pm \frac{1}{4} \sfb^2\partial_a}
\left(
\prod_\pm
\frac{
{ \cos ( 2\pi a \pm \pi {\mathsf m})
}}
{
 \sin( 2\pi a \pm  \frac{\pi}{2}\sfb^2)
}
\right)^{1/2}
 e^{\pm \frac{1}{4} \sfb^2\partial_a}\,.
\label{2star-L10}
\end{align}
This is indeed  related to
the 't Hooft operator vev 
(\ref{SU22starL10})
by the Weyl transform above.

\subsection{$SU(N)$ $\mathcal N=2^*$
}

The Verlinde operator corresponding to the 't Hooft operator
with charge $B=(1,0^{N-1})$, acting on the Toda conformal block
for the one-punctured torus, was computed in 
\cite{Gomis:2010kv} in the standard  normalization.
In Appendix \ref{app:Toda-2star} we convert it
to a difference operator acting on the block $\mathcal B$
in the normalization that absorbs the square root of
the three-point function.
We find
\begin{equation}
  \mathcal L_{B=(1,0^{N-1})}=
 \sum_{l=1}^N 
e^{-\frac{\mathsf b^2}{2} h_l\cdot \partial_a}
\left(
\prod_\pm
\prod_{j\neq l}
\frac{
\cos\pi (a_{lj}\pm \mathsf m)
}{
\sin\pi (a_{lj}\pm\lambda/2)
}
\right)
e^{-\frac{\mathsf b^2}{2} h_l\cdot \partial_a}
\,,
\label{Ver-SUN-star-thooft}
\end{equation}
where $(h_l)_j=\delta_{jl}-1/N$.
Note that $h_l$ are the coweights that correspond to
the weights in the fundamental representation
of the Langlands dual group.
The Verlinde operator (\ref{Ver-SUN-star-thooft}) 
is the Weyl transform of the vev
(\ref{UN2starBpm10N-1}) on $S^1\times\mathbb R^3$ as expected.

\subsection{$SU(2)$ $N_\text{F}=4$}

To compare with gauge theory calculations,
we relate $a$ and ${\mathsf m}_f$ to $\alpha$ and $\alpha_e$ by
\begin{equation}
  \begin{aligned}
&  \alpha=\frac{Q}{2} + \frac{a}{{\mathsf b}}\,,
\quad
  \alpha_1=\frac{Q}{2} + \frac{{\mathsf m}_1-{\mathsf m}_2}{2{\mathsf b}}\,,
\quad
  \alpha_2=\frac{Q}{2} + \frac{{\mathsf m}_1+{\mathsf m}_2}{2{\mathsf b}}\,,
\\
&
\hspace{15mm}
  \alpha_3=\frac{Q}{2} + \frac{{\mathsf m}_3+{\mathsf m}_4}{2{\mathsf b}}\,,
\quad
  \alpha_4=\frac{Q}{2} + \frac{{\mathsf m}_3-{\mathsf m}_4}{2{\mathsf b}}\,.
  \end{aligned}
\end{equation}
For the minimal Wilson operator,
the corresponding Verlinde operator is 
\begin{equation}
\mathcal L_{0,1}=e^{2\pi i a}+e^{-2\pi i a}\,.
\end{equation}
In Appendix (\ref{app:Toda-NF=2N}), we show that
\begin{equation}
\begin{aligned}
\mathcal L_{2,0}
&=
\sum_\pm e^{\pm \frac{1}{2} {\mathsf b}^2\partial_a}
\left(
 \frac{
\prod_{f=1}^4\prod_{s=\pm}
\cos (\pi a +s \pi  {\mathsf m}_f)
}{
\sin(2\pi a + \pi  {\mathsf b}^2)
\sin^2 (2\pi a )
\sin( 2\pi  a- \pi {\mathsf b}^2)
}
\right)^{1/2}
 e^{\pm \frac{1}{2} {\mathsf b}^2\partial_a}
\\
&\quad
\quad\quad
-\frac 1 2
\cos\pi( \sfb^2-\sum_f  {\mathsf m}_f)
+
\sum_{s=\pm} \frac{
\prod_{f=1}^4 \cos \pi \left(-\sfb^2/2 +s  a +  {\mathsf m}_f\right)
}{
\sin( 2\pi a) \sin \pi (s \sfb^2 -2 a)
}\,.
\label{Ver20}
\end{aligned}
\end{equation}
In view of  (\ref{VZZ}) and (\ref{L-S1R3-conj}),
this is related to
the 't Hooft operator vev (\ref{U2NF4})
in the $U(2)$ theory with $a_1=-a_2=a$
by the Weyl transform, 
up to an $a$-independent term $-\frac12\cos\pi(\sfb^2-\sum  {\mathsf m}_f)$.
The whole expression is invariant under $a\ra -a$ 
as well as under the action of the $SO(8)$ Weyl group.\footnote{Four generators of the  Weyl group of the $SO(8)$ flavor group
 act on the masses
as ${\mathsf m}_1\leftrightarrow {\mathsf m}_2$, 
${\mathsf m}_2\leftrightarrow {\mathsf m}_3$, 
${\mathsf m}_3\leftrightarrow {\mathsf m}_4$, 
and
${\mathsf m}_3\leftrightarrow - {\mathsf m}_4$ respectively.
Agreement up to an additive constant 
is almost as much as one can hope for.
Without $SO(8)$ Weyl invariance, however, 
(\ref{U2NF4}) with $a_1=a=-a_2$
cannot be the answer for $SU(2)$ gauge theory.
}
We see that Liouville theory in fact fixes the $a$-independent
additive constant.
This constant is such that the whole second line 
of (\ref{Ver20}) vanishes in the limit $a\rightarrow \pm i\infty$.
By the argument given above, then,
Liouville theory predicts that the $SU(2)$ theory with $N_\text{F}=4$
has
\begin{equation}
  Z_\text{mono}(a,m_f;2,0)=
-\frac{1}{2}\cos\pi( \lambda-\sum_f  m_f)
-\sum_{s=\pm} \frac{
\prod_{f=1}^4 \sin \pi \left (s  a -  m_f+\lambda/2\right)
}{
\sin( 2\pi a) \sin \pi (s \lambda +2 a)
}\,.
\label{Zmono-SU2NF4}
\end{equation}
Thus the minimal 't Hooft operator vev 
 on $S^1\times \mathbb R^3$
should be%
\footnote{%
Essentially the same expression has been obtained
purely from quantization of the Hitchin system \cite{Teschner-KITP}.
}
\begin{equation}
  \begin{aligned}
\langle L_{2,0}\rangle
=&
(e^{4\pi i b}+e^{-4\pi i b})
\left(
\frac{
\prod_\pm \prod_{f=1}^4 \sin\pi(a\pm m_f)}
{
\sin^22\pi a \prod_\pm \sin\pi(2a\pm\lambda)
}
\right)^{1/2}
\\
&
\quad\quad
-\frac{1}{2}\cos\pi( \lambda-\sum_f  m_f)
-\sum_{s=\pm} \frac{
\prod_{f=1}^4 \sin \pi \left (s  a -  m_f+\lambda/2\right)
}{
\sin( 2\pi a) \sin \pi (s \lambda +2 a)
}\,.
  \end{aligned}
\label{minimal-thooft-from-liouville}
\end{equation}
In Appendix \ref{sec:cl-holo}, we show that
the $\lambda=0$ limit of this expression coincides with
the classical holonomy on the four-punctured sphere
written in terms of the complexified Fenchel-Nielsen coordinates.

\subsection{$SU(N)$ $N_\text{F}=2N$}

The Verlinde operator in the $SU(N)$ superconformal QCD
corresponding to the 't Hooft operator
with $B=(1,-1,0^{N-2})$ was computed in 
 \cite{Gomis:2010kv}.
This was done in the standard normalization,
and in Appendix \ref{app:Toda-NF=2N}, we convert the operator to
the difference operator acting on the block $\mathcal B$.
It is given up to a multiplicative constant by
\begin{eqnarray}
\hspace{-10mm}
&&
\hspace{6mm}  \mathcal L_{B=(1,-1,0^{N-2})}
\nonumber\\
\hspace{-10mm}
&&=
\hspace{-2mm}
\mathop{\sum_{1\leq j,k\leq N}}_{j\neq k}
\hspace{-2mm}
e^{-\frac{{\mathsf b}^2}{4}
e_{jk}
\cdot\partial_a}
\frac{
\left[\prod_{f=1}^N 
\cos \pi (a_j-{\mathsf m}_f)
\cos\pi (a_k-{\mathsf m}_f)
\right]^{\frac 1 2}
}{
\sin\pi a_{jk}
{\displaystyle \prod_\pm }
\bigg[
\sin \pi (a_{jk}\pm{\mathsf b}^2)
{\displaystyle
\prod_{i\neq j,k}}
\sin\pi(a_{ji}\pm \frac{{\mathsf b}^2}{2})
\sin\pi(a_{ik}\pm \frac{{\mathsf b}^2}{2})
\bigg]^{\frac 1 2}
}
e^{-\frac{{\mathsf b}^2}{4} e_{jk}
\cdot\partial_a}
\nonumber\\
\hspace{-10mm}&&
\quad\quad\quad\quad
+\sum_{k=1}^N
  \frac{
\prod_{f=1}^{2N}
\cos \pi
\left(
a_k-{\mathsf m}_f +\frac{{\mathsf b}^2}{2}
\right)
}
{
\prod_{i\neq k} 
\sin \pi a_{ki}
\sin \pi\left(- a_{ki}-{\mathsf b}^2\right)
}
\\
\hspace{-10mm}&&
\quad\quad\quad\quad
+(-1)^{N-1}
\frac{e^{N \pi i   (\sum_{f>N} \mathsf m_f-\sum_{f\leq N} \mathsf m_f)/N}
\sin\pi \mathsf b^2}
{\sin(\pi(N-2)\mathsf b^2)}
-\frac 1 2 \cos \pi (\mathsf b^2 -\sum_f \mathsf m_f)
\,.
\nonumber
\end{eqnarray}
Here $e_{jk}\equiv h_j-h_k$ are the coroots.
This implies that for the minimal 't Hooft operator
in the $SU(N)$ theory with $N_\text{F}=2N$,
the vev on $S^1\times\mathbb R^3$ is given by
\begin{eqnarray}
&&  \langle T_{B=(1,-1,0^{N-2})}\rangle
\nonumber \\
&=&
\mathop{\sum_{1\leq j,k\leq N}}_{j\neq k}
 e^{\pi i(b_j-b_k)}
\frac{
\left[
\prod_{f=1}^N 
\sin\pi (a_j-m_f)
\sin\pi (a_k-m_f)
\right]^{1/2}
}{
\sin\pi a_{jk}
{\displaystyle \prod_\pm }
\left[
\sin \pi (a_{jk}\pm\lambda)
{\displaystyle
\prod_{i\neq j,k}}
\sin\pi(a_{ji}\pm \lambda/2)
\sin\pi(a_{ik}\pm \lambda/2)
\right]
^{1/2}
}
\nonumber
\\
&&
\quad\quad\quad\quad
+\sum_{k=1}^N
  \frac{
\prod_{f=1}^{2N}
\sin \pi
\left(
a_k- m_f +\lambda/2
\right)
}
{
\prod_{i\neq k} 
\sin \pi a_{ki}
\sin \pi\left(- a_{ki}-\lambda\right)
}
\label{SUN-SQCD}
\\
&&
\quad\quad\quad\quad
+(-1)^{N-1}
\frac{e^{\pi i   (\sum_{f>N}  m_f-\sum_{f\leq N}  m_f)}
\sin\pi \lambda}
{\sin(\pi(N-2)\lambda )}
+\frac{(-1)^{N-1}} 2 \cos \pi (\lambda -\sum_f  m_f)
\,.\nonumber
\end{eqnarray}
This is identical to the $U(N)$  result (\ref{UNSQCD}) up to the terms
independent of $a$ and $b$.
The expression (\ref{SUN-SQCD}) is a prediction of Toda theory
for the $SU(N)$ gauge theory.

\section{Discussion}
\label{sec:discussion}

Let us conclude with remarks on future directions and related works.

We focused on conformal $\mathcal N=2$ gauge theories because localization
calculations are the cleanest for them.
Line operators in non-conformal asymptotically free theories also exhibit
rich dynamics \cite{Gaiotto:2010be} and the spectrum of BPS states is often simpler.
The easiest way to compute correlation functions in such theories would be
to start with a conformal theory and decouple some matter fields
by sending their mass to infinity.
It would be interesting to study this limit in detail.

Our calculation of the line operator vevs, or the supersymmetric index (\ref{trace-def}),
in terms of the complexified Fenchel-Nielsen coordinates made use of
the equivariant index theorem.
It is amusing to note that the calculation of the supersymmetric index
in terms of the Fock-Goncharov coordinates can also be formulated
in terms of an index theorem,
but applied to the moduli space constructed from the Seiberg-Witten
prepotential governing the IR dynamics \cite{Lee:2011ph,Kim:2011sc} .

In our computational scheme for the monopole screening contributions $Z_\text{mono}(B,v)$
in $\mathcal N=2^*$ theory,
the 't Hooft operator is S-dual to the Wilson operator
in a product of fundamental representations.
As such the 't Hooft operator is reducible, i.e., it can be written as a
linear combination of other line operators with positive coefficients.
Related to this is the fact that the 't Hooft operator vev (\ref{U2NF4}) in the $G=U(2)$ theory
with $N_\text{F}=4$ fundamental hypermultiplets 
becomes $SO(8)_\text{F}$ Weyl-invariant not just by substituting 
$(a_1,a_2)\ra (a,-a)$,
but only after adding an $a$-independent term in (\ref{Ver20}).
It is important to develop a method  intrinsic to irreducible
line operators for gauge group $SU(N)$ rather than $U(N)$.
This may involve decomposing the cohomology
of the monopole moduli space into irreducible representations
of the Langlands dual gauge group \cite{Witten:2009mh}
and incorporate the computation of operator product expansions
 \cite{Kapustin:2006pk,Kapustin:2007wm,Saulina:2011qr}.

We found that the line operators in $\mathcal N=2$ theories
on $S^1\times \mathbb R^3$
realize a deformation quantization of the Hitchin moduli space.
We expect that this can be explained in the framework of
\cite{Nekrasov:2010ka}, by dimensionally reducing the theory on the
circle parametrized by $\tau$ as well as the one
parametrized by the polar angle in the 12-plane.
We would obtain a $(4,4)$ sigma model on a half plane whose target space
is the Hitchin moduli space, and the 
boundary condition would correspond to
the canonical coisotropic brane.
Liouville/Toda conformal blocks arise as open string states
by including another boundary mapped to the brane of opers.
It would be interesting to study these systems in more detail and
understand the appearance of the Weyl transform.

Some of the results in \cite{Gaiotto:2009hg}
obtained by the wall-crossing formula
can be reproduced from our results that are obtained
directly by localization calculations.
It would be interesting to further explore the relation between
the UV and IR theories as well as the integrability aspects of the
line operators.

\section*{Acknowledgements}
We are grateful to Nadav Drukker, Jaume Gomis, Anton Kapustin,
Yoichi Kazama, Satoshi Nawata, Andy Neitzke, 
Vasily Pestun,
Natalia Saulina, and Masahito Yamazaki  for fruitful discussions.
T.O. and M.T. also thank the KITP for providing us with 
a stimulating environment where part of this work was done.
The research at KITP was supported in part by DARPA under Grant No.
HR0011-09-1-0015 and by National Science Foundation under Grant
No. PHY05-51164.  
T.O. thanks the Perimeter Institute for warm hospitality.
The research of Y.I is supported in part by
a JSPS Research Fellowship for Young Scientists.
The research of T.O. is supported in part
by Grant-in-Aid for Young Scientists (B) No. 23740168,
Grant-in-Aid for Scientific Research (B) No. 20340048,
and Institutional Program for Young Researcher Overseas Visits 
No. R10 from the Japan Society for the Promotion of Science.
The research of M.T. is  supported in part
 by JSPS Grant-in-Aid for Creative Scientific Research No. 19GS0219.

  \vfill\eject

\appendix

\section{Spinors and gamma matrices}
\label{sec:spinor-gamma}

Chiral spinors $\Psi$ and $\ep$ transform 
in a representation of $Spin(10)$, 
 whose generators are constructed from $32\times 32$ matrices
$\boldsymbol \Gamma^M$ obeying
\begin{equation}
\{\boldsymbol\Gamma^M,\boldsymbol\Gamma^N\}=2\eta^{MN}\quad ~~M=0,1,\ldots,9\,.
\end{equation} 
We use the Euclidean signature $\eta^{MN}=\delta^{MN}$.
We can take $\boldsymbol \Gamma^M$ in the form
  \begin{equation}
 \boldsymbol\Gamma^M=\left(
\begin{array}{cc}
 0& \tilde\Gamma^M \\
\Gamma^M&0\end{array}
\right)\,,
 \end{equation}
where $\Gamma^{M}\equiv(\Gamma^0,\Gamma^1,\ldots,\Gamma^9)$
and
 $\tilde\Gamma^M\equiv(-\Gamma^0,\Gamma^1,\ldots,\Gamma^9)$
 are $16\times 16$ matrices that satisfy
\begin{equation}
\tilde\Gamma^M\Gamma^N+\tilde\Gamma^N\Gamma^M=2\delta^{MN}\,,\qquad \Gamma^M\tilde\Gamma^N+\Gamma^N\tilde\Gamma^M=2\delta^{MN}\,.
\end{equation}
We also use notation $\Gamma^{MN}\equiv \tilde \Gamma^{[M} \Gamma^{N]}$,
$\tilde \Gamma^{MN}\equiv  \Gamma^{[M} \tilde \Gamma^{N]}$,
and $\Gamma^{MNPQ}\equiv \tilde \Gamma^{[M}\Gamma^N\tilde\Gamma^P\Gamma^{Q]}$.
Our spinors have positive chirality with respect to the chirality matrix
\begin{equation}
\boldsymbol\Gamma\equiv-i\boldsymbol\Gamma^1\ldots\boldsymbol\Gamma^9
\boldsymbol\Gamma^0=\left(
\begin{array}{cc}
-i \tilde\Gamma^1\Gamma^2\ldots \tilde\Gamma^9\Gamma^0& 0 \\
0&-i \Gamma^1\tilde\Gamma^2\ldots \Gamma^9\tilde\Gamma^0\end{array}
\right)=
\left(
\begin{array}{cc}
1& 0 \\
0&-1\end{array}
\right)
\,.
\end{equation}

In ten dimensions with Euclidean signature
the chiral spinor representation is complex.
We take $\Gamma^{1},\ldots, \Gamma^9$ to be real and $\Gamma^0=i$
pure imaginary. 
As in \cite{Gomis:2011pf}, for the explicit expressions
we use matrices
as defined in appendix A of  \cite{Pestun:2007rz} with a 
permutation of spacetime indices. 
Let $\underline\Gamma^{M}$ be the gamma matrices
in \cite{Pestun:2007rz}.  Then our $\Gamma^{M}$ are given by
\begin{equation}
\begin{aligned}
  \Gamma^M = \underline \Gamma^{M+1} \quad \text{for} \quad  M = 1,2,3,5,6,7\,, \\
  \Gamma^{4} = \underline \Gamma^{1}\,, \quad  \Gamma^{8} =
  \underline \Gamma^{5}\,, \quad \Gamma^{9} = \underline \Gamma^{9}\,,
  \quad \Gamma^{0} = i \underline \Gamma^{0}.
\end{aligned}
\end{equation}
The factor of $i$ in the relation to $\Gamma^{0}$ arises because our
present conventions   use the Euclidean metric $\eta^{MN} = \delta^{MN}$, while 
\cite{Pestun:2007rz} used the  Lorentz metric with $\eta^{00} = -1$.

For off-shell supersymmetry, we need
a set of spinors $\nu^i$ ($i=1,\ldots, 7$) that satisfy the relations 
\cite{Pestun:2007rz,Berkovits:1993zz}
\begin{eqnarray}
  \ep \Gamma^M \nu_i&=&0\,,\nonumber
\\
\frac 12 (\ep\Gamma_N \ep)\tilde \Gamma^N_{\alpha\beta}&=& \nu^i_\alpha\nu^i_\beta
+\ep_\alpha\ep_\beta\,,
\\
\nu_i\Gamma^M\nu_j&=&\delta_{ij} \ep\Gamma^M\ep\,.
\nonumber
\end{eqnarray}
Explicitly, we take
\begin{equation}
  \begin{aligned}
  \nu_j=&\Gamma^{8,j+4}\ep\quad\quad j=1,2,3\,,\\
  \nu_4=&\Gamma^{89}\ep\,,\\
\nu_j=&\Gamma^{8,j-4}\ep\quad\quad j=5,6,7\,.
  \end{aligned}
\end{equation}

We also use the standard Pauli matrices $\sigma^i$, $i=1,2,3$,
defined as
\begin{equation}
\sigma_1=
  \begin{pmatrix}
    &1\\
1&
  \end{pmatrix}
\,,\quad
\sigma_2=
  \begin{pmatrix}
    &-i\\
i&
  \end{pmatrix}
\,,\quad
\sigma_3=
  \begin{pmatrix}
1    &\\
& -1
  \end{pmatrix}
\,.
\end{equation}

\section{Differential operators for the one-loop determinants}
\label{sec:diff-op}

In this appendix we derive the differential operators whose indices
enter the one-loop calculations.

We will need the 
relations inverse to (\ref{X-def}) and (\ref{Xprime-def}):
\begin{equation}
  \begin{aligned}
\tilde A_M&=X_{0M}\,\quad\quad M=1,\ldots,9\,,
\\
  \tilde\Phi_0&\equiv \tilde A_0= i X'_{18} +i X_{04}\,
\\
\Psi_M&= X'_{0M}-D_{(0)M} X_{18}\quad\quad M=1,\ldots,9\,,
\\
K_j &=X_{1j}' +i \sum_{M=1}^9\sum_{N=1}^9 (\bar \nu_j \Gamma^{MN}\ep) D_{(0)M} X_{0N}
\quad\quad j=1,\ldots,7\,,
\\
\Upsilon_j &= X_{1j}\quad\quad j=1,\ldots,7
\\
c& =X_{18}\,,\quad\quad
\tilde c= X_{19}\,,\quad
\tilde b= X'_{19}\,.
  \end{aligned}
\end{equation}
Then the quadratic part of
$\hat V$ is given by
\begin{equation}
  \begin{aligned}
\hat V^{(2)}
=&
\int d^4x \,
\Tr\left(
\sum_{M=1,2,3,9}
(X'_{0M}-D_{(0)M} X_{18})
(-D_{(0)\tau} X_{0M} +D_{(0)M}X_{04})
\right.
\\
&
\hspace{-2mm}
\left.
  +\sum_{j=1}^7 
X_{1j}
\left(X'_{1j} +2i ( \nu_j \tilde \Gamma^{NP}\bar \ep) D_{(0)N} X_{0P}
\right)
+X_{19}\left(
i
\hspace{-2mm}
 \sum_{M=1,2,3,9}
\hspace{-2mm}
 D_{(0)}^M X_{0M}
+\frac{\xi}2
X'_{19}
\right)
\right)\,.
  \end{aligned}
\end{equation}
From this we read off $D_{10}$:
\begin{equation}
  \begin{aligned}
&(D_{10} \cdot X_0)_{j=1}^7
\nonumber\\
=& 
2i \sum_{k=1}^3 \sum_{l=1}^3(\nu_j \tilde\Gamma^{kl} \bar\ep)D_{(0)k} X_{0l}
+ 2i \sum_{k=1}^3 \sum_{l=5}^8(\nu_j \tilde\Gamma^{kl} \bar\ep)D_{(0)k} X_{0l}
+ 2i \sum_{k=1}^3 (\nu_j \tilde\Gamma^{k9} \bar\ep)D_{(0)k} X_{09}
\\&
+ 2i\sum_{l=1}^3 (\nu_j \tilde\Gamma^{9l} \bar\ep)D_{(0)9} X_{0l}
+ 2i\sum_{l=5}^8 (\nu_j \tilde\Gamma^{9l} \bar\ep)D_{(0)9} X_{0l}\,.
  \end{aligned}
\label{D10X0}
\end{equation}
The differential operator $D_{10}$ splits into the 
vector and hypermultiplet parts.
Let us begin with the vector multiplet.
For $j=1,2,3$, we have
\begin{equation}
  \begin{aligned}
&(D_{10} \cdot X_0)_{j=1}^3
\\
=&
-2i \ep_{jkl}D_{(0)k}X_{0l}+2i D_{(0)j}X_{09}-2i D_{(0)9} X_{0j}
\\
=&-2i (D_\text{Bogo}\cdot X_0)_j
  \end{aligned}
\label{DBogo}
\end{equation}
where we used that 
$\nu_j \tilde\Gamma^{kl}\bar\ep=-\epsilon_{jkl}$ for $j,k,l\in\{1,2,3\}$,
$\nu_j \tilde\Gamma^{kl}\bar\ep=0$ for $j,k\in\{1,2,3\}$ and $l\in\{5,6,7,8\}$,
$\nu_j \tilde\Gamma^{k9}\bar\ep=\delta_{jk}$ for $j,k\in\{1,2,3\}$,
$\nu_j \tilde\Gamma^{9l}\bar\ep=0$ for $j\in\{1,2,3\}$ and $l\in\{5,6,7,8\}$.
The differential operator $D_\text{Bogo}$ is the linearization
of the Bogomolny equations.
For $j=9$, we get
\begin{equation}
  (D_{10}\cdot X_0)_9= i\sum_{M=1,2,3,9} D_{(0)}^M X_{0M}\,.
\label{dual-Bogo}
\end{equation}
This is the conjugate of the linearized gauge transformation and
has its origin in the gauge-fixing condition.
We also have
\begin{equation}
  \begin{aligned}
 ( D_{10}\cdot X_0)_8=&
\sum_{M=0}^9 D_{(0)}^M\left(
D_{(0)M} X_{04}-D_{(0)\tau} X_{0M}
\right)\,.
  \end{aligned}
\label{D10X08}
\end{equation}
As in \cite{Pestun:2007rz}, the computation of the symbol shows 
that (\ref{D10X08}) can be dropped by neglecting $X_{04}$ and $X_{18}$, and 
that $D_{10}$ acting on the vector multiplet
fails to be elliptic, though
we have checked that $D_{10}$ is transversally elliptic,
i.e., it is elliptic in the directions other than $\tau$.
Since we work in a non-compact space, our application of the localization
formula for the index is done formally, as in the  calculation
of the instanton partition function.

For the hypermultiplet, we need to consider the components $j=4,5,6,7$
of (\ref{D10X0}):
\begin{equation}
  \begin{aligned}
&(D_{10} \cdot X_0)_{j=4}^7
\nonumber\\
=& 
 2i \sum_{k=1}^3 \sum_{l=5}^8(\nu_j \tilde\Gamma^{kl} \bar\ep)D_{(0)k} X_{0l}
+ 2i\sum_{l=5}^8 (\nu_j \tilde\Gamma^{9l} \bar\ep)D_{(0)9} X_{0l}\,.
  \end{aligned}
\end{equation}
This differential operator is the ``realification''
of the Dirac-Higgs operator
\begin{equation}
D_\text{DH}\equiv  \sigma^i D_{(0)i}+
 [\Phi_{(0)9},
\,\cdot\,]
\label{D-DH}
\end{equation}
acting on the ``spinor'' 
$2^{-1/2}(X_{05}-iX_{06}+iX_{07}+X_{08}, iX_{05}-X_{06}-X_{07}-iX_{08})^T$
and mapping to another
$2^{-1/2}(i X_{11}+i X_{12}+X_{13}-X_{14}, X_{11}-X_{12}-i X_{13}-iX_{14})^T$.

\section{Monopoles  on $\mathbb R^3$ and instantons on Taub-NUT}
\label{sec:mono-inst}
Let us
review Kronheimer's correspondence
\cite{Kronheimer:MTh} 
between several singular monopoles on $\mathbb R^3$
and $U(1)_K$-invariant instantons on
a multi-centered Taub-NUT space, which has the metric
\begin{equation}
  ds^2=V d\vec x^2+V^{-1}(d\psi+\omega)^2\,,
\quad\quad
V=l+\sum_j\frac{1}{2|\vec x-\vec x_{j}|}\,,\quad\quad
d\omega=-*_3 dV\,,
\label{TN-metric}
\end{equation}
where $l>0$ is a constant.
From the three-dimensional fields $(A,\Phi)$ with singularities
\begin{equation}
  A\sim \frac {B_j}2 \cos\theta d\varphi\,,
\quad
\Phi \sim \frac{B_j}{2r}\,
\quad
\text{ near } \vec x=\vec x_j\,,
\end{equation}
where $(r,\theta, \varphi)$ are the spherical coordinates
on a 3-ball centered at $\vec x=\vec x_j$,
we construct a four-dimensional gauge connection
\begin{equation}
  \mathcal A \equiv  
g
\left(A+\Phi \frac{d\psi+\omega}{V}\right)
g^{-1}
-i g d g^{-1}
\label{calA}
\end{equation}
and its curvature $\mathcal F=d\mathcal A+i\mathcal A\wedge\mathcal A$.
The singularities in $A$ and $\Phi$ cancel in (\ref{calA})
to define a smooth four-dimensional gauge field $\mathcal A$.
Here $g$ is a suitable singular gauge transformation
that locally behaves as $g\sim e^{i B_j \psi}$
near $\vec x=\vec x_j$ so that $\mathcal A$ is smooth there.\footnote{For a single singular monopole we can take
$g=e^{iB\psi}$.
In the present paper this is all we need 
even when there is more than one singularity because
the index calculation is local. }
The four-dimensional field $\mathcal A$ is invariant 
under the $U(1)_K$ action $\psi \ra \psi+\nu$,
which rotates the circle fiber as well as acts on the gauge bundle
as a gauge transformation.
The claim is that the Bogomolny equations
\begin{equation}
  *_3 F=D\Phi
\end{equation}
are equivalent to the anti-self-dual equations
\begin{equation}
  *_4\mathcal F+\mathcal F=0\,.
\end{equation}
To show this, let us use the fact that $\mathcal A$ is obtained
by a singular gauge transformation from
\begin{equation}
\tilde{\mathcal A}=A+\Phi \frac{d\psi+\omega}{V}\,,
\end{equation}
therefore $\mathcal F=g\tilde{\mathcal F} g^{-1}$.
Then, for the orientation 
(volume form)$\propto (d\psi+\omega)dx^1 dx^2 dx^3$,
\begin{eqnarray}
   \mathcal F&=&
g
\left(F+D\Phi\wedge \frac{d\psi+\omega}{V}
-\Phi \frac{*_3dV}V
+\Phi  (d\psi+\omega)\wedge \frac{dV}{V^2}
\right)
g^{-1}
\,,
\end{eqnarray}
and
\begin{eqnarray}
*_4   \mathcal F&=&
g
\left(
-*_3F \wedge \frac {d\psi+\omega} V
- *_3D\Phi
-\Phi \frac{(d\psi+\omega) \wedge dV}{V^2}
+\Phi \frac{*_3 dV}{V}
\right)
g^{-1}
\,,
\end{eqnarray}
so $\mathcal F+*_4\mathcal F=0$  if and only if
$F=*_3 D\Phi$.

The holonomy of the four-dimensional field
at infinity $|\vec x|= \infty$ is related to the
scalar expectation value as
\begin{equation}
  P e^{-i \oint \mathcal A} \rightarrow e^{- 2\pi i  \Phi^{(\infty)}/l}
\text{ as } |\vec x |\ra \infty
\end{equation}
up to conjugation.

In the single-center case, the metric (\ref{TN-metric}) approaches
twice the metric of $\mathbb C^2$ in the limit $l\ra 0$,
$ds^2\ra 2ds^2_{\mathbb C^2}$, where
\begin{equation}
  \begin{aligned}
ds^2_{\mathbb{C}^2} =
&(2r)^{-1}\left[dr^2+r^2(d\theta^2+\sin^2\theta d\varphi^2)\right]
+2 r(d\psi+\omega)^2
\\  
=&  
|dz_1|^2+|dz_2|^2 
  \end{aligned}
\end{equation}
and
\begin{equation}
  z_1=r^{1/2}\cos\frac \theta 2 e^{-i\psi+i\varphi/2}\,,
\quad
  z_2=r^{1/2}\sin\frac \theta 2 e^{i\psi+i\varphi/2}\,,
\quad
\omega=-\frac 1 2 \cos\theta d\varphi\,.
\label{z1z2}
\end{equation}
For general $l$, Taub-NUT space  is isomorphic as
a complex manifold to $\mathbb C^2=\{(z_1,z_2)\}$
with the same parametrization.
See for example \cite{Gibbons:1996nt}.

\section{$Z_\text{1-loop}$ 
from Liouville/Toda
 theories}
\label{sec:CFT-details}

In this appendix we rewrite the Verlinde operators for 't Hooft
loops  \cite{Alday:2009fs,Drukker:2009id,Gomis:2010kv}
into the form (\ref{Ver-on-B}) acting on the normalized conformal block 
(\ref{B-block}).
We will first do this in the simplest case $G=SU(2)$ $\mathcal N=2^*$ theory
 for illustration,
and then perform more complicated calculations for $\mathcal N=2^*$ Yang-Mills
and super conformal QCD with gauge group $G=SU(N)$.

\subsection{$SU(2)$ $\mathcal N=2^*$ 
}
\label{app:Liou-2star}

We shift the argument in (\ref{tHooftLiou2star}) 
as $\alpha \to \alpha \mp {\mathsf b}/4$.
Then the expectation value of the Verlinde operator for 
the minimal 't Hooft loop $T=L_{1,0}$ becomes
\begin{align}
\langle
T\rangle
=&
\sum_\pm
\int d \alpha\,
C(\alpha\mp {\mathsf b}/4, \alpha_e,Q-\alpha \pm {\mathsf b}/4)
\overline{\mathcal F(\alpha \pm {\mathsf b}/{4})}
H_+(\alpha \mp {\mathsf b}/4)
\mathcal F\left(\alpha \pm {\mathsf b}/{4}\right)\,.
\end{align}
The monodromy factors are
\begin{align}
H_\pm\left(\alpha \mp \frac{\mathsf b}{4}\right)
=
\frac{\Gamma(\pm 2 a-\mathsf b^2/2)\Gamma({\mathsf b}Q\pm 2a-{\mathsf b}^2/2)}
{\prod_{s=\pm} \Gamma(\frac{1}{2}\pm 2a +s  m)}.
\end{align}
The $a$-dependent part of the
three-point function $C(\alpha, \alpha_e,Q-\alpha)$ reads
\begin{align}
C(\alpha,\alpha_e,Q-\alpha)
\propto
\frac{\prod_{s_1,s_2=\pm}\Gamma_{\mathsf b}(Q/2+s_1 2a/{\mathsf b}+s_2 m/{\mathsf b})}
{\prod_{s=\pm}\Gamma_{\mathsf b}(Q+ 2s a/{\mathsf b})
\Gamma_{\mathsf b}( 2sa/{\mathsf b})}\,,
\end{align}
where $\Gamma_{\mathsf b}(z)$ is the double gamma function.
For its full definition, see for example \cite{Drukker:2009id}.
For the present purpose we only need the relations
\begin{equation}
  \Gamma_{\mathsf b}(z)=\Gamma_{1/{\mathsf b}}(z)\,,
\quad\quad
\Gamma_{\mathsf b}(z+{\mathsf b})=\frac{\sqrt{2\pi} {\mathsf b}^{{\mathsf b}z-1/2}}{\Gamma({\mathsf b}x)}\Gamma_{\mathsf b}(z)\,.
\end{equation}
According to (\ref{Ver-on-B}) and (\ref{ver-general}) 
the equator contribution is
\begin{align}
Z_\text{equator}
=
&\left(
\frac{C(\alpha \mp {\mathsf b}/4,\alpha_e, Q-\alpha\pm {\mathsf b}/4)}
{C(\alpha \pm {\mathsf b}/4,\alpha_e, Q-\alpha\mp {\mathsf b}/4)}
\right)^{1/2}
H_\pm(\alpha\mp {\mathsf b}/4)
\nonumber\\
=&
\left(
\frac
{\prod_{\pm}\cos (2\pi a\pm \pi m)}
{\prod_\pm \sin ( 2\pi a\pm \pi {\mathsf b}^2/2)}
\right)^{1/2}\,.
\end{align}
Thus the Verlinde operator (\ref{ver-general}) acting on 
$\mathcal B(\alpha;\alpha_e)$ 
is given by (\ref{2star-L10}).

\subsection{$SU(N)$ $\mathcal N=2^*$
}
\label{app:Toda-2star}

Let us generalize the calculation for $SU(2)$ above
to $SU(N)$.
The weights in the fundamental representation are given by
\begin{equation}
h_i=(0,\cdots,0,1,0,\cdots,0)-\frac{1}{N}(1,\cdots,1) 
\,,
\quad
N=1,\ldots, N\,.
\end{equation}
The roots are 
\begin{equation}
 e_{ij} = h_i-h_j\,,\quad \quad 1\leq i\,,j \leq N \,,
\end{equation}
and the simple roots are $e_i:=e_{i,i+1}$, $i=1,\ldots, N-1$.
The fundamental weights $\omega_i$ ($i=1,\ldots,N-1$)
are defined as the dual basis,
$\omega_i\cdot e_j=\delta_{ij}$,
since we identify roots and coroots by the metric.
Let $\rho=\sum_l \omega_l$ be the Weyl vector,
and keep the same notation $Q=\mathsf b+\mathsf b^{-1}$
as in the Liouville case.
The three-point function 
with two generic momenta $\alpha_1=i\hat a_1+Q\rho,\ \alpha_2=i\hat a_2+Q\rho$
 and one semi-degenerate momentum $\alpha_3=\kappa \omega_{N-1}$
is given by
\begin{align}
C^{(1)}(\alpha_1,\alpha_2,\kappa):=C(\alpha_1,\alpha_2,\alpha_3=\kappa \omega_{N-1})\propto\frac{\prod_{i<j} \Upsilon(-i \hat a_1\cdot e_{ij} )
 \Upsilon(-i\hat a_2\cdot e_{ij})}{\prod_{i,j=1}^N 
\Upsilon(\kappa/N +i\hat a_1\cdot h_i + i\hat a_2\cdot h_j)} \, .
\end{align}
When $\alpha_1=i\hat a_1+Q \rho,\ \alpha_2=i\hat a_2+Q\rho ,\ \alpha_3=\kappa \omega_1$, 
the three-point function is
\begin{align}
C^{(2)}(\alpha_1,\alpha_2,\kappa):=C(\alpha_1,\alpha_2,\alpha_3=\kappa \omega_1)\propto\frac{\prod_{i<j} \Upsilon(-i\hat a_1\cdot e_{ij} ) 
\Upsilon(-i\hat a_2\cdot e_{ij})}{\prod_{i,j=1}^N
 \Upsilon(\kappa/N -i\hat a_1\cdot h_i -i\hat a_2\cdot h_j)} \, .
\end{align}

The two-dimensional theory corresponding to $\mathcal N=2^*$ is
the $SU(N)$ Toda theory on the torus with one semi-degenerate puncture.
With the parametrization
\begin{align}
\alpha=Q+i\hat a,\ \ \ \
\alpha_e=\left( \frac{Q}{2} + i\hat{m} \right) N\omega_{N-1}\, ,
\end{align}
the vev of the Verlinde operator corresponding to the
minimal 't Hooft operator $T=T_{B=(1,0^{N-1})}$ is
\begin{equation}
\langle T\rangle = \int d\alpha \, C(2Q\rho-\alpha, 
\alpha_e, \alpha) \ \bar{\mathcal{F}(\alpha)}\sum_{l\neq k} H_{k}(\alpha) \mathcal{F}(\alpha-\mathsf b h_{k})\,, 
\end{equation}
where
\begin{equation}
H_{k}(\alpha)=  \prod_{j \neq k}  \frac{
\Gamma(i\mathsf b \hat a_{jk})
\Gamma(\mathsf b Q +i\mathsf b \hat a_{jk})
}{
\Gamma(\mathsf b Q/2 +i\mathsf b \hat a_{jk} -i \mathsf b\hat{m})
\Gamma(\mathsf b Q/2 +i\mathsf b \hat a_{jk} +i\mathsf b\hat{m})
} \, .
\end{equation}
In order to relate this to 
the vev of 't Hooft operator on $S^1\times \mathbb R^3$ we  set
\begin{align}
&\alpha=Q\rho-\frac{a}{\mathsf{b}}\, , \ \ \ \ \  \ 
\alpha_e=\left( \frac{Q}{2}-\frac{\mathsf{m}}{\mathsf{b}} \right) N \omega_{N-1}\,.
\end{align}
Let us define
\begin{equation}
\tilde  \Upsilon(x):=\frac{\Upsilon(x+{\mathsf b})}{\Upsilon(x)}
=
\frac{\Gamma({\mathsf b}x)}{\Gamma(1-{\mathsf b}x)}
\sfb^{1-2\sfb x}\,.
\end{equation}
Then
\begin{align}
&Z_k(\alpha):=\left(
\frac{
C(\alpha+\mathsf{b}h_k/2, \alpha_e,2Q\rho-\alpha-\mathsf{b}h_k/2)
}{
C(\alpha-\mathsf{b}h_k/2, \alpha_e,2Q\rho-\alpha+\mathsf{b}h_k/2)
}
 \right)^{1/2} H_k(\alpha+\mathsf{b}h_k/2) 
\nonumber\\
 &
=  
 \Bigg( 
\displaystyle\prod_{j<l, \pm}
\frac{
 \Upsilon\Big( \pm ( a_{jl}/\mathsf{b} -\mathsf{b}(\delta_{jk}-\delta_{kl})/2 ) \Big)
 }{
 \Upsilon\Big( \pm (a_{jl}/\mathsf{b} +\mathsf{b}(\delta_{jk}-\delta_{kl})/2) \Big)
 }
 \displaystyle\prod_{j\neq l}
 \frac{
 \Upsilon \left(Q/2 -m/\mathsf{b} -a_{jl}/\mathsf{b} -\mathsf{b}(\delta_{jk}-\delta_{kl})/2 \right)
 }{
 \Upsilon \left(Q/2 -m/\mathsf{b} - a_{jl}/\mathsf{b} +\mathsf{b}(\delta_{jk}-\delta_{kl})/2 \right)
 }
 \Bigg)^{1/2} \nonumber\\
&\ \ \  \ \ \ \times \displaystyle\prod_{j\neq k} 
\frac{\Gamma(-a_{jk}-\mathsf{b}^2/2) \Gamma(\mathsf{b}Q-a_{jk}-\mathsf{b}^2/2)}{
\prod_{\pm} \Gamma (\mathsf{b}Q/2 -a_{jk}-\mathsf{b}^2/2 \pm \mathsf{m})
} 
\end{align}
\begin{align}
&=
\Bigg(
\prod_{j<k} \frac{
\tilde{\Upsilon}(a_{jk}/\mathsf{b}-\mathsf{b}/2)
}
{
\tilde{\Upsilon}(-a_{jk}/\mathsf{b}-\mathsf{b}/2)
}
\prod_{k<l} \frac{
\tilde{\Upsilon}(-a_{kl}/\mathsf{b}-\mathsf{b}/2)
}
{
\tilde{\Upsilon}(a_{kl}/\mathsf{b}-\mathsf{b}/2)
} 
\frac{
\prod_{j\neq k} \tilde{\Upsilon}(Q/2 -\mathsf{m}/\mathsf{b} -a_{jk}/\mathsf{b} -\mathsf{b}/2 )
}{
\prod_{k\neq l} \tilde{\Upsilon}(Q/2 -\mathsf{m}/\mathsf{b} -a_{kl}/\mathsf{b} -\mathsf{b}/2 )
}
\Bigg)^{1/2}
\nonumber\\
&\ \  \ \ \  \ \times \displaystyle\prod_{j\neq k} 
\frac{
\Gamma(-a_{jk}-\mathsf{b}^2/2) \Gamma(1-a_{jk}+\mathsf{b}^2/2) 
}{
\prod_{\pm} \Gamma (1/2 - a_{jk}  \pm \mathsf{m})
}
\nonumber\\
&=
\Bigg(
\prod_{j\neq k} \frac{
\tilde{\Upsilon}(a_{jk}/\mathsf{b}-\mathsf{b}/2)
}
{
\tilde{\Upsilon}(-a_{jk}/\mathsf{b}-\mathsf{b}/2)
}
\frac{
 \tilde{\Upsilon}(Q/2 -\mathsf{m}/\mathsf{b} -a_{jk}/\mathsf{b} -\mathsf{b}/2 )
}{
\tilde{\Upsilon}(Q/2 -\mathsf{m}/\mathsf{b} +a_{jk}/\mathsf{b} -\mathsf{b}/2 )
}
\Bigg)^{1/2}
\nonumber\\
& \ \ \ \ \ \times \displaystyle\prod_{j\neq k} 
\frac{
\Gamma(-a_{jk}-\mathsf{b}^2/2) \Gamma(1-a_{jk}+\mathsf{b}^2/2) 
}{
\prod_{\pm} \Gamma (1/2 - a_{jk}  \pm \mathsf{m})
}
\nonumber\\
&=
\Bigg(
\prod_{j\neq k} \frac{
\Gamma (a_{jk}-\mathsf{b}^2/2)
\Gamma(1+a_{jk}+\mathsf{b}^2/2)
\prod_{\pm} \Gamma(1/2 \pm \mathsf{m} -a_{jk})
}
{
\Gamma(1-a_{jk}+\mathsf{b}^2/2)
\Gamma (-a_{jk}-\mathsf{b}^2/2)
\prod_{\pm} \Gamma(1/2\pm \mathsf{m} +a_{jk})
}
\Bigg)^{1/2}
\nonumber\\
& \ \ \ \ \ \times \displaystyle\prod_{j\neq k} 
\frac{
\Gamma(-a_{jk}-\mathsf{b}^2/2) \Gamma(1-a_{jk}+\mathsf{b}^2/2)
}{
\prod_{\pm} \Gamma (1/2 - a_{jk}  \pm \mathsf{m})
}
\nonumber \\
&=
\Bigg(
\prod_{j\neq k} \frac{
\Gamma (a_{jk}-\mathsf{b}^2/2)
\Gamma(1+a_{jk}+\mathsf{b}^2/2)
\Gamma(-a_{jk}-\mathsf{b}^2/2) \Gamma(1-a_{jk}+\mathsf{b}^2/2)
}
{
\prod_{s_1,s_2=\pm} \Gamma(1/2+ s_1 \mathsf{m}+s_2 a_{jk})
}
\Bigg)^{1/2}
\nonumber\\
&
= \left( \prod_{j\neq k} \prod_{\pm} \frac{
\cos\pi(a_{jk}\pm \mathsf{m})
}{\sin\pi(a_{jk}\pm\mathsf{b}^2/2)}\right)^{1/2} \, .
 \end{align}
It follows that the Verlinde operator is given by
(\ref{Ver-SUN-star-thooft}).

\subsection{$SU(N)$ $N_\text{F}=2N$
}
\label{app:Toda-NF=2N}
We use the notation in Section \ref{app:Toda-2star}
for Toda theory.
For the $SU(N)$ theory with $N_\text{F}=2N$ fundamentals
corresponding to the sphere with two full and two semi-degenerate
punctures,
 we set the parameters as 
\begin{align}
&\alpha=i\hat a+Q\rho \nonumber\\
&\hat{\rm m}_2 = \left(\frac{Q}{2}+ i\hat{m}_2 \right) N\omega_{N-1} \, , \hspace{0.4cm} \hat{\rm m}_3^{\ast} = \left(\frac{Q}{2}+i \hat{m}_3 \right)N\omega_{1}
\,, \nonumber \\
&{\rm m}_1=Q\rho +i\tilde m_1\, ,  \hspace{2cm} {\rm m}_4^{\ast}=Q\rho
+i \tilde m_4^{\ast}\,, \\
&\tilde m_f=
\left\{
\begin{array}{ll}
\hat{m}_2 + i\tilde m_1\cdot h_i \hspace{1cm} & {\rm for}\ \  f=i=1,\ldots,N\,, \\
  \hat{m}_3 + i\tilde m_4 \cdot h_i \hspace{1cm} & {\rm for}\ \  f=N+i=N+1,
\ldots, 2N  \, , \nonumber\\
\end{array}
\right.
\end{align}
where $h_i^{\ast}:=-h_{N+1-i}\,$.
We slightly abuse notation; $\tilde m_1$ and $\tilde m_4$
differ from $\tilde m_{f=1}$ and $\tilde m_{f=4}$.
Similar remarks apply for $\mathsf m_f$ below.
The nonzero coweight term in the vev of 't Hooft operator on $S^4$ is
given as
\begin{align}
\langle T\rangle =& \int d\alpha \, C({\rm m}_4^{\ast},\hat{\rm m}_3^{\ast},\alpha)
C(2Q-\alpha, \hat{\rm m}_2, {\rm m}_1) \bar{\mathcal{F}(\alpha)}\sum_{l\neq k} H_{l,k}(\alpha) \mathcal{F}(\alpha- \mathsf b h_{lk})
\nonumber \\
&
\quad +\text{ zero-coweight terms}\, ,
\end{align}
where
\begin{align}
&H_{l,k}(\alpha)
=\pi^2 \frac{\prod_{j\neq l} \Gamma(i \mathsf b\hat a_{jl} ) 
\Gamma(\mathsf b Q+i\mathsf b \hat a_{jl}) 
\prod_{j\neq k} \Gamma(\mathsf b^2\delta_{jl} +i\mathsf b \hat a_{kj})  
\Gamma(\mathsf b Q+\mathsf b^2\delta_{lj} +i\mathsf b \hat a_{kj}) }{\prod_f \Gamma(\mathsf b Q/2 -i\mathsf b \hat a_l +i\mathsf b \tilde m_f)  
\Gamma(\mathsf b Q/2 +i\mathsf b \hat a_k- i\mathsf b \tilde m_f)} 
\end{align}
and $h_{lk}=h_l-h_k\,$, $\widehat{a}_i\equiv \widehat{a}\cdot h_i$ and $\widehat{a}_{ij}\equiv \widehat{a}\cdot (h_i-h_j)\, $.
In order to relate this to the vev of the 't Hooft operator on $S^1\times \mathbb R^3$, we introduce a slightly different parametrization
\begin{align}
&\alpha=Q \rho -\frac{a}{\mathsf{b}}\,, \nonumber\\
&\alpha_2= \left(\frac{Q}{2}-\frac{\mathsf m_2}{\mathsf{b}} \right) N\omega_{N-1} \, , \hspace{0.4cm} \alpha_3 = \left(\frac{Q}{2} -\frac{\mathsf m_3}{\mathsf{b}} \right)N\omega_{1}\,, \nonumber \\
&\alpha_1=Q\rho-\frac{\mathsf m_1}{\mathsf{b}}\, ,  \hspace{2.2cm} 
\alpha_4=Q\rho-\frac{\mathsf m_4}{\mathsf{b}} \,,\\
&\mathsf{m}_f=
\left\{
\begin{array}{ll}
\mathsf m_2  + \mathsf m_1 \cdot h_i \equiv \mathsf m_2  + \mathsf m_{1,i}  \hspace{1cm} & {\rm for}\ \  f=i \,,\\
\mathsf m_3  - \mathsf m_4  \cdot h_i  \equiv \mathsf m_3  - \mathsf m_{4,i}  & {\rm for}\ \  f=N+i\,. \nonumber\\
\end{array}
\right.
\end{align}
We define $(h_{lk})_{ij}:=(h_l-h_k)\cdot(h_i-h_j)=\delta_{li}-\delta_{lj}-\delta_{ki}+\delta_{kj}$ and $\tilde{\Upsilon}^{(2)}:= \frac{\Upsilon(x+2\mathsf{b})}{\Upsilon(x)}=\frac{\Gamma(\mathsf{b}x) \Gamma(\mathsf{b}x+\mathsf{b}^2)}{\Gamma(1-\mathsf{b}x)\Gamma(1-\mathsf{b}x-\mathsf{b}^2)}\, $, which is analogous to $\tilde{\Upsilon}(x):=\frac{\Upsilon(x+\mathsf b)}{\Upsilon(x)}=\frac{\Gamma(\mathsf{b}x)}{\Gamma(1-\mathsf{b}x)}$.
Let us calculate
\begin{align}
&Z_{l,k}(\alpha):=\left(  \frac{C(\alpha_4, \alpha_3, \alpha+\mathsf{b} h_{lk}/2 )
C(2Q \rho-\alpha-\mathsf{b} h_{lk}/2 , \alpha_2, \alpha_1)  }{ C(\alpha_4, \alpha_3, \alpha-\mathsf{b} h_{lk}/2 )
C(2Q \rho-\alpha + \mathsf{b} h_{lk}/2 , \alpha_2, \alpha_1 ) } \right)^{1/2} H_{l,k}(\alpha+\mathsf{b} h_{lk}/2) \nonumber \\
&
=\Bigg( \frac{C^{(2)}(Q \rho-\mathsf m_4/\mathsf{b},\ Q\rho-a/\mathsf{b}+\mathsf{b}h_{lk}/2,\ \kappa=Q\rho/2-\mathsf m_3/\mathsf{b} ) }{C^{(2)}(Q\rho-\mathsf m_4/\mathsf{b},
\ Q\rho-a/\mathsf{b}-\mathsf{b}h_{lk}/2,\ \kappa=Q\rho/2-\mathsf m_3/\mathsf{b} )} \nonumber\\
&\ \ \ \ \ \ \  \times
\frac{C^{(1)}(Q \rho-\mathsf m_1/\mathsf{b},\ Q\rho+a/\mathsf{b}-\mathsf{b}h_{lk}/2,\ \kappa=Q/2-\mathsf m_2/\mathsf{b} ) }{C^{(1)}(Q\rho-\mathsf m_1/\mathsf{b},
\ Q\rho+a/\mathsf{b}+\mathsf{b}h_{lk}/2,\ \kappa=Q/2-\mathsf m_2/\mathsf{b} )} \Bigg)^{1/2} H_{l,k}(\alpha+\mathsf{b} h_{lk}/2)
\nonumber
 \end{align}
 \begin{align}
&=
\pi^2
\Bigg(
\prod_{i<j}
\frac{
\Upsilon (a_{ij}/\mathsf{b} -\mathsf{b}(h_{lk})_{ij}/2 )
}{
\Upsilon (a_{ij}/\mathsf{b} +\mathsf{b}(h_{lk})_{ij}/2 )
}
  \prod_{i,j} 
 \frac{
 \Upsilon(Q/2 -\mathsf m_3/\mathsf{b} +\mathsf m_{4,i}/\mathsf{b} +a_j/\mathsf{b} +\mathsf{b}(\delta_{lj}-\delta_{kj})/2) 
   }{
  \Upsilon(Q/2 -\mathsf m_3/\mathsf{b} +\mathsf m_{4,i}/\mathsf{b} +a_j/\mathsf{b} -\mathsf{b}(\delta_{lj}-\delta_{kj})/2)
    }
    \nonumber\\
    &\ \ \ \ \ \ \  \times 
    \prod_{i<j}
\frac{
\Upsilon ( - a_{ij}/\mathsf{b} +\mathsf{b}(h_{lk})_{ij}/2 )
}{
\Upsilon (-a_{ij}/\mathsf{b} -\mathsf{b}(h_{lk})_{ij}/2 )
}
\prod_{i,j}
     \frac{
  \Upsilon(Q/2 -\mathsf m_2/\mathsf{b} - \mathsf m_{1,i}/\mathsf{b} +a_j/\mathsf{b} +\mathsf{b}(\delta_{lj}-\delta_{kj})/2 ) 
  }{
\Upsilon(Q/2 -\mathsf m_2/\mathsf{b} -\mathsf m_{1,i}/\mathsf{b} +a_j/\mathsf{b} -\mathsf{b}(\delta_{lj}-\delta_{kj})/2) 
}
\Bigg)^{1/2}
 \nonumber\\
&\ \ \ \ \quad \times  \frac{
\prod_{j\neq l} \Gamma(-a_{jl}-\mathsf{b}^2(1+\delta_{kj})/2 )
 \Gamma(\mathsf b Q-a_{jl}-\mathsf{b}^2(1+\delta_{kj})/2) 
}{
\prod_f \Gamma \left( \mathsf{b}Q/2+a_l-\mathsf{b}^2/2 -\mathsf{m}_f \right) \Gamma \left( \mathsf{b} Q/2-a_k-\mathsf{b}^2/2 + \mathsf{m}_f \right) }
\nonumber\\
&\ \ \ \ \ \ \ \times \prod_{j\neq k}
\Gamma(-a_{kj}+\mathsf{b}^2(\delta_{lj}-1)/2) \Gamma(\mathsf b Q-a_{kj}+\mathsf{b}^2(\delta_{jl}-1)/2 ) )  \nonumber
\\
&=
\pi^2
\Bigg(
\prod_{i\neq j}
\frac{
\Upsilon (a_{ij}/\mathsf{b} -\mathsf{b}(h_{lk})_{ij}/2 )
}{
\Upsilon (a_{ij}/\mathsf{b} +\mathsf{b}(h_{lk})_{ij}/2 )
}
  \prod_{i,j} 
 \frac{
 \Upsilon(Q/2 -\mathsf m_3/\mathsf{b} +\mathsf m_{4,i}/\mathsf{b} +a_j/\mathsf{b} +\mathsf{b}(\delta_{lj}-\delta_{kj})/2) 
   }{
  \Upsilon(Q/2 -\mathsf m_3/\mathsf{b} +\mathsf m_{4,i}/\mathsf{b} +a_j/\mathsf{b} -\mathsf{b}(\delta_{lj}-\delta_{kj})/2)
    }
    \nonumber\\
    &\ \ \ \ \ \ \  \times 
   \prod_{i,j}
     \frac{
  \Upsilon(Q/2 -\mathsf m_2/\mathsf{b} -\mathsf m_{1,i}/\mathsf{b} +a_j/\mathsf{b} +\mathsf{b}(\delta_{lj}-\delta_{kj})/2 ) 
  }{
\Upsilon(Q/2 -\mathsf m_2/\mathsf{b} -\mathsf m_{1,i}/\mathsf{b} +a_j/\mathsf{b} -\mathsf{b}(\delta_{lj}-\delta_{kj})/2) 
}
\Bigg)^{1/2}
 \nonumber\\
 &\quad  \ \ \ \ \times \frac{\Gamma (-a_{kl} -\mathsf{b}^2) \Gamma(\mathsf b Q-a_{kl}-\mathsf{b}^2) \Gamma(-a_{kl}) \Gamma(\mathsf b Q-a_{kl})}
 {
 \prod_f \Gamma \left( 1/2+a_l-\mathsf{m}_f \right) \Gamma \left( 1/2-a_k + \mathsf{m}_f \right) 
 } \nonumber\\
 &\ \ \ \ \quad \times \prod_{j\neq l,k} \Gamma(-a_{jl} -\mathsf{b}^2/2)
 \Gamma(\mathsf b Q-a_{jl}-\mathsf{b}^2/2)
 \Gamma(-a_{kj}-\mathsf{b}^2/2) 
\Gamma(\mathsf b Q-a_{kj}-\mathsf{b}^2/2) 
\nonumber
\\
&=
\pi^2\Bigg(
\frac{
\Upsilon (a_{lk}/\mathsf{b} -\mathsf{b} )
\Upsilon (a_{kl}/\mathsf{b} +\mathsf{b} )
}{
\Upsilon (a_{lk}/\mathsf{b} +\mathsf{b} )
\Upsilon (a_{kl}/\mathsf{b} -\mathsf{b} )
} 
\nonumber\\
&\ \ \ \ \   \times  \prod_{i\neq l,k}
\frac{
\Upsilon (a_{il}/\mathsf{b} +\mathsf{b}/2 )
\Upsilon (a_{ik}/\mathsf{b} -\mathsf{b}/2 )
}{
\Upsilon (a_{il}/\mathsf{b} -\mathsf{b}/2 )
\Upsilon (a_{ik}/\mathsf{b} +\mathsf{b}/2 )
} 
\prod_{j\neq l,k}
\frac{
\Upsilon (a_{lj}/\mathsf{b} -\mathsf{b}/2 )
\Upsilon (a_{kj}/\mathsf{b} +\mathsf{b}/2 )
}{
\Upsilon (a_{lj}/\mathsf{b} +\mathsf{b}/2 )
\Upsilon (a_{kj}/\mathsf{b} -\mathsf{b}/2 )
} 
\nonumber\\
&\  \ \ \ \ \times
 \prod_{i}  
\frac{ 
\Upsilon\left(Q/2 -\mathsf m_3/\mathsf{b} + \mathsf m_{4,i}/\mathsf{b} +a_l/\mathsf{b} +\mathsf{b}/2\right)  \Upsilon\left(Q/2 -\mathsf m_3/\mathsf{b} + \mathsf m_{4,i}/\mathsf{b} +a_k/\mathsf{b} -\mathsf{b}/2\right) 
}{
 \Upsilon\left(Q/2 -\mathsf m_3/\mathsf{b} +\mathsf m_{4,i}/\mathsf{b} +a_l/\mathsf{b} -\mathsf{b}/2\right)
 \Upsilon\left(Q/2 -\mathsf m_3/\mathsf{b} +\mathsf m_{4,i}/\mathsf{b} +a_k/\mathsf{b} +\mathsf{b}/2\right)
 } 
 \nonumber\\
&\   \ \  \ \ \times \frac{ \Upsilon\left(Q/2 -\mathsf m_2/\mathsf{b} -\mathsf m_{1,i}/\mathsf{b} +a_l/\mathsf{b} +\mathsf{b}/2\right)  \Upsilon\left(Q/2 -\mathsf m_2/\mathsf{b} -\mathsf m_{1,i}/\mathsf{b} +a_k/\mathsf{b} -\mathsf{b}/2\right) 
}{
 \Upsilon\left(Q/2 -\mathsf m_2/\mathsf{b} -\mathsf m_{1,i}/\mathsf{b} +a_l/\mathsf{b} -\mathsf{b}/2\right)
 \Upsilon\left(Q/2 -\mathsf m_2/\mathsf{b} -\mathsf m_{1,i}/\mathsf{b} +a_k/\mathsf{b} +\mathsf{b}/2\right) 
 } \Bigg)^{1/2} \nonumber\\
 &\ \ \ \ \ \times   \frac{\Gamma (-a_{kl} -\mathsf{b}^2) \Gamma(1-a_{kl}) \Gamma(-a_{kl}) \Gamma(\mathsf b\mathsf Q-a_{kl})}
 {
 \prod_f \Gamma \left( 1/2+a_l-\mathsf{m}_f \right) \Gamma \left( 1/2-a_k + \mathsf{m}_f \right) 
 }
 \prod_{j\neq l,k}
\prod_\pm
 \Gamma(-a_{jl} \pm \mathsf{b}^2/2)
\Gamma(-a_{kj}\pm\mathsf{b}^2/2) 
 \nonumber \\
&
=
\pi^2
\Bigg( 
\frac{
\tilde{\Upsilon}^{(2)}(a_{kl}/\mathsf{b}-\mathsf{b})
}{
\tilde{\Upsilon}^{(2)}(a_{lk}/\mathsf{b}-\mathsf{b})
}
    \ \ \prod_{i\neq l,k}
\frac{
\tilde{\Upsilon} (a_{il}/\mathsf{b} -\mathsf{b}/2 )
}{
\tilde{\Upsilon} (a_{ik}/\mathsf{b} -\mathsf{b}/2 )
} 
\prod_{j\neq l,k}
\frac{
\tilde{\Upsilon} (a_{kj}/\mathsf{b} -\mathsf{b}/2 )
}{
\tilde{\Upsilon} (a_{lj}/\mathsf{b} -\mathsf{b}/2 )
} 
\nonumber\\
&  \ \ \ \times \prod_{i} 
\frac{ \tilde{\Upsilon}\left(Q/2 -\mathsf m_3/\mathsf{b} + \mathsf m_{4,i}/\mathsf{b} +a_l/\mathsf{b} -\mathsf{b}/2\right)  
}{
 \tilde{\Upsilon}\left(Q/2 -\mathsf m_3/\mathsf{b} + \mathsf m_{4,i}/\mathsf{b} +a_k/\mathsf{b} -\mathsf{b}/2\right) 
 }
 \frac{ \tilde{\Upsilon}\left(Q/2 -\mathsf m_2/\mathsf{b} -\mathsf m_{1,i}/\mathsf{b} +a_l/\mathsf{b} -\mathsf{b}/2\right)  
}{
 \tilde{\Upsilon}\left(Q/2 -\mathsf m_2/\mathsf{b} -\mathsf m_{1,i}/\mathsf{b} +a_k/\mathsf{b} -\mathsf{b}/2\right) 
 } \Bigg)^{1/2} \nonumber\\
 &\ \ \ \ \  \times \frac{\Gamma (-a_{kl} -\mathsf{b}^2) \Gamma(1-a_{kl}) \Gamma(-a_{kl}) \Gamma(\mathsf b Q-a_{kl})}
 {
 \prod_f \Gamma \left( 1/2+a_l -\mathsf{m}_f \right) \Gamma \left( 1/2-a_k+ \mathsf{m}_f \right) 
 }
  \prod_{j\neq l,k}
\prod_\pm \Gamma(-a_{jl} \pm\mathsf{b}^2/2) 
\Gamma(-a_{kj}\pm\mathsf{b}^2/2) 
\nonumber
 \end{align}
 \begin{align}
 &
=
\pi^2
  \Bigg(
 \frac{
 \Gamma(a_{kl}-\mathsf{b}^2) \Gamma(a_{kl})
   \Gamma(1-a_{lk}+\mathsf{b}^2)
   \Gamma(1-a_{lk})
 }{
  \Gamma(1-a_{kl}+\mathsf{b}^2)
   \Gamma(1-a_{kl})
      \Gamma(a_{lk}-\mathsf{b}^2) \Gamma(a_{lk})
 }
\prod_{f}  
\frac{ \Gamma\left(1/2 -\mathsf{m}_f +a_l \right)  \Gamma\left(1/2 +\mathsf{m}_f -a_k \right) 
}{
\Gamma\left(1/2 +\mathsf{m}_f -a_l \right)  \Gamma\left(1/2 -\mathsf{m}_f +a_k \right) 
 }
 \nonumber\\
 &\ \ \ \ \ \times \prod_{i \neq l,k}  \frac{
 \Gamma(a_{il}-\mathsf{b}^2/2) \Gamma(1-a_{ik}+\mathsf{b}^2/2)
  \Gamma(a_{ki}-\mathsf{b}^2/2) \Gamma(1-a_{li}+\mathsf{b}^2/2)
 }{
 \Gamma(1-a_{il}+\mathsf{b}^2/2) \Gamma(a_{ik}-\mathsf{b}^2/2)
  \Gamma(1-a_{ki}+\mathsf{b}^2/2) \Gamma(a_{li}-\mathsf{b}^2/2)
  }
  \Bigg)^{1/2}
 \nonumber\\
 &\ \  \  \ \ \times 
\frac{\Gamma (-a_{kl} -\mathsf{b}^2) \Gamma(1-a_{kl}) \Gamma(-a_{kl}) \Gamma(\mathsf b Q-a_{kl})}
 {
 \prod_f \Gamma \left( 1/2+a_l -\mathsf{m}_f \right) \Gamma \left( 1/2-a_k + \mathsf{m}_f \right) 
 }
 \nonumber\\
 &\ \ \ \ \ \times
  \prod_{j\neq l,k}
 \Gamma(-a_{jl} -\mathsf{b}^2/2) \Gamma(1-a_{jl}+\mathsf{b}^2/2) \Gamma(-a_{kj}-\mathsf{b}^2/2) \Gamma(1-a_{kj}+\mathsf{b}^2/2) 
\nonumber
\\ &
 =
\frac{\prod_f [ \cos\pi(a_l-\mathsf{m}_f) \cos\pi(a_k-\mathsf{m}_f) ]^{\frac 12}
 }{
{\displaystyle \prod_{\pm}}
 \Big[
\sin\pi(\pm a_{lk}) \sin \pi (\pm a_{lk} -\mathsf{b}^2) 
{\displaystyle\prod_{j\neq l,k}}
 \sin\pi(\pm a_{jl}-{\mathsf{b}^2}/2)
\sin\pi(\pm a_{jk} -{\mathsf{b}^2}/2)
\Big]^{\frac 12}
 } \, .
\label{Z1loopSQCDToda}
\end{align}
This gives the one-loop factors for the terms with
non-zero coweights  in (\ref{minimal-thooft-from-liouville})
and (\ref{SUN-SQCD}).
The terms with zero coweight 
given in \cite{Gomis:2010kv}
appear in
(\ref{minimal-thooft-from-liouville})
and (\ref{SUN-SQCD})
without modification because their expressions
are independent of the normalization of the conformal block.

\section{$SU(2)$ holonomies on the four-punctured sphere}
\label{sec:cl-holo}

The Hitchin moduli space on the four-punctured sphere
as a complex manifold
is described by
 four $SL(2,\mathbb C)$ holonomy matrices $M_e$ ($e=1,\ldots, 4$)
satisfying $  M_1 M_2 M_3 M_4=1 $ up to conjugation
with fixed conjugacy classes for $M_e$.
We set
\begin{eqnarray}
  W=\Tr M_1 M_2,\,,~~~T= \Tr M_1 M_4\,,~~~~ D=\Tr M_1 M_3\,.
\end{eqnarray}
They satisfy the identity
\begin{equation}
  \begin{aligned}
0=&  D^2+ (W T-\Tr M_1 \Tr M_3- \Tr M_2 \Tr M_4) D
\\
&
+(W-\Tr M_1 \Tr M_2)(W-\Tr M_3 \Tr M_4)
+(T-\Tr M_2 \Tr M_3)(T-\Tr M_1 \Tr M_4)
\\
&
\quad\quad
+\sum_{e=1}^4 (\Tr M_e)^2
-\prod_{e=1}^4 \Tr M_e
-4\,.
  \end{aligned}
\label{quad-eq-D}
\end{equation}
We expect that the quantities $W$, $T$, and $D$
correspond to Wilson, 't Hooft, and dyonic operators \cite{Drukker:2009tz}
in the $SU(2)$ theory with $N_\text{F}=4$ fundamental hypermultiplets.
Anticipating  a match with the results of localization,
we make an ansatz 
\begin{eqnarray}
  W=x+1/x\,,
\quad
T=-(y^2+1/y^2) Z(x)+ C_1\,,
\quad
D=\left( x y^2+\frac 1{x y^2}\right)
Z(x)
+C_2\,,
\end{eqnarray}
where $Z(x)$ is a function of $x\equiv e^{2\pi i a}$, and $C_1$ and $C_2$ are independent of  $x$ and $y\equiv e^{2\pi i b}$.
The ansatz is motivated by the localization computation,
where we expect a common one-loop factor $Z(x)$
for $T$ and $D$.
Let us substitute these into (\ref{quad-eq-D})
and organize the equation in powers of $y$.
The minus sign in the first term in $T$ was
put by hand to ensure that there are no terms
proportional to $y^4$ or $1/y^4$.
We can choose $C_1$ and $C_2$ such that
terms proportional to $y^2$ and $1/y^2$
also vanish.
Then $y$ drops out of the equation (\ref{quad-eq-D}), which 
can then be solved for $Z$.
The result is
\begin{equation}
  \begin{aligned}
Z=&
4 \frac{
\prod_\pm 
\prod_{ f=1}^4
\sin^{1/2}\pi( a\pm m_f)
}{\sin^2 2\pi a}\,,
\\
C_1=&
2\frac{\prod_{f=1}^4\cos \pi m_f }{\cos^2 \pi a}
+
2\frac{\prod_{f=1}^4\sin\pi m_f}{\sin^2 \pi a}
\,,\\
C_2=&
2\frac{\prod_{f=1}^4\cos \pi m_f}{\cos^2 \pi a}
-
2\frac{\prod_{f=1}^4\sin\pi m_f}{\sin^2 \pi a}
\,,
  \end{aligned}
\end{equation}
where $\Tr M_e= e^{2\pi i\gamma_e}+e^{-2\pi i\gamma_e}$
and
\begin{equation}
2\gamma_1=m_1-m_2\,,
\quad
2\gamma_2=m_1+m_2\,,
\quad
2\gamma_3=m_3+m_4\,,
\quad
2\gamma_4=m_3-m_4\,.
\end{equation}
Then $-T/4$ is precisely the $\lambda=0$ limit of
(\ref{minimal-thooft-from-liouville}).\footnote{It would be nice to understand the origin of several minus signs
that seem unavoidable.
}
These expressions for $W$, $T$, and $D$ were given
in \cite{Nekrasov:2011bc} as the definition of Darboux coordinates
$a$ and $b$.

\bibliography{refs}

\end{document}